\documentclass[reprint,aps,superscriptaddress,floatfix]{revtex4-2}

\usepackage[english]{babel}
\usepackage[utf8]{inputenc}
\usepackage{mathtools,amssymb}
\usepackage{mleftright}\mleftright
\usepackage{array}
\usepackage{verbatim}
\usepackage[export]{adjustbox}

\usepackage[unicode]{hyperref}
\usepackage{graphicx}
\usepackage{setspace}
\usepackage{xcolor}
\usepackage{url}

\hypersetup{
  bookmarksopen,
  colorlinks=true,
  linkcolor=[rgb]{0,0,0.8},
  citecolor=[rgb]{0.5,0.5,0.5},
  urlcolor=[rgb]{0,0,0.8},
}
\usepackage[capitalise,nameinlink,nosort]{cleveref}

\usepackage[range-phrase=--,range-units=single,per-mode=symbol]{siunitx}
\DeclareSIUnit\field{field}

\usepackage{bm}
\newcommand{\nonum}{\nonumber\\}
\newcommand{\ve}[1]{\bm{\mathbf{#1}}}

\newcommand{\dd}{\mathrm{d}}

\newcommand{\ee}{\mathrm{e}}
\newcommand{\ii}{\mathrm{i}}
\newcommand{\Tr}{\mathrm{Tr}}

\newcommand{\sig}{\mathop{\mathrm{sig}}\nolimits}

\newcommand{\erf}{\mathop{\mathrm{erf}}\nolimits}

\newcommand{\cond}[1]{\quad\textrm{ #1 }\quad}

\newcommand{\parpar}[2]{\frac{\partial #1}{\partial #2}}

\newcommand{\ar}{{\alpha r}}
\newcommand{\arp}{{\alpha' r'}}
\newcommand{\mn}{{\mu\nu}}
\newcommand{\mo}{{\mu\omega}}
\newcommand{\no}{{\nu\omega}}
\newcommand{\rs}{{\rho\sigma}}

\newcommand{\erfy}{\erf\frac{y}{\sqrt{2}}}
\newcommand{\expy}{\ee^{-y^2/2}}
\newcommand{\expyy}{\sqrt\frac{2}{\pi}y\mspace{1.5mu}\expy}

\newcommand{\mcH}{\mathcal{H}}
\newcommand{\rms}{\textrm{s}}
\newcommand{\rmc}{\textrm{c}}
\newcommand{\rmd}{\textrm{d}}

\DeclareFontFamily{OMX}{MnSymbolE}{}
\DeclareFontShape{OMX}{MnSymbolE}{m}{n}{
    <-6>  MnSymbolE5
   <6-7>  MnSymbolE6
   <7-8>  MnSymbolE7
   <8-9>  MnSymbolE8
   <9-10> MnSymbolE9
  <10-12> MnSymbolE10
  <12->   MnSymbolE12}{}
\DeclareSymbolFont{mnlargesymbols}{OMX}{MnSymbolE}{m}{n}
\SetSymbolFont{mnlargesymbols}{bold}{OMX}{MnSymbolE}{b}{n}
\DeclareMathDelimiter{\llangle}{\mathopen}{mnlargesymbols}{'164}{mnlargesymbols}{'164}
\DeclareMathDelimiter{\rrangle}{\mathclose}{mnlargesymbols}{'171}{mnlargesymbols}{'171}

\newcommand{\add}[1]{#1}

\allowdisplaybreaks

\begin{document}

\title{A Hopfield-like network with complementary encodings of memories}

\author{Louis Kang}
\email[Corresponding author: ]{louis.kang@riken.jp}
\affiliation{Neural Circuits and Computations Unit, RIKEN Center for Brain Science}
\affiliation{Graduate School of Informatics, Kyoto University}

\author{Taro Toyoizumi}
\affiliation{Laboratory for Neural Computation and Adaptation, RIKEN Center for Brain Science}
\affiliation{Graduate School of Information Science and Technology, University of Tokyo}

\begin{abstract}
  We present a Hopfield-like autoassociative network for memories representing examples of concepts. Each memory is encoded by two activity patterns with complementary properties. The first is dense and correlated across examples within concepts, and the second is sparse and exhibits no correlation among examples. The network stores each memory as a linear combination of its encodings. During retrieval, the network recovers sparse or dense patterns with a high or low activity threshold, respectively. As more memories are stored, the dense representation at low threshold shifts from examples to concepts, which are learned from accumulating common example features. Meanwhile, the sparse representation at high threshold maintains distinctions between examples due to the high capacity of sparse, decorrelated patterns. Thus, a single network can retrieve memories at both example and concept scales and perform heteroassociation between them. We obtain our results by deriving macroscopic mean-field equations that yield capacity formulas for sparse examples, dense examples, and dense concepts. We also perform network simulations that verify our theoretical results and explicitly demonstrate the capabilities of the network.
\end{abstract}

\maketitle

\section{Introduction}

Autoassociation is the ability for a network to store patterns of activity and to retrieve complete patterns when presented with incomplete cues. Autoassociative networks are widely used as models for neural phenomena, such as episodic memory \cite{McNaughton.1987, O'Reilly.2001, Rolls.20069yq}, and also have applications in machine learning \cite{Hopfield.1985, Barra.2012}. It is well-known that properties of the stored patterns can influence the computational capabilities of the network. Sparse patterns, in which a small fraction of the neurons is active, can be stored at higher capacity compared to dense patterns \cite{Marr.1971, Tsodyks.1988, Kanerva.1988, Nadal.1990, Rolls.1990, Treves.1991, Palm.2013}. Correlated patterns can be merged by the network to represent shared attributes \cite{Fontanari.1990, Stariolo.1992, Dominguez.1998}. Previous autoassociation models have largely considered the storage of patterns with a single set of statistics, which requires trade-offs among computational features. \add{For example, the ability to learn categories with correlated patterns may be desired, but correlations decrease the capacity for retrieving patterns distinctly.}

We consider the possibility that a network can store two types of patterns with different properties, and thus, different computational roles. This idea is inspired by the architecture of the hippocampus in mammalian brains \cite{Kang.2023b}. The hippocampal subfield CA3 is the presumptive autoassociative network that stores memories of our daily experiences \cite{Marr.1971, McNaughton.1987}, and it receives sensory information from two parallel pathways with complementary properties \cite{Amaral.2006}. The mossy fibers present sparser, decorrelated patterns to CA3 for storage, and the perforant path presents denser, correlated patterns. Both pathways originate from the same upstream region, the entorhinal cortex, so they presumably encode the same sensory experiences. \add{However, based on theoretical studies described above, their computational capabilities may differ. We expect the denser, correlated patterns to build representations of \emph{concepts} through the accumulation of \emph{examples}; meanwhile, we expect the sparser, decorrelated patterns to represent distinct examples at high capacity.} We wish to explore whether an autoassociative network can store and retrieve memory encodings from each pathway. \add{Doing so could enable information representation at different scales, enabling the network to simultaneously discriminate between examples and generalize across them.}

To address this aim, we implement a Hopfield-like network \cite{Hopfield.1982} that stores memories, each of which is an example $\mu$ of a concept $\nu$. Each example is encoded as both a sparse pattern $\ve\xi_\mn$ and a dense pattern $\ve\psi_\mn$. The former is generated independently and exhibits no correlation with other sparsely encoded examples. The latter is generated from a dense encoding $\ve\psi_\mu$ of the concept $\mu$ with correlations among examples within the same concept. The model is defined in \cref{sec:model}, along with an outline of the derivation of its mean-field equations.

In \cref{sec:capacities}, we present our major results regarding pattern retrieval. We can use a high or low activity threshold to retrieve sparse or dense patterns, respectively. The network has a high capacity for sparse examples $\ve\xi_\mn$ and a low capacity for dense examples $\ve\psi_\mn$. As the number of examples stored increases beyond the dense example capacity, a critical load is reached above which the network instead retrieves dense concepts $\ve\psi_\mu$. This critical load can be smaller than the sparse example capacity, which means that the network can recover both $\ve\xi_\mn$\!'s as distinct memories and $\ve\psi_\mu$\!'s as generalizations across them.

In \cref{sec:hetero}, we show that the network can perform heteroassociation between sparse and dense encodings of the same memory. Their respective energies can predict regimes in which heteroassociation is possible. We discuss our results and their significance in \cref{sec:discussion}. Mean-field equations governing network behavior are derived in \cref{sec:mean}, and capacity formulas for $\ve\xi_\mn$, $\ve\psi_\mn$, and $\ve\psi_\mu$ are derived in \cref{sec:se,sec:de,sec:dc}.

\section{\label{sec:model}The model}

\subsection{Patterns and architecture}

We consider a Hopfield network with neurons $i = 1, \ldots, N$ that are either inactive ($S_i = 0$) or active ($S_i = 1$). The network stores $\nu = 1, \ldots, s$ examples for each of $\mu = 1, \ldots, p$ concepts. The concept load per neuron is $\alpha = p/N$. Examples are encoded both sparsely as $\ve\xi_\mn$ and densely as $\ve\psi_\mn$. Following Ref.~\onlinecite{Tsodyks.1988}, sparse examples are generated independently with \add{density} $a$:
\begin{equation}
  \xi^i_\mn =
    \begin{cases} 0 & \textrm{with probability } 1-a \\
                  1 & \textrm{with probability } a.  \end{cases}
  \label{eq:model-xi}
\end{equation}
\add{While the term \emph{sparsity} has also been used in the literature, we use \emph{density} for $a$ because higher $a$ implies lower sparsity.} Following Ref.~\onlinecite{Fontanari.1990}, dense examples within a concept are correlated in the following way. Each concept corresponds to a dense pattern $\ve\psi_\mu$, generated independently with \add{density} $\frac{1}{2}$:
\begin{equation}
  \psi^i_\mu =
      \begin{cases} 0 & \textrm{with probability } \frac{1}{2}  \\
                    1 & \textrm{with probability } \frac{1}{2}. \end{cases}
  \label{eq:model-psimu}
\end{equation}
Dense examples are then generated from these concepts, with the correlation parameter $c > 0$ controlling the likelihood that example patterns match their concept:
\begin{equation}
  \psi^i_\mn =
    \begin{cases} \psi^i_\mu   & \textrm{with probability } \frac{1+c}{2}  \\
                  1-\psi^i_\mu & \textrm{with probability } \frac{1-c}{2}. \end{cases}
  \label{eq:model-psi}
\end{equation}
The average Pearson correlation coefficient between $\ve\psi_\mn$ and $\ve\psi_\mu$ is $c$, and that between $\ve\psi_\mn$ and $\ve\psi_\mo$ for $\nu\neq\omega$ is $c^2$. The average overlaps are
\begin{align}
  \bigl\langle \psi^i_\mn \psi^i_\mu \bigr\rangle &= \tfrac{1}{4} + \tfrac{c}{4} \nonum
  \bigl\langle \psi^i_\mn \psi^i_\mo \bigr\rangle &= \tfrac{1}{4} + \tfrac{c^2}{4},
  \label{eq:model-overlaps}
\end{align}
where angle brackets indicate averaging over patterns.

During storage, the parameter $2\gamma$ sets the relative strength of dense encodings compared to sparse encodings. The factor of 2 is for theoretical convenience. Linear combinations of $\ve\xi_\mn$ and $\ve\psi_\mn$ are stored in a Hopfield-like fashion with symmetric synaptic weights
\begin{alignat}{3}
  J_{ij} &={}& \frac{1}{N} \sum_\mn & \Bigl[(1 - 2\gamma)\bigl(\xi^i_\mn - a\bigr) + 2\gamma\bigl(\psi^i_\mn - \tfrac{1}{2}\bigr) \Bigr] \nonum
  &&& {}\times \Bigl[(1 - 2\gamma)\bigl(\xi^j_\mn - a\bigr) + 2\gamma\bigl(\psi^j_\mn - \tfrac{1}{2}\bigr) \Bigr] \nonum
  &\mathrlap{ {}={} \frac{1}{N} \sum_\mn (\eta^i_\mn + \zeta^i_\mn) (\eta^j_\mn + \zeta^j_\mn) }
  \label{eq:model-J}
\end{alignat}
for $i \neq j$, and $J_{ii} = 0$. The second expression uses rescaled sparse and dense patterns
\begin{alignat}{2}
  \eta^i_\mn &{}\equiv{}& (1 - 2\gamma)&\bigl(\xi^i_\mn - a\bigr) \nonum
  \zeta^i_\mn &{}\equiv{}& 2\gamma&\bigl(\psi^i_\mn - \tfrac{1}{2}\bigr).
  \label{eq:model-rescaled}
\end{alignat}

After initializing the network with a cue, neurons are asynchronously and stochastically updated via Glauber dynamics \cite{Amit.1985mb}. That is, at each timestep $t$, one neuron $i$ is randomly selected, and the probability that it becomes active is given by
\begin{equation}
  P[S_i(t+1) = 1] = \frac{1}{1 + \exp\bigl\{-\beta \bigl[\sum_j J_{ij} S_j(t) - \theta\bigr]\bigr\}}.
  \label{eq:model-glauber}
\end{equation}
Thus, activation likely occurs when the total synaptic input $\sum_j J_{ij} S_j(t)$ is greater than the activity threshold $\theta$. The inverse temperature $\beta = 1/T$ sets the width of the threshold, with $\beta \rightarrow 0$ corresponding to chance-level activation and $\beta \rightarrow \infty$ corresponding to a strict, deterministic threshold. We shall see that $\theta$ plays a key role in selecting between sparse and dense patterns; a higher $\theta$ suppresses activity and favors recovery of sparse patterns, and vice versa for lower $\theta$ and dense patterns.

\subsection{Overview of mean-field equations}

Network behavior in the mean-field limit is governed by a set of equations relating macroscopic order parameters to one another. Their complete derivation following Refs. \onlinecite{Tsodyks.1988}, \onlinecite{Fontanari.1990}, \onlinecite{Amit.1985mb}, and \onlinecite{Hertz.2018} is provided in \cref{sec:mean}, but we will outline our approach here. The first task is calculating the replica partition function $\langle Z^n \rangle$, where the angle brackets indicate averaging over rescaled patterns $\ve\eta_\mn$ and $\ve\zeta_\mn$ and $n$ is the number of replica systems. By introducing auxiliary fields via Hubbard-Stratonovich transformations and integrating over interactions with off-target patterns, we obtain
\begin{align}
  \langle Z^n \rangle \propto \int &\biggl[\prod_{\nu\rho} \dd m^\rho_{1\nu} \biggl(\frac{\beta N}{2\pi}\biggr)^{\!\frac{1}{2}}\biggr] \biggl[\prod_\rs \dd q^\rs\,\dd r^\rs\biggr] \nonum
  & {}\times \exp[-\beta N f],
\end{align}
where $\rho$ and $\sigma$ are replica indices, $m^\rho_{1\nu}$, $r^\rs$, and $q^\rs$ are order parameters, and
\begin{widetext}
\begin{align}
  f ={}& \frac{1}{2} \sum_{\nu\rho} (m^\rho_{1\nu})^2 + \frac{\alpha}{2\beta} \Tr\log\bigl[ \delta_\no\delta^\rs - \beta \Gamma^2\bigl((1-\kappa^2)\delta_\no + \kappa^2\bigr) q^\rs \bigr] + \frac{\beta\alpha}{2} \sum_\rs q^\rs r^\rs \nonum
  &{}- \frac{1}{\beta} \Biggl\langle \log\Tr_S \exp\Biggl\{ \beta \biggl[ \sum_{\nu\rho} m^\rho_{1\nu} \chi_{1\nu} S^\rho - \biggl(\theta + \frac{\alpha s \Gamma^2}{2}\biggr) \sum_\rho S^\rho + \frac{\beta\alpha}{2} \sum_\rs r^\rs S^\rho S^\sigma \biggr] \Biggr\} \Biggr\rangle.
  \label{eq:model-f}
\end{align}
\end{widetext}
$\delta$ is the Kronecker delta and
\begin{align}
  \Gamma^2 &\equiv (1-2\gamma)^2a(1-a) + \gamma^2, \nonum
  \kappa^2 &\equiv \frac{\gamma^2 c^2}{(1-2\gamma)^2a(1-a) + \gamma^2}.
  \label{eq:model-gammakappa}
\end{align}
\Cref{eq:model-f} assumes a successful retrieval regime in which the network overlaps significantly with either one sparse example $\ve\eta_{11}$ or dense, correlated examples $\ve\zeta_{1\nu}$ of one concept. We capture these two possibilities by introducing $\ve\chi_{1\nu}$, where $\chi^i_{1\nu} = \eta^i_{11}\delta_{1\nu}$ or $\zeta^i_{1\nu}$ respectively for retrieval of sparse or dense patterns. Through self-averaging, we have replaced averages over neurons $i$ with averages over entries $\chi_{1\nu}$ at a single neuron. Thus, the index $i$ no longer appears in \cref{eq:model-f}.

Then, we use the replica symmetry ansatz and saddle-point method to obtain the following mean-field equations in terms of the replica-symmetric order parameters $m_{1\nu}$, $r$, and $Q$:
\begin{align}
  m_{1\nu} &= \bigl\llangle \chi_{1\nu} \sig[\beta h] \bigr\rrangle, \nonum
  r &= s\Gamma^4 \frac{\bigl(1 - Q(1-\kappa^2)(1+s_0\kappa^2)\bigr)^{\!2} + s_0\kappa^4}{\bigl(1 - Q(1-\kappa^2)\bigr)^{\!2}\bigl(1 - Q(1+s_0\kappa^2)\bigr)^{\!2}} \nonum
  &\phantom{{}={}} {}\times \bigl\llangle \sig[\beta h]^2 \bigr\rrangle, \nonum
  Q &= \beta \Gamma^2 \bigl\llangle \sig[\beta h]^2 - \sig[\beta h] \bigr\rrangle,
  \label{eq:model-meanfield}
\end{align}
where the double angle brackets indicate averages over $\chi_{1\nu}$ and $z$, an auxiliary random field with a standard normal distribution. Meanwhile, $s_0 \equiv s-1$, $\sig(x) \equiv 1/(1+\ee^{-x})$, 
\begin{align}
  h &\equiv \sum_\nu m_{1\nu} \chi_{1\nu} - \phi + \sqrt\ar z, \nonum
  \phi &\equiv \theta - \frac{Q\alpha s\Gamma^2}{2} \cdot \frac{1+s_0\kappa^4 - Q(1-\kappa^2)(1+s_0\kappa^2)}{\bigl(1 - Q(1-\kappa^2)\bigr) \bigl(1 - Q(1+s_0\kappa^2)\bigr)}.
\end{align}
As derived in \cref{sec:mean}, $m_{1\nu}$'s are network overlaps with the target pattern and other patterns correlated with it, $r$ represents noise due to overlap with off-target patterns, $Q$ is related to the overall neural activity. $h$ is the local field in the mean-field limit, which encapsulates the mean network interaction experienced by each neuron. $\phi$ is the shifted threshold, which is empirically very similar to the original threshold $\theta$.

\Cref{eq:model-meanfield} applies to all target pattern types $\ve\chi_{1\nu}$ that we wish to recover. We now simplify the mean-field equations for either sparse targets with $\chi_{1\nu} = \eta_{11}\delta_{1\nu}$ or dense patterns with $\chi_{1\nu} = \zeta_{1\nu}$. In the latter case, we will perform further simplifications corresponding to recovery of either one dense example $\ve\zeta_{11}$ or one dense concept $\ve\zeta_1$, in which case the network overlaps equally with all dense examples $\ve\zeta_{1\nu}$ belonging to it. We also take the $T \rightarrow 0$ limit, which implies a strict threshold without stochastic activation. The full derivations are provided in \cref{sec:mean,sec:se,sec:de,sec:dc}, but the results for each target type are provided below.
\begin{enumerate}

  \item Sparse example $\ve\eta_{11}$: \Cref{eq:model-meanfield} becomes
    \begin{alignat}{2}
      m_{11} &\mathrlap{ {}= \frac{(1-2\gamma)a}{2} \Biggl\{\erf\frac{\phi}{\sqrt{2\ar}} + \erf\frac{(1-2\gamma)m_{11}-\phi}{\sqrt{2\ar}} \Biggr\}, }\nonum
      r &= \frac{s \bigl(1+s_0\kappa^4\bigr) \Gamma^4}{2} \Biggl\{ && 1 - \erf\frac{\phi}{\sqrt{2\ar}} \nonum
      &&&{}+ a \erf\frac{(1-2\gamma)m_{11}-\phi}{\sqrt{2\ar}} \Biggr\}.
      \label{eq:model-A}
    \end{alignat}

  \item Dense example $\ve\zeta_{11}$: If we call $m_0 \equiv m_{1\nu}$ the overlap with other dense examples $\nu > 1$ of the same concept, \cref{eq:model-meanfield} becomes
    \begin{align}
      m_{11} &= \frac{\gamma}{2} \biggl\{\frac{1+c}{4} \Bigl[\erf Y_{++}+\erf Y_{+-}\Bigr] \nonum
      &\phantom{{}= \frac{\gamma}{2} \biggl\{} {}+ \frac{1-c}{4} \Bigl[\erf Y_{-+}+\erf Y_{--}\Bigr] \biggr\}, \nonum
      m_0 &= \frac{\gamma c}{2\Bigl(1-Q\frac{\gamma^2}{\Gamma^2}(1-c^2)\Bigr)} \nonum
      &\phantom{{}={}} {}\times \biggl\{ \frac{1+c}{4} \Bigl[\erf Y_{++}+\erf Y_{+-}\Bigr] \nonum
      &\phantom{{}={} {}\times \biggl\{} {}- \frac{1-c}{4} \Bigl[\erf Y_{-+}+\erf Y_{--}\Bigr] \biggr\}, \nonum
      r &= \frac{s\Gamma^4}{2} \cdot \frac{\bigl(1 - Q(1-\kappa^2)(1+s_0\kappa^2)\bigr)^{\!2} + s_0\kappa^4}{\bigl(1 - Q(1-\kappa^2)\bigr)^{\!2}\bigl(1 - Q(1+s_0\kappa^2)\bigr)^{\!2}}  \nonum
      &\phantom{{}={}} {}\times \biggl\{1 - \frac{1+c}{4} \Bigl[\erf Y_{++}-\erf Y_{+-}\Bigr] \nonum
      &\phantom{{}={} {}\times \biggl\{} - \frac{1-c}{4} \Bigl[\erf Y_{-+}-\erf Y_{--}\Bigr] \biggr\}, \nonum
      Q &= \frac{\Gamma^2}{\sqrt{2\pi}\sigma_0} \biggl\{ \frac{1+c}{4} \Bigl[\ee^{-Y_{++}^2}+\ee^{-Y_{+-}^2}\Bigr] \nonum
      &\phantom{{}= \frac{\Gamma^2}{\sqrt{2\pi}\sigma_0} \biggl\{} {}+ \frac{1-c}{4} \Bigl[\ee^{-Y_{-+}^2}+\ee^{-Y_{--}^2}\Bigr] \biggr\},
      \label{eq:model-0}
    \end{align}
    where
    \begin{align}
      \sigma_0^2 &\equiv s_0\gamma^2(1-c^2)m_0^2 + \ar \nonum
      Y_{\pm\pm} &\equiv \frac{\gamma m_{11} \pm s_0\gamma cm_0 \pm \phi}{\sqrt{2}\sigma_0}.
    \end{align}
    Sign choices in $Y_{\pm\pm}$ correspond to respective signs on the right-hand side of the equation.

  \item Dense concept $\ve\zeta_1$: If we call $m_1$ the overlap with the target dense concept and $m_\rms \equiv m_{1\nu}$ the overlap with all of its dense examples $\nu$, \cref{eq:model-meanfield} becomes
    \begin{align}
      m_1 ={}& \frac{\gamma}{4} \biggl\{ \erf Y_++\erf Y_- \biggr\}, \nonum 
      m_\rms ={}& \frac{\gamma c}{4\Bigl(1-Q\frac{\gamma^2}{\Gamma^2}(1-c^2)\Bigr)} \biggl\{ \erf Y_++\erf Y_- \biggr\}, \nonum
      r ={}& \frac{s\Gamma^4}{2} \cdot \frac{\bigl(1 - Q(1-\kappa^2)(1+s_0\kappa^2)\bigr)^{\!2} + s_0\kappa^4}{\bigl(1 - Q(1-\kappa^2)\bigr)^{\!2}\bigl(1 - Q(1+s_0\kappa^2)\bigr)^{\!2}} \nonum
      &{}\times \biggl\{ 1 - \frac{1}{2} \Bigl[\erf Y_+-\erf Y_-\Bigr] \biggr\}, \nonum
      Q ={}& \frac{\Gamma^2}{\sqrt{8\pi}\sigma_\rms} \biggl\{ \ee^{-Y_+^2}+\ee^{-Y_-^2} \biggr\},
      \label{eq:model-S}
    \end{align}
    where
    \begin{align}
      \sigma_\rms^2 &\equiv s\gamma^2(1-c^2)m_\rms^2 + \ar \nonum
      Y_\pm &\equiv \frac{s\gamma cm_\rms \pm \phi}{\sqrt{2}\sigma_\rms}.
    \end{align}
    The sign choice in $Y_\pm$ corresponds to the sign on the right-hand side of the equation.

\end{enumerate}

\section{\label{sec:capacities}\texorpdfstring{$T=0$}{T = 0} capacities}

\subsection{Retrieval regimes}

\begin{figure}[t!]
  \centering
  \includegraphics{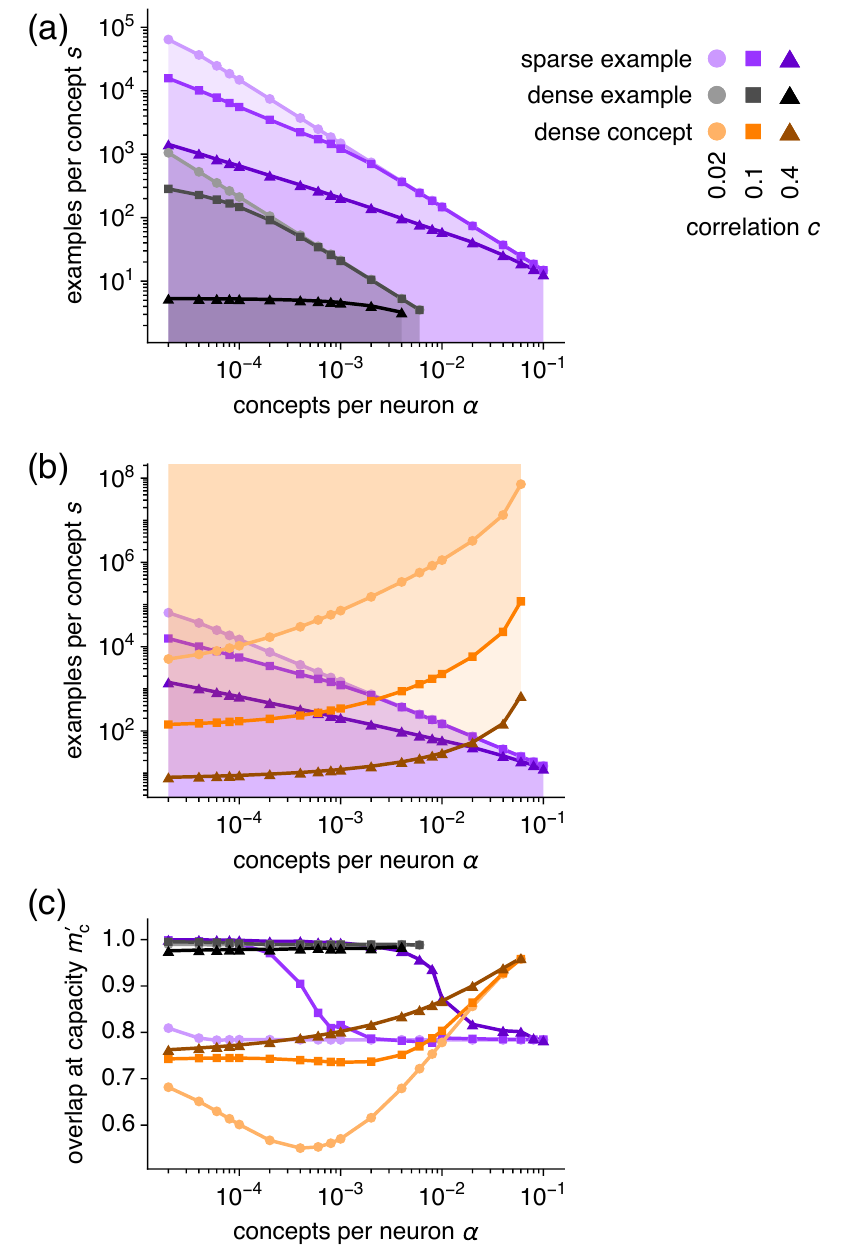}
  \caption{
    \label{fig:combined}
    Retrieval properties for sparse examples, dense examples, and dense concepts.
    (a),~(b) Retrieval regimes (shaded regions) obtained by numerically solving the mean-field equations. Their boundaries correspond to capacities $s_\rmc$. Sparse patterns are recovered at high threshold and dense patterns at low threshold.
    (a) More examples can be retrieved sparsely than they can be densely.
    (b) For small enough concept loads $\alpha$ and intermediate example loads $s$, both sparse examples and the dense concepts can be retrieved.
    (c) Network overlap with target patterns at capacity [\cref{eq:results-mprime}]. The curves for the three dense example conditions closely follow one another.
    Sparse patterns have \add{density} $a = 0.01$ and the dense storage strength is $\gamma = 0.1$.
  }
\end{figure}

Large values for the overlaps $m_{11}$ and $m_1$ in \cref{eq:model-A,eq:model-0,eq:model-S} signal that retrieval of target patterns is possible. To be more precise, we derive in \cref{sec:mean} that for $T=0$,
\begin{equation}
  m_{1\nu} = \langle \chi_{1\nu} S \rangle,
\end{equation}
where $\chi_{1\nu}$ and $S$ are respectively the pattern entry and activity for a single neuron and angle brackets indicate an average over $\chi_{1\nu}$. Again, the neuron index $i$ does not appear due to self-averaging. Successful retrieval means that the network activity $\ve S$ is similar to the original, unscaled patterns $\ve\xi_{11}$, $\ve\psi_{11}$, and $\ve\psi_1$ with 0/1 entries. With the rescalings in \cref{eq:model-rescaled}, this condition implies $m_{11} \sim (1-2\gamma)a(1-a)$ for sparse example targets, $m_{11} \sim \gamma/2$ for dense example targets, and $m_1 \sim \gamma/2$ for dense concept targets. For ease of comparison, we define a rescaled overlap
\begin{equation}
  m' = 
  \begin{cases}
    m_{11} / (1-2\gamma)a(1-a) & \textrm{sparse example}, \\
    m_{11} / (\gamma/2)        & \textrm{dense example}, \\
    m_1    / (\gamma/2)        & \textrm{dense concept},
  \end{cases}
  \label{eq:results-mprime}
\end{equation}
so $m' \sim 1$ corresponds to the retrieval phase, as an order-of-magnitude estimate.

To determine the extent of retrieval phase, we numerically solve the mean-field equations for a given set of network parameters. Phase boundaries are found by adjusting the number of examples stored per concept $s$ and looking for the appearance or disappearance of nontrivial solutions. These boundaries will change as a function of the number of concepts per neuron $\alpha$, the sparse pattern \add{density} $a$, the dense pattern correlation $c$, the relative dense storage strength $\gamma$. We treat the shifted activity threshold $\phi$ as a free parameter that can be adjusted to maximize $m_{11}$ and $m_1$.

\Cref{fig:combined}(a) shows that for a given concept load $\alpha$, the network can retrieve sparse and dense examples below critical example loads $s_\rmc$, which we call the capacities. Above the capacities, catastrophic interference between the target and off-target patterns prevents successful retrieval. \Cref{fig:combined}(b) shows that the network can retrieve dense concepts above a critical $s_\rmc$. Thus, it builds concepts, which are not directly stored, through accumulating shared features among dense examples. With greater correlation $c$, fewer examples are required to appreciate commonalities, so $s_\rmc$ is lower. Note that for low enough $\alpha$, the network can recover both sparse examples and dense concepts at intermediate values of $s$. Thus, our network is capable of retrieving both example and concept representations of the same memories by tuning an activity threshold.

Optimal retrieval of dense patterns occurs at threshold $\phi = 0$ and of sparse patterns at $\phi/(1-2\gamma)^2a \approx 0.6$. These values which match results for classic Hopfield networks that store only dense or only sparse patterns \cite{Weisbuch.1985, Tsodyks.1988}. At $s_\rmc$, the rescaled overlap $m'_\rmc$ takes values above 0.5 over the parameters explored [\cref{fig:combined}(c)] before jumping discontinuously to a much lower value immediately outside the retrieval regime. Such a first-order transition has also been observed in classic Hopfield networks \cite{Amit.1985mb, Tsodyks.1988, Fontanari.1990}.

\subsection{Overview of capacity formulas}

\begin{figure}[t!]
  \centering
  \includegraphics{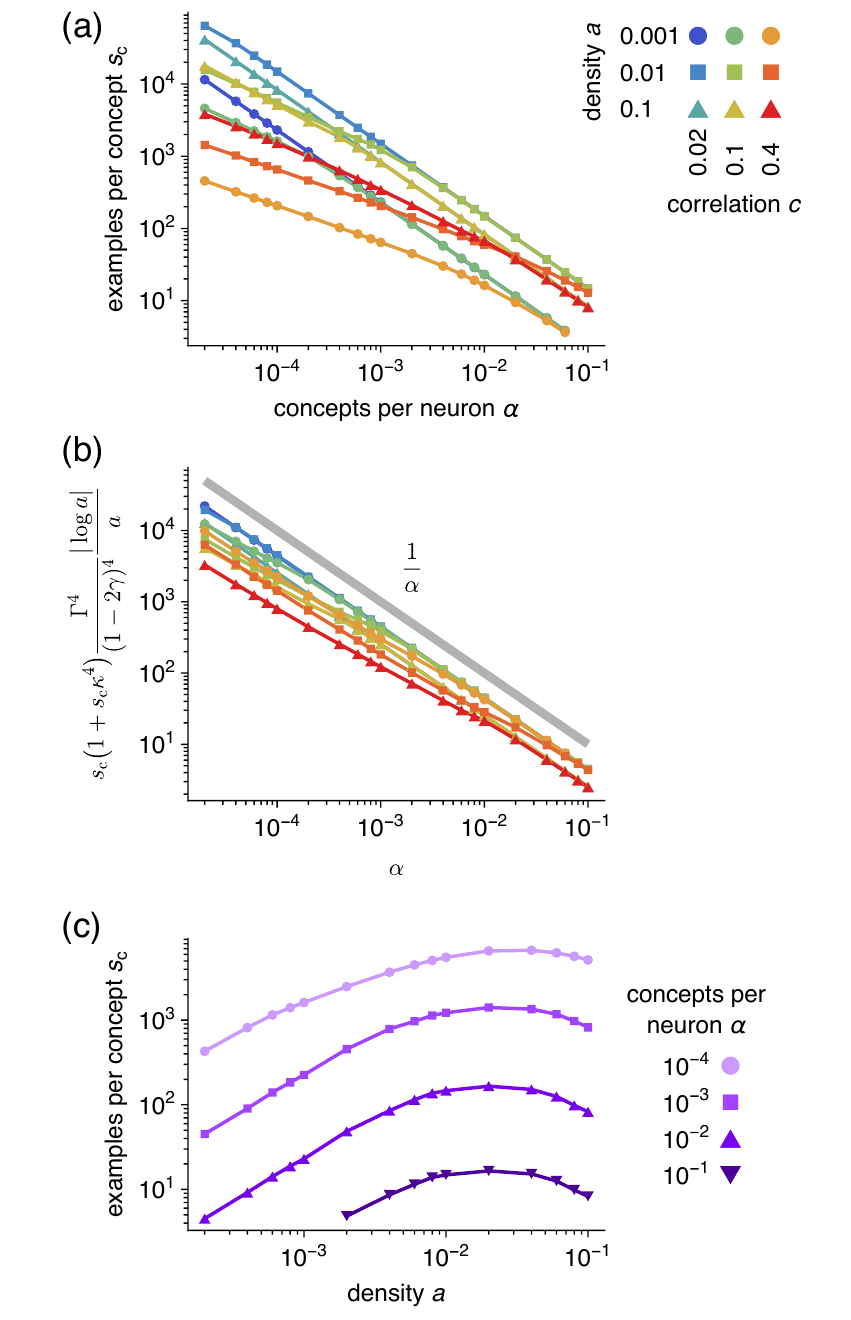}
  \caption{
    \label{fig:A}
    (a) Capacity $s_\rmc$ for sparse examples. Connected points indicate numerical analysis of \cref{eq:model-A}.
    (b) Collapse of $s_\rmc$ curves under rescaled variables. Gray line indicates theoretical formula \cref{eq:results-A1}.
    (c) $s_\rmc$ is maximized at intermediate values of \add{density} $a$. Dense patterns have correlation $c = 0.1$.
    The dense storage strength is $\gamma = 0.1$.
  }
\end{figure}

\begin{figure}[t!]
  \centering
  \includegraphics{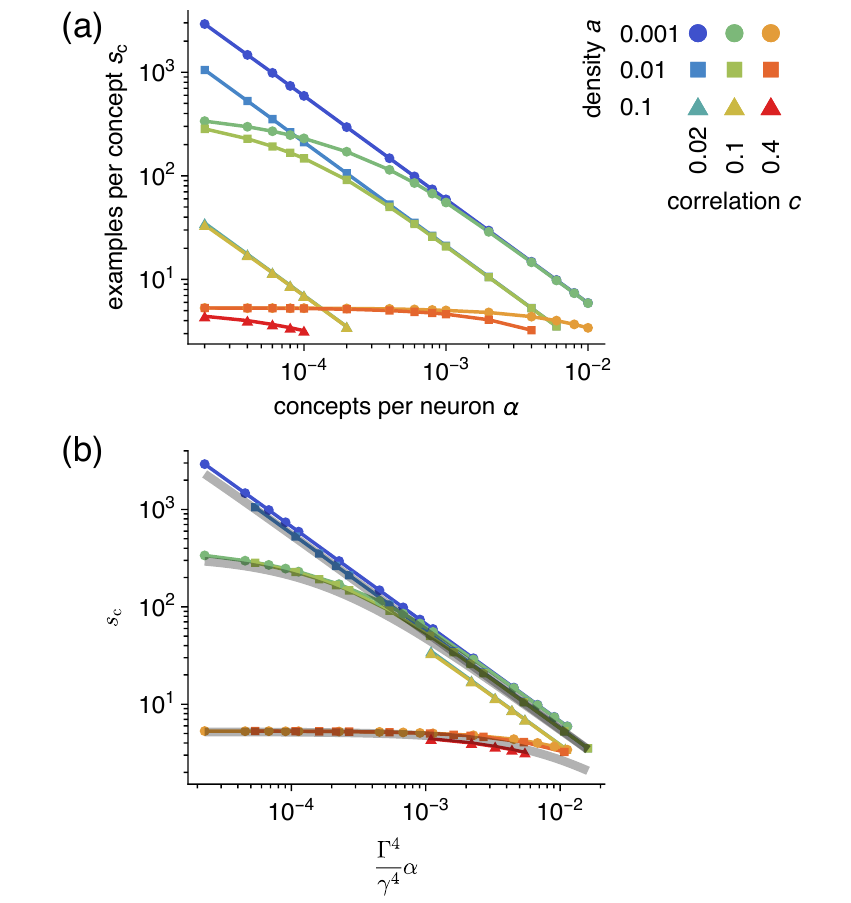}
  \caption{
    \label{fig:0}
    (a) Capacity $s_\rmc$ for dense examples. Connected points indicate numerical analysis of \cref{eq:model-0}.
    (b) Collapse of $s_\rmc$ curves under rescaled variables. Gray lines indicate theoretical formula \cref{eq:results-0}.
    The dense storage strength is $\gamma = 0.1$.
  }
\end{figure}

\begin{figure*}[t!]
  \centering
  \includegraphics{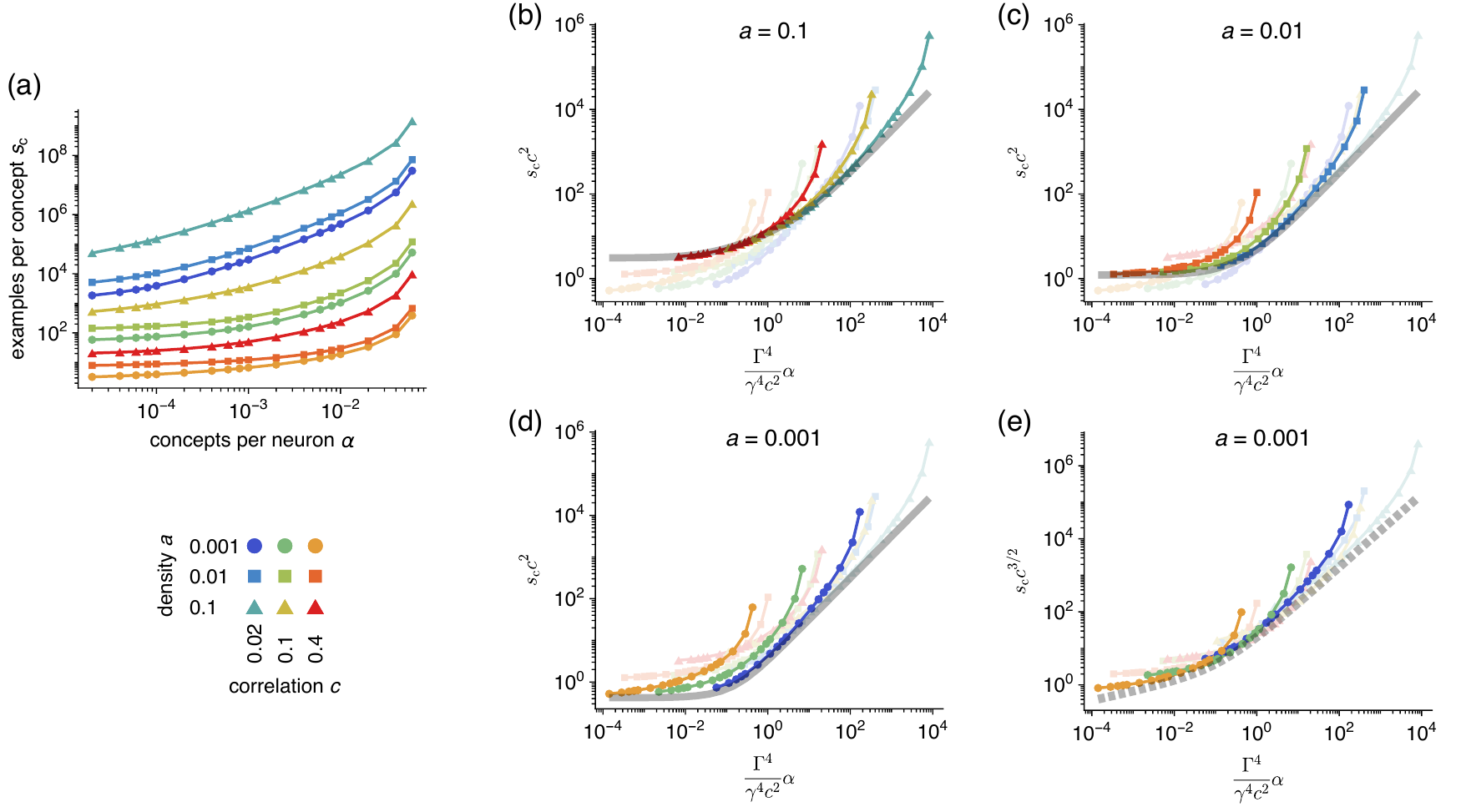}
  \caption{
    \label{fig:S}
    (a) Capacity, or critical example load, $s_\rmc$ for dense concepts. Connected points indicate numerical analysis of \cref{eq:model-S}.
    (b)--(d) \add{Approximate} collapse of $s_\rmc$ curves under rescaled variables. Gray \add{solid} lines indicate theoretical formula \cref{eq:results-S1}.
    (e) For the sparsest patterns, $s_\rmc$ curves exhibit better collapse under \add{the rescaling $s_\rmc c^{3/2}$ compared to $s_\rmc c^2$ in (d)}. Gray \add{dotted} line indicates theoretical formula \cref{eq:results-S2}, which better matches the numerical results. It exhibits weak dependence on dense correlation $c$, and we only show its behavior for $c = 0.02$.
    The dense storage strength is $\gamma = 0.1$.
  }
\end{figure*}

We then seek to obtain mathematical formulas for the capacity, or critical example load, $s_\rmc$ of each type of pattern. Not only would these formulas provide a direct way of determining whether pattern retrieval is possible for a given set of network parameters, they would offer mathematical insight into network behavior. As detailed in \cref{sec:se,sec:de,sec:dc}, we apply various approximations to the mean-field equations \cref{eq:model-A,eq:model-0,eq:model-S} to derive the following formulas for $s_\rmc$, which match well with numerical solutions over a wide range of parameters (\cref{fig:A,fig:0,fig:S}). 
\begin{enumerate}

  \item Sparse example $\ve\eta_{11}$ (\cref{fig:A}): The capacity is
    \begin{equation}
      \frac{1}{\alpha} \sim s_\rmc \bigl(1+s_\rmc\kappa^4\bigr) \frac{\Gamma^4}{(1-2\gamma)^4} \frac{|\!\log a|}{a},
      \label{eq:results-A1}
    \end{equation}
    which means that
     \begin{equation}
       s_\rmc \sim \sqrt{ \frac{1}{4\kappa^8} + \frac{(1-2\gamma)^4}{\gamma^4c^4} \cdot \frac{a}{|\!\log a|} \cdot \frac{1}{\alpha} } - \frac{1}{2\kappa^4}.
      \label{eq:results-A2}
    \end{equation}
    In sparse Hopfield networks without dense patterns, the capacity always increases for sparser patterns \cite{Tsodyks.1988}. In contrast, our capacity for sparse examples peaks at intermediate \add{densities} $a$ [\cref{fig:A}(c)]. While sparser patterns interfere less with one another, their smaller basins of attraction are more easily overwhelmed by those of dense patterns, whose \add{density} is always 0.5. We can quantitatively understand the tradeoff between these two factors in the $c^2 \rightarrow 0$ limit, where \cref{eq:results-A1} becomes 
    \begin{equation}
      \alpha s_\rmc \sim \frac{a}{(a+a_\rmd)^2 |\!\log a|}
      \label{eq:results-A3}
    \end{equation}
    for $a_\rmd \equiv \gamma^2/(1-2\gamma)^2$. $a_\rmd$ represents interference from dense patterns and acts as the crossover point in the tradeoff. For $a \gg a_\rmd$, $\alpha s_\rmc \sim 1 / a |\!\log a|$, recovering the classic sparse Hopfield scaling in which sparser patterns exhibit higher capacity \cite{Tsodyks.1988}. However, for $a \ll a_\rmd$, $a_\rmd$ dominates the denominator and $\alpha s_\rmc \sim a / |\!\log a|$, disfavoring sparser patterns. If we ignore the slowly varying logarithm in \cref{eq:results-A3}, $s_\rmc$ is exactly maximized at $a = a_\rmd$. Using the value $\gamma = 0.1$ in \cref{fig:A}(c), $a_\rmd \approx 0.016$, which agrees well with the numerically obtained maxima.

  \item Dense example $\ve\zeta_{11}$ (\cref{fig:0}): The capacity is
    \begin{equation}
      s_\rmc \sim \frac{1}{3c^3+18\frac{\Gamma^4}{\gamma^4}\alpha}.
      \label{eq:results-0}
    \end{equation}
    At large $\alpha$, this critical number of examples per concept $s_\rmc$ is inversely proportional to the number of concepts per neuron $\alpha$, indicating that the total number of examples stored per neuron $\alpha s_\rmc$ saturates at a constant value. When examples are distributed into many concepts, concept identity becomes insignificant, so only the total number of stored patterns matters. At small $\alpha$, $s_\rmc$ itself saturates at a constant value determined by the dense correlation $c$. When concepts are few, interference with other concepts becomes less important than interference within the same concept, so only the number of stored patterns per concept matters.

  \item Dense concept $\ve\zeta_1$ (\cref{fig:S}): There are two cases. For larger \add{densities} $a$, the critical example load approximately collapses as a function of $s_\rmc c^2$ \add{[\cref{fig:S}(b), (c)]}. This function can be obtained by numerically inverting the following first equation for $y$ and substituting it into the second:
    \begin{align}
      \frac{2\Gamma^4}{\gamma^4c^2}\alpha ={}& \frac{(1-2\gamma)^2a(1-a)}{\gamma^2} \frac{\Bigl[\expyy\Bigr]^3}{y^2\Bigl[\erfy-\expyy\Bigr]} \nonum
      & {}- \Bigl[\textstyle\expyy\Bigr]^2 \nonum
      s_\rmc c^2 \approx{}& \frac{(1-2\gamma)^2a(1-a)}{\gamma^2} \frac{\expyy}{\erfy-\expyy}.
      \label{eq:results-S1}
    \end{align}
    The solution is unique for any parameter values because the right-hand side of the first equation always monotonically decreases as a function of $y$ over its positive range. \add{For smaller $a$, the critical example load does not collapse so tightly as a function of $s_\rmc c^2$ for different values of $c$ [\cref{fig:S}(d)].} We calculate that it instead approximately collapses as a function of $s_\rmc c^{3/2}$ \add{[\cref{fig:S}(e)]}:
    \begin{equation}
      s_\rmc c^{3/2} \approx 3 \bigl(\tfrac{3\pi}{4}\bigr)^{\!1/4} \Bigl(\tfrac{\Gamma^4}{\gamma^4c^2}\alpha\Bigr)^{\!1/4} + \tfrac{3\pi}{4} c^{-1/2} \Bigl(\tfrac{\Gamma^4}{\gamma^4c^2}\alpha\Bigr).
      \label{eq:results-S2}
    \end{equation}
    The second term contains a factor of $c^{-1/2}$, which changes relatively slowly compared to the other powers of $c$ found in the rescaled concept load $\alpha\Gamma^4/\gamma^4c^2$. The two terms capture the behavior of $s_\rmc$ at low and high rescaled concept load, respectively. Nevertheless, more universal scaling relationships have yet to be found for the dense concept $s_\rmc$, indicating that many network features may independently govern concept building.
    
\end{enumerate}

\subsection{\label{sec:sim}Capacities of simulated networks}

We perform simulations to verify our capacity calculations. For each simulation condition, we construct replicate networks that store different randomly generated patterns. When generating sparse patterns of \add{density} $a$, we fix the number of active neurons to $Na$ to reduce finite-size effects. Neural dynamics proceed asynchronously in cycles wherein every neuron is updated once in random order. We use $N = \num{10000}$ neurons and dense strength $\gamma = 0.1$, unless otherwise noted. Retrieval is assessed by the following definition of overlap between network activity $\ve S$ and the unscaled target pattern $\ve\omega$, which is a sparse example $\ve\xi_\mn$, a dense concept $\ve\psi_\mn$, or a dense concept $\ve\psi_\mu$:
\begin{equation}
  \hat m = \frac{1}{N a_\omega (1-a_\omega)}\sum_i (\omega_i - a_\omega) S_i,
  \label{eq:results-mhat}
\end{equation}
where $a_\omega = a$ for sparse patterns and $a_\omega = 1/2$ for dense patterns. Based on \cref{eq:model-xi,eq:model-psimu,eq:model-psi}, we expect $\hat m \approx 1$ to indicate successful retrieval. For random activity, $\hat m \approx 0$. This overlap $\hat m$ is similar to $m'$ in \cref{eq:results-mprime}, which concerned the \emph{scaled} target patterns $\ve\chi = \ve\eta_{11}$, $\ve\zeta_{11}$, and $\ve\zeta_1$.

\begin{figure}[t!]
  \centering
  \includegraphics{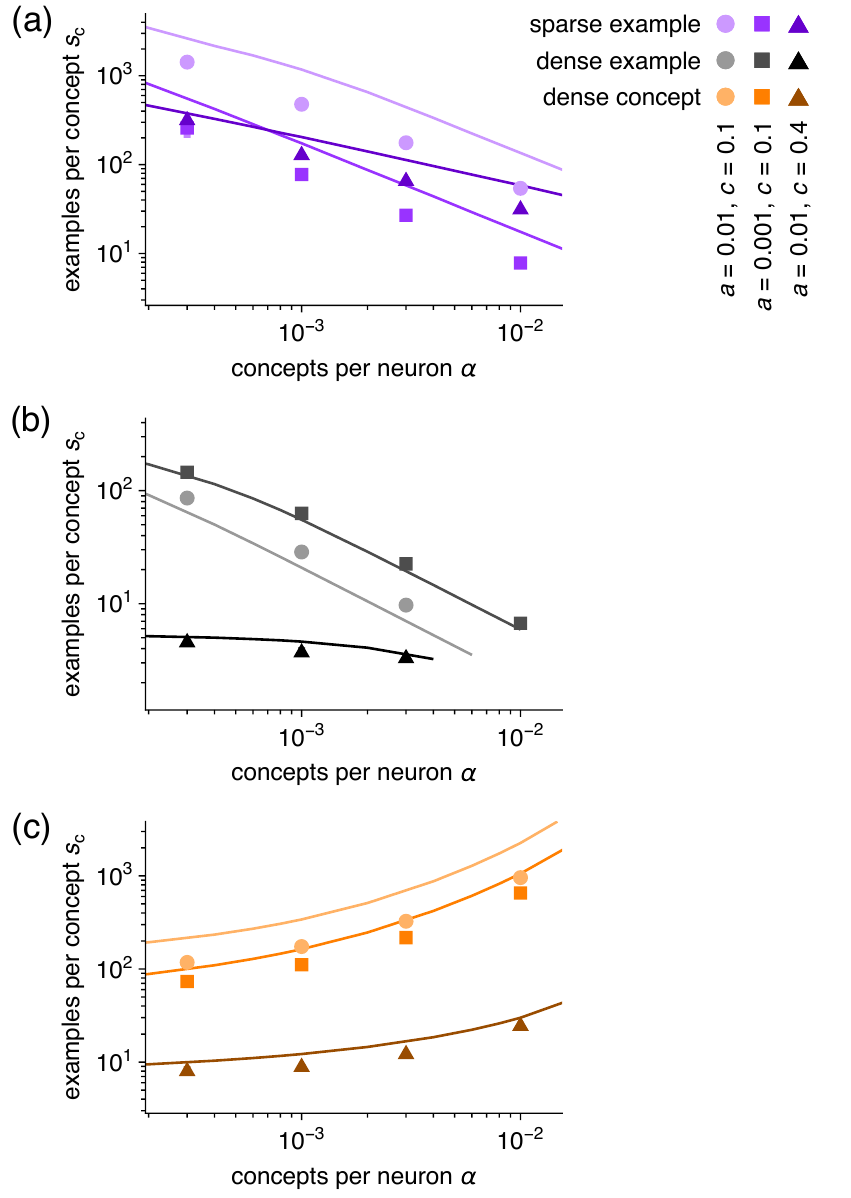}
  \caption{
    \label{fig:sim}
    Capacities $s_\rmc$ for (a) sparse examples, (b) dense examples, and (c) dense concepts obtained by numerical calculations (lines) and simulations (points). Lines indicate analysis of the mean-field equations \cref{eq:model-A,eq:model-0,eq:model-S}. Points indicate means over 8 replicate simulated networks, and vertical bars indicate standard deviations which are often obscured by the points. In each replicate network, 20 cues are tested with simulations lasting 10 update cycles.
    The dense storage strength is $\gamma = 0.1$.
  }
\end{figure}

Capacities are assessed by using the true target patterns as cues; in other words, our simulations probe the stability of the target patterns. For sparse examples, we optimize over the threshold $\theta$ by numerical search. For dense patterns, we use $\theta = 0$. We use $\beta \rightarrow \infty$ in \cref{eq:model-glauber} because our theoretical calculations were performed for $T \rightarrow 0$. We define successful retrieval as $\hat m > (1 + \hat m_0)/2$, where $\hat m_0$ is the overlap expected for off-target patterns within the same concept. Using \cref{eq:model-overlaps}, $\hat m_0 = 0$ for sparse examples, $\hat m_0 = c^2$ for dense examples, and $\hat m_0 = c$ for dense concepts. 

\Cref{fig:sim} reveals good agreement between simulations and numerical analysis of the mean-field equations for capacities of all target types. This supports the validity of our derivations and the simplifications we invoked to perform them.

\section{\label{sec:hetero}Heteroassociation}

\subsection{Performance of simulated networks}

Our network stores linear combinations of sparse and dense patterns, and its connectivity matrix contains interactions between the two [\cref{eq:model-J}]. Thus, we suspect that in addition to autoassociation for each target type, it can perform heteroassociation between them. We run simulations to test this ability. We use $p = 10$ concepts and store either $s = 20$ examples per concept during retrieval of sparse examples $\ve\xi_\mn$ and dense concepts $\ve\psi_\mu$ or $s = 3$ during retrieval of dense examples $\ve\psi_\mn$. Sparse patterns have \add{density} $a = 0.01$ and dense patterns have correlation parameter $c = 0.4$. We \add{initialize} the network state to a noisy version of a sparse example, dense example, or dense concept, and attempt to retrieve each type as the target pattern. \add{We create these noisy cues by randomly flipping a fraction 0.01 of the cue pattern between inactive and active. We then asynchronously evolve the network similarly as in the previous section.} With theoretical motivation in \cref{sec:se}, we define the rescaled parameters
\begin{equation}
  \theta' = \theta/(1-2\gamma)^2a \cond{and} \beta' = \beta\cdot(1-2\gamma)^2a,
\end{equation}
with rescaled temperature $T' = 1/\beta'$. To retrieve sparse examples, we apply a threshold $\theta' = 0.6$, and to retrieve dense examples and concepts, we apply $\theta' = 0$; \add{these thresholds are immediately applied from the start}. We use inverse temperature $\beta' = 50$. \add{Finally, we assess the overlap between the final network activity and the target pattern.} If concepts are used as cues and examples are desired as targets, the highest overlap with any example within the cued concept is reported. Successful retrieval is defined via the overlap $\hat m$ as described above [\cref{eq:results-mhat}].

\begin{figure}[t!]
  \centering
  \includegraphics{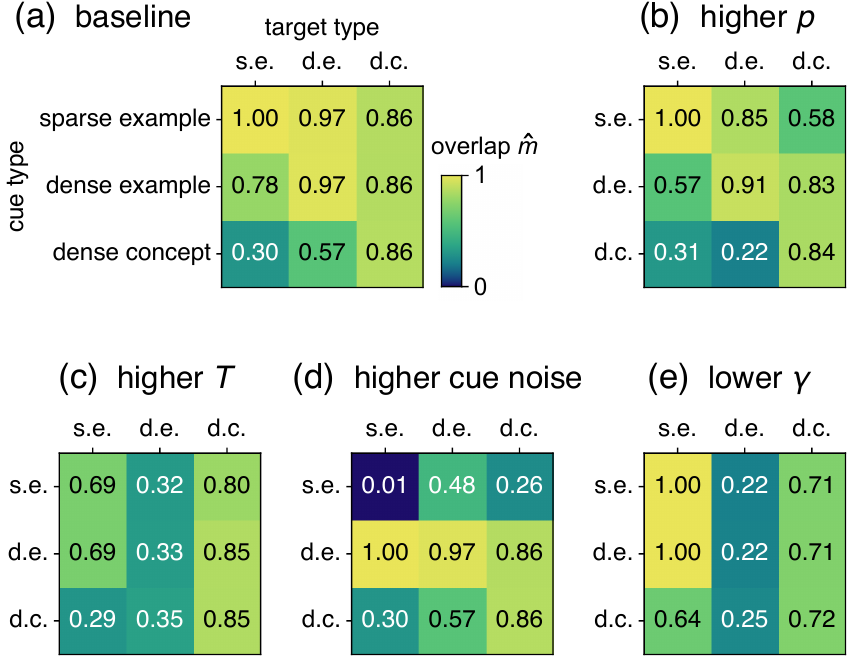}
  \caption{
    \label{fig:hetero}
    Auto- and heteroassociation among sparse and dense patterns demonstrated by network simulations.
    (a) Baseline to which various conditions are compared.
    (b) The number of concepts is increased from $p = 5$ to $30$.
    (c) The rescaled temperature is increased from $T' = 1/50$ to $1/5$.
    (d) The fraction of the cue pattern flipped is increased from 0.01 to 0.2.
    (e) The dense pattern storage strength is decreased from $\gamma = 0.1$ to $0.055$. 
    For dense example and concept targets, we use rescaled threshold $\theta' = 0$. For sparse example targets, we use $\theta' = 0.6$. Overlaps $\hat m$ reported are averages over 8 replicate networks, with 1 corresponding to perfect retrieval and 0 corresponding to random activity. In each, 20 cues are tested with simulations lasting 20 update cycles.
  }
\end{figure}

\Cref{fig:hetero}(a) shows that the network is generally capable of heteroassociation using the parameters described above, which define the baseline condition. By increasing the number of concepts, heteroassociative performance is largely preserved, but note that the retrieval of dense concepts from sparse examples is impaired [\cref{fig:hetero}(b)]. We next amplify noise by either raising the temperature, which introduces more randomness during retrieval, or randomly flipping more neurons during cue generation. Sparse example targets are more robust than dense example targets with respect to higher temperature [\cref{fig:hetero}(c)]; meanwhile, dense example cues are more robust than sparse example cues with respect to cue corruption [\cref{fig:hetero}(d)]. These observations encompass autoassociation as well as heteroassociation. Thus, the dual encoding of memories with not only allows for retrieval of both examples and concepts, as noted in \cref{fig:combined}(b), but it also mitigates the impact of noise since sparse and dense patterns are more robust to retrieval and cue noise, respectively.

\subsection{Bidirectional heteroassociation and \texorpdfstring{$\gamma$}{γ}}

Notice in \cref{fig:hetero}(a) that while dense concept targets can be retrieved from sparse example cues, the reverse is not possible. The ability to perform bidirectional heteroassociation between a concept and its examples is of computational significance, so we seek to find network parameters that achieve it. Intuitively, lowering the storage strength of dense patterns $\gamma$ should bias the network towards retrieving sparse patterns. Indeed, doing so improves retrieval of sparse examples from dense concepts [\cref{fig:hetero}(e)]. Moreover, the network is still capable of the reverse process, albeit with some decrease in performance.

\begin{figure}[t!]
  \centering
  \includegraphics{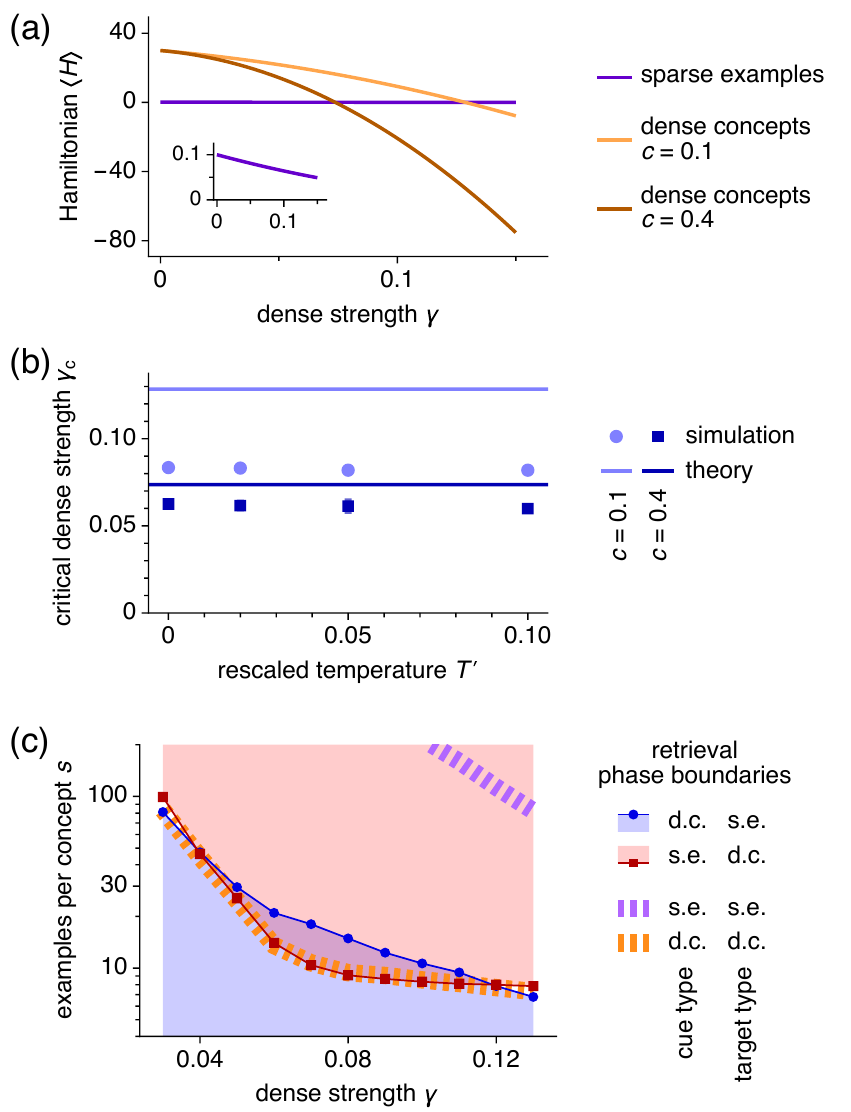}
  \caption{
    \label{fig:hamiltonian}
    The dense pattern storage strength $\gamma$ controls the ability to retrieve sparse examples from dense concepts by changing their relative energies.
    (a) Hamiltonian energies for rescaled threshold $\theta' = 0.6$ and network size $N = \num{10000}$ [\cref{eq:results-H}]. For $c = 0.1$, we store $s = 80$ patterns per concept, and for $c = 0.4$, $s = 20$. Inset shows sparse example energy in detail.
    (b) Critical dense strength $\gamma_\rmc$ below which sparse examples can be retrieved by dense concept cues. Theoretical predictions are the locations of energy crossovers in (a).
    (c) Phase diagram for auto- and heteroassociation among sparse examples and dense concepts in simulated networks. Dense patterns have correlation parameter $c = 0.4$, and the temperature is $T = 0$. Blue and red shaded regions exhibit unidirectional heteroassociation, and the doubly shaded region exhibits bidirectional heteroassociation. Autoassociation occurs below the purple dashed line and above the orange dashed line; for clarity, these regions are not shaded.
    We use $p = 5$ concepts, and sparse patterns have \add{density} $a = 0.01$. Simulations are performed without cue noise. For dense concept targets, we use $\theta' = 0$, and for sparse example targets, we use $\theta' = 0.6$. Points indicate means over 8 replicate networks, and vertical bars indicate standard deviations which are often obscured by the points. In each replicate network, 20 cues are tested with simulations lasting 20 update cycles.
  }
\end{figure}

The value of $\gamma$ appears critical to the ability to retrieve sparse examples from dense concepts. We hypothesize that this connection is mediated by the relative energy of different pattern types. As described in \cref{sec:mean}, the Hamiltonian of our network is
\begin{equation}
  H = -\frac{1}{2N}\sum_\mn \sum_{i \neq j} (\eta^i_\mn + \zeta^i_\mn) (\eta^j_\mn + \zeta^j_\mn) S_i S_j + \theta \sum_i S_i,
\end{equation}
where, again, $\ve\eta_\mn$ and $\ve\zeta_\mn$ are rescalings of sparse examples $\ve\xi_\mn$ and dense examples $\ve\psi_\mn$ [\cref{eq:model-rescaled}]. We set the network activity $\ve S$ to a sparse example $\ve\xi_\mn$ or dense concept $\ve\psi_\mu$ and calculate the average over patterns $\langle H \rangle$. Using \cref{eq:model-overlaps}, we obtain
\begin{equation}
  \frac{\langle H \rangle}{N} \approx \begin{dcases}
    -\frac{(1-2\gamma)^2 a^2(1-a)^2}{2} + \theta a & \textrm{sparse example}, \\
    -\frac{s \gamma^2 c^2}{8} + \frac{\theta}{2} & \textrm{dense concept}.
  \end{dcases}
  \label{eq:results-H}
\end{equation}
\Cref{fig:hamiltonian}(a) shows \cref{eq:results-H} calculated in the retrieval regime for sparse examples with $\theta' = 0.6$. The Hamiltonian for dense concepts decreases with $\gamma$ and eventually crosses the value for sparse examples, which remains relatively constant. To connect these results with heteroassociative performance, first consider $c = 0.4$, which is the correlation value used in the simulations in \cref{fig:hetero}. Recall that baseline networks experience difficulty in retrieving sparse examples from dense concepts [\cref{fig:hetero}(a)]. These networks have $\gamma = 0.1$, for which dense concepts exhibit lower energy than sparse examples do [\cref{fig:hamiltonian}(a)], even with the high threshold $\theta' = 0.6$ intended to retrieve the latter. The increase in energy required to proceed from cue to target may explain the failure to perform this heteroassociation. It can be performed for $\gamma = 0.055$ [\cref{fig:hetero}(e)], and here, the energy of dense concepts at high threshold increases above that of sparse examples [\cref{fig:hamiltonian}(a)]. Thus, the progression from cue to target is energetically favored.

The crossover point $\gamma_\rmc$ between the high-threshold energies of dense concepts and sparse examples appears to define the phase boundary for heteroassociation from the former to the latter. To test this prediction, we evaluate simulated networks at varying values of $\gamma$. Successful retrieval of sparse examples is assessed through the overlap $\hat m$ with the same cutoff values as described above [\cref{eq:results-mhat}]. \Cref{fig:hamiltonian}(b) demonstrates that the energy crossover indeed predicts $\gamma_\rmc$ for $c = 0.4$. The $c = 0.1$ case shows lower quantitative agreement between simulation and theory, although the qualitative observation of a higher $\gamma_\rmc$ is captured. Finite-size effects, higher energies of intermediate states along possible transition paths, and trapping in local energy minima may account for the discrepancy. For $T > 0$, the disregard of entropic contributions in our Hamiltonian analysis may also contribute to the disparity, although the lack of significant temperature dependence in our simulations makes this consideration less important [\cref{fig:hamiltonian}(b)].

For the $c = 0.4$ and $T = 0$ case, we construct a heteroassociation phase diagram by simulating networks with various dense strengths $\gamma$ and example loads $s$ [\cref{fig:hamiltonian}(c)]. At intermediate values of $\gamma$ and $s$, there is a regime for successful bidirectional heteroassociation between sparse examples and dense concepts. At lower values of either $\gamma$ or $s$, only unidirectional heteroassociation from dense concept cues to sparse example targets is possible, and at higher values, only the reverse unidirectional heteroassociation is possible. For comparison, autoassociation capacities for sparse examples and dense concepts are also shown. The phase boundary for retrieving sparse examples is much higher with identical cues than with dense concept cues, reflecting our observations that even below capacity, this heteroassociation direction is only granted for certain $\gamma$. In contrast, the phase boundary for retrieving dense concepts is similar with either type of cue, indicating an easier heteroassociation direction.

\begin{figure*}[t!]
  \centering
  \includegraphics{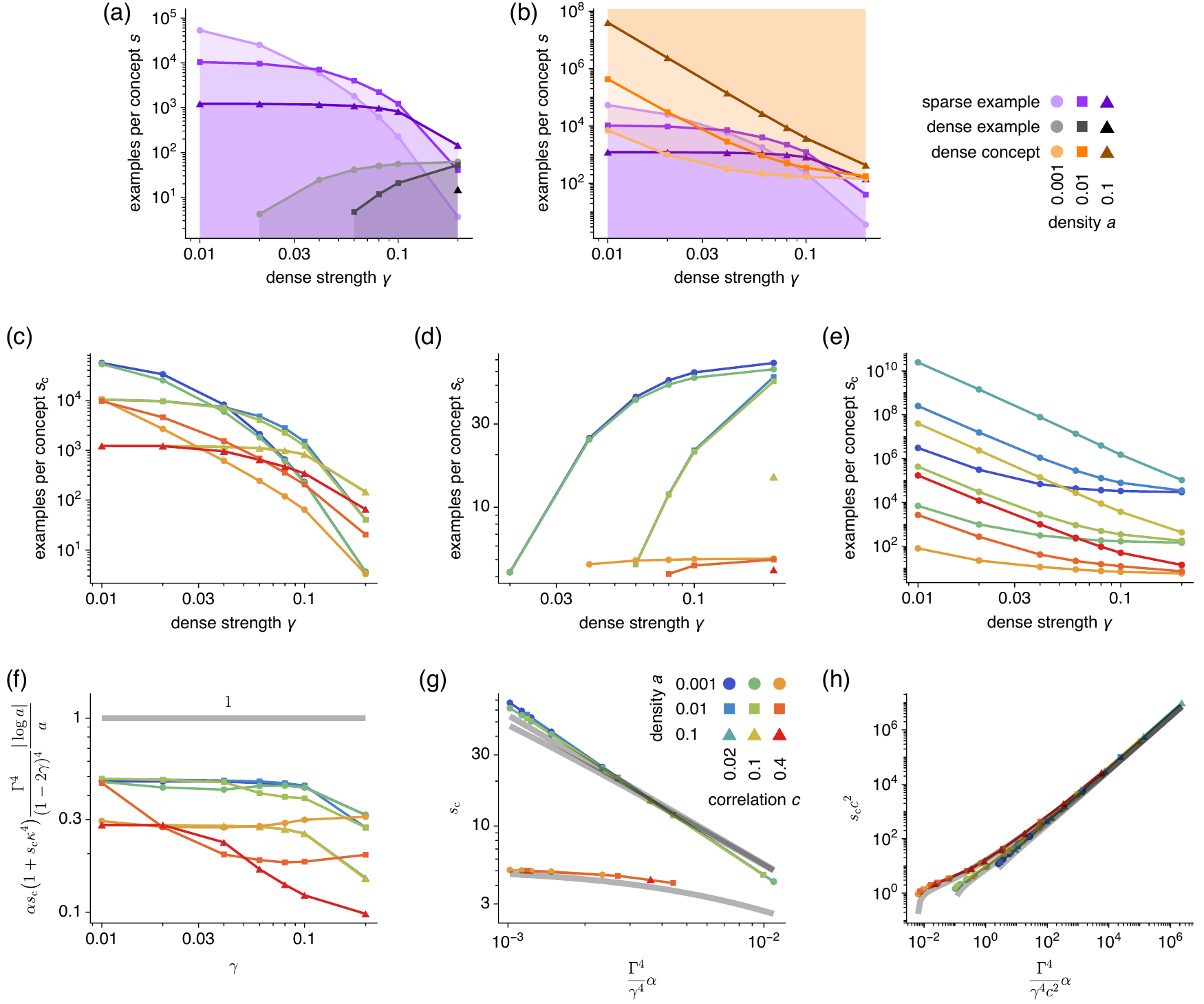}
  \caption{
    \label{fig:gamma}
    Capacities $s_\rmc$ as a function of dense pattern storage strength $\gamma$.
    (a), (b) Retrieval regimes (shaded regions) for sparse examples, dense examples, and dense concepts obtained by numerically solving the mean-field equations. Their boundaries correspond to capacities $s_\rmc$. Dense patterns have correlation parameter $c = 0.1$.
    Capacities $s_\rmc$ for (c) sparse examples, (d) dense examples, and (e) dense concepts.
    Collapse of $s_\rmc$ curves for (f) sparse examples, (g) dense examples, and (e) dense concepts under rescaled variables. Gray lines indicate theoretical formulas \cref{eq:results-A1,eq:results-0,eq:results-S1}, respectively.
    The concept load is $\alpha = 0.001$ concepts per neuron.
  }
\end{figure*}

Due to the importance of $\gamma$, we present additional mean-field capacity results in which it is systematically varied [\cref{fig:gamma}]. For low \add{density} $a$, there is a range of intermediate $\gamma$ and $s$ in which both sparse examples and dense concepts are stable [\cref{fig:gamma}(b)]. \Cref{fig:gamma}(c)--(h) illustrates that our theoretical capacity formulas are still valid as functions over $\gamma$.

\section{\label{sec:discussion}Discussion}

In summary, we present a Hopfield-like network that stores memories as both sparse patterns with low correlation and dense patterns with high correlation. By adjusting the activity threshold, the network can retrieve patterns of either sparsity. The capacity for sparse patterns is large, so many distinct memories can be retrieved. In contrast, as more dense patterns are stored, they merge according to their correlation structure such that concepts are built through the accumulation of examples. We derive mean-field equations that govern the retrieval of sparse examples, dense examples, and dense concepts, and we calculate capacity formulas for each type of retrieved pattern. We observe that the network can retrieve one type of target pattern from its corresponding cue of a different type, and we explain that regimes of successful heteroassociation can be predicted by the relative energies of cue and target patterns.

Our network offers an alternative paradigm for building memory hierarchies in autoassociative networks. Ultrametric networks have been previously explored as an architecture for storing and retrieving memories at different scales \cite{Mezard.1985, Dotsenko.1985, Cortes.1987, Virasoro.1988, Gutfreund.1988, Krogh.1988}. Their structure resembles a tree spanning multiple levels. Each pattern at one level serves as a concept-like trunk from which correlated branches are generated to form the next, more example-like level. While these models are insightful and influential, they possess certain disadvantages that our network can address. They typically use an activity threshold or, equivalently, an external field to move between levels, which is also the case in our work. In one ultrametric model, the field is inhomogeneous and proportional to the pattern retrieved \cite{Gutfreund.1988}. Our activity threshold is homogeneous and does not require memory of the pattern retrieved, though implementing such a feature may improve retrieval performance. In another hierarchical model, coarser representations are stored more sparsely and retrieved at higher threshold \cite{Krogh.1988}. This arrangement prevents the network from leveraging the higher capacity of sparser patterns to store finer representations, which are more numerous. Moreover, ultrametric Hopfield networks often require complex storage procedures that require a priori knowledge of concepts or other examples \cite{Dotsenko.1985, Cortes.1987, Gutfreund.1988, Krogh.1988}. They do not permit the unsupervised learning of concepts through the accumulation of examples over time, which is achieved by our simple Hebbian learning rule and strengthens the biological significance of our model. Meanwhile, our model's requirement for sparse, decorrelated patterns in addition to dense, correlated patterns can be implemented by neural circuits which are thought to naturally perform decorrelation through sparsification \citep{Marr.1971, Treves.1992, O'Reilly.1994, Vinje.2000, Wiechert.2010, Pitkow.2012, Cayco-Gajic.2017, Kang.2023b}.

\add{While the two pattern types are linearly summed in our model to facilitate mathematical derivations, it is possible to implement nonlinear summation, which may better reflect how inputs are combined in biological neurons \citep{Kim.2012, Makara.2013, Kaifosh.2016}. In Ref.~\onlinecite{Kang.2023b}, we have shown through simulations that the central capabilities of this model can be preserved under sublinear and superlinear summation. Ref.~\onlinecite{Kaifosh.2016} explores multiplicative integration with the storage of only one pattern per memory formed by the neurons commonly activated through both pathways.

Returning to biological motivation for our model, our results offer a mechanistic explanation for how complementary pathways within the hippocampus can underlie its observed ability to recall memories at different resolutions. The hippocampus has long been known to mediate episodic memory, the ability to recall specific, personally experienced events \citep{Scoville.1957, Squire.1992}. It is thought to be capable of pattern separation, a process that accentuates differences between similar memories \citep{Leutgeb.2007, Aimone.2011}. Meanwhile, more recent research has uncovered that the hippocampus is also involved in generalizing over episodes through statistical learning \citep{Knowlton.1993, Zeithamova.2008, Schapiro.2014, Mack.2016, Covington.2018}. The observation of neurons that respond to many representations of a single celebrity or personal acquaintance is one striking case of learning concepts through the accumulation of individual experiences \citep{Quiroga.2005, Quiroga.2009}.

Because of a difference in sparsity, memory types in our model are retrieved at different activity thresholds, which may correspond biologically to different levels of inhibition in CA3. In many mammals, including rodents and primates, the hippocampus exhibits a \emph{theta} oscillation, in which inhibition is modulated with subsecond periodicity \citep{Buzsaki.2002}. By analyzing neural encoding properties as a function of theta phase in Ref.~\onlinecite{Kang.2023b}, we indeed find experimental support for the theta oscillation to serve as the activity threshold in our model. The selection between example-like and concept-like representations by theta inhibition has certain computational advantages. A downstream network that serves to integrate information across the two representational scales can access both over subsecond timescales. Meanwhile, the activation of only one encoding at a time may prevent interference or overshadowing between them. In addition, certain tasks may better performed with either more example-like or more concept-like representations. Such preferential recall could be accomplished by adjusting the average inhibitory tone.

It is conceivable that our model may apply to brain regions other than the hippocampus if they receive converging inputs with different sparsities and correlation structures. In particular, the \emph{Drosophila} olfactory system and the mammalian cerebellum contain circuits that start at a common upstream region; branch into two pathways, one of which undergoes decorrelation through sparsification; and converge at a common downstream region. In the former system, the antennal lobe is the upstream region, Kenyon cells perform decorrelation, and the lateral horn is the downstream region \citep{Litwin-Kumar.2017, Jeanne.2018}. In the latter, the three components are mossy fibers (different from the ones in hippocampus), granule cells (also different from the ones in hippocampus), and deep cerebellar nuclei \citep{Litwin-Kumar.2017, Medina.2002}. Note that these circuits have an additional population between the decorrelation and downstream regions that is believed to perform rich computations through highly plastic synapses: the mushroom body output neurons in \emph{Drosophila} and the Purkinje cells in the cerebellum \citep{Hige.2015, Medina.2002}. More investigation is required to assess the applicability of our model to these systems.}

\section*{Acknowledgments}

LK is supported by JSPS KAKENHI for Early-Career Scientists (22K15209) and has been supported by the Miller Institute for Basic Research in Science and a Burroughs Wellcome Fund Collaborative Research Travel Grant. TT is supported by Brain/MINDS from AMED (JP19dm0207001) and JSPS KAKENHI (JP18H05432).

%

\appendix

\begin{widetext}

\section{\label{sec:mean}Mean-field equations}

\subsection{Replica partition function}

This derivation of mean-field equations governing the macroscopic behavior of our network is strongly influenced by Refs.~\onlinecite{Tsodyks.1988}, and \onlinecite{Fontanari.1990}, and \onlinecite{Hertz.2018}. All of our calculations will be performed in the thermodynamic limit where the network size $N \rightarrow \infty$. Our network, presented in \cref{sec:model}, is described by a Hamiltonian
\begin{align}
  H &= -\frac{1}{2N}\sum_\mn \sum_{i \neq j} (\eta^i_\mn + \zeta^i_\mn) (\eta^j_\mn + \zeta^j_\mn) S_i S_j + \theta \sum_i S_i \nonum
  &= -\frac{1}{2N}\sum_\mn \biggl[\sum_i (\eta^i_\mn + \zeta^i_\mn) S_i\biggr]^2 + \frac{1}{2N}\sum_\mn \sum_i \Bigl[(\eta^i_\mn + \zeta^i_\mn) S_i\Bigr]^2 + \theta \sum_i S_i.
\end{align}
To reiterate, $\ve S$ is the network activity and $\theta$ is the activity threshold. $\ve\eta_\mn$ and $\ve\zeta_\mn$ are rescaled sparse and dense patterns, respectively, for $\nu = 1, \ldots, s$ examples in each of $\mu = 1, \ldots, \alpha N$ concepts [\cref{eq:model-rescaled}]. Each rescaled pattern entry is randomly generated as follows:
\begin{align}
  \eta^i_\mn &=
      \begin{cases} (1-2\gamma)(1-a) & \textrm{with probability } a   \\
                    -(1-2\gamma)a    & \textrm{with probability } 1-a \end{cases} \nonum
  \zeta^i_\mn &=
      \begin{cases} \mathrlap{\hphantom{-}\zeta^i_\mu}\hphantom{(1-2\gamma)(1-a)} & \textrm{with probability } \frac{1+c}{2} \\
                    -\zeta^i_\mu & \textrm{with probability } \frac{1-c}{2} \end{cases} \nonum
  \zeta^i_\mu &=
      \begin{cases}  \mathrlap{\hphantom{-}\gamma}\hphantom{(1-2\gamma)(1-a)} & \textrm{with probability } \frac{1}{2}  \\
                    -\gamma & \textrm{with probability } \frac{1}{2} \end{cases}
  \label{eq:patterns}
\end{align}
for sparse pattern density $a$, dense pattern correlation $c$, and dense pattern storage strength $2 \gamma$. Their average values are 0, and the average overlaps between them are also 0 except for
\begin{align}
  \bigl\langle \zeta^i_\mn \zeta^i_\mu \bigr\rangle &= \gamma^2 c \nonum
  \bigl\langle \zeta^i_\mn \zeta^i_\mo \bigr\rangle &= \gamma^2 c^2.
  \label{eq:overlaps}
\end{align}

We will forgo introducing external fields. By averaging over examples and concepts,
\begin{equation}
  \frac{1}{2N}\sum_\mn \sum_i \Bigl[(\eta^i_\mn + \zeta^i_\mn) S_i\Bigr]^2
  \approx \frac{\alpha s}{2} \Bigl[ (1-2\gamma)^2 a(1-a) + \gamma^2 \Bigr] \sum_i S_i,
\end{equation}
If we define
\begin{equation}
  \Gamma^2 \equiv (1-2\gamma)^2 a(1-a) + \gamma^2,
\end{equation}
we obtain
\begin{equation}
  H = -\frac{1}{2N}\sum_\mn \biggl[\sum_i (\eta^i_\mn + \zeta^i_\mn) S_i\biggr]^2 + \biggl(\theta + \frac{\alpha s\Gamma^2}{2}\biggr) \sum_i S_i.
\end{equation}

To understand this system, we would like to calculate its free energy averaged over instantiations of the patterns: $F = -(1/\beta) \langle \log Z \rangle$. Here, $Z$ is the partition function, $\beta = 1/T$ is inverse temperature, and angle brackets indicate averages over $\eta^i_\mn$ and $\zeta^i_\mn$. Since we cannot directly average over the logarithm of the partition function $Z$, we use the replica trick by writing formally:
\begin{equation}
  \frac{F}{N} = -\frac{1}{\beta N} \langle \log Z \rangle = -\frac{1}{\beta N} \lim_{n\rightarrow 0} \frac{\langle Z^n \rangle - 1}{n} = -\frac{1}{\beta N} \lim_{n\rightarrow 0} \frac{1}{n} \log \langle Z^n \rangle.
  \label{eq:trick}
\end{equation}

We interpret $Z^n$ as a partition function for a set of replica networks $\rho = 1, \ldots, n$ with the same parameter values and stored patterns, but the neural activities $S^\rho_i$ may vary across replicas. The Hamiltonian of each replica is
\begin{equation}
  H^\rho = -\frac{1}{2N}\sum_\mn \biggl[\sum_i (\eta^i_\mn + \zeta^i_\mn) S^\rho_i\biggr]^2 + \biggl(\theta + \frac{\alpha s \Gamma^2}{2}\biggr) \sum_i S^\rho_i,
\end{equation}
and the replica partition function, averaged over patterns, is
\begin{equation}
  \langle Z^n \rangle = \left\langle \Tr_S \prod_\rho \exp \bigl[-\beta H^\rho\bigr] \right\rangle.
\end{equation}
The trace is evaluated over all neurons $i$ and replicas $\rho$. We invoke the standard Gaussian integral identity
\begin{equation}
  \int \dd m\,\ee^{-A m^2 + B m}
  = \sqrt{\frac{\pi}{A}} \ee^{B^2/4A}
  \label{eq:gaussian}
\end{equation}
to obtain
\begin{align}
  \langle Z^n \rangle = \Biggl\langle \Tr_S \prod_\rho \Biggl\{
    &\exp\biggl[-\beta \biggl(\theta + \frac{\alpha s \Gamma^2}{2}\biggr) \sum_i S^\rho_i\biggr] \nonum
    &\times \prod_\mn \int \dd m^\rho_\mn \biggl(\frac{\beta N}{2\pi}\biggr)^{\!\!\frac{1}{2}} \exp\biggl[ -\frac{\beta N}{2} (m^\rho_\mn)^2 + \beta m^\rho_\mn \sum_i (\eta^i_\mn + \zeta^i_\mn) S^\rho_i \biggr] \Biggr\} \Biggr\rangle.
  \label{eq:Zn-partition}
\end{align}

\subsection{Uncondensed patterns}

We search for a retrieval regime in which the network successfully recovers a sparse example $\ve\eta_{11}$, a dense example $\ve\zeta_{11}$, or a dense concept $\ve\zeta_1$. All stored patterns in other concepts $\mu > 1$ are called \emph{uncondensed} and will not significantly overlap with the network activity. We seek to expand in these small overlaps and integrate over them. First,
\begin{equation}
  \left\langle \prod_{\substack{\mu>1\\\nu\rho}} \exp\biggl[\beta m^\rho_\mn \sum_i (\eta^i_\mn + \zeta^i_\mn) S^\rho_i\biggr] \right\rangle
  = \prod_{\substack{i\\\mu>1}} \left\langle \prod_\nu \exp\biggl[\beta (\eta^i_\mn + \zeta^i_\mn) \sum_\rho m^\rho_\mn S^\rho_i\biggr] \right\rangle.
\end{equation}
Using
\begin{equation}
  Y_\nu \equiv \beta \sum_\rho m^\rho_\mn S^\rho_i,
\end{equation}
where we have suppressed dependence on $\mu$ and $i$ for convenience, we can write
\begin{equation}
  \biggl\langle \prod_\nu \exp\bigl[(\eta^i_\mn + \zeta^i_\mn) Y_\nu\bigr] \biggr\rangle
  = \biggl[ \prod_\nu \bigl\langle \exp[\eta^i_\mn Y_\nu] \bigr\rangle \biggr] \biggl\langle \prod_\nu \exp[\zeta^i_\mn Y_\nu] \biggr\rangle.
\end{equation}
For uncondensed patterns $\mu > 1$, $m^\rho_\mn \ll 1$ because, as we will derive later, it is the overlap between $\ve S^\rho$ and $\ve\eta_\mn + \ve\zeta_\mn$. Thus, we can crucially expand in $Y_\nu \ll 1$ and average over the uncondensed patterns:
\begin{equation}
  \bigl\langle \exp[\eta^i_\mn Y_\nu] \bigr\rangle
  \approx 1 + \frac{1}{2}(1-2\gamma)^2 a(1-a) Y_\nu^2 
  \approx \exp \biggl[ \frac{1}{2}(1-2\gamma)^2 a(1-a) Y_\nu^2 \biggr].
  \label{eq:Y-expand}
\end{equation}
Continuing,
\begin{align}
  \biggl\langle \prod_\nu \exp[\zeta^i_\mn Y_\nu] \biggr\rangle
  &\approx \frac{1}{2}\prod_\nu \biggl\{1 + \gamma cY_\nu + \frac{\gamma^2}{2}Y_\nu^2\biggr\} + \frac{1}{2}\prod_\nu \biggl\{1 - \gamma cY_\nu + \frac{\gamma^2}{2}Y_\nu^2\biggr\} \nonum
  &= 1 + \frac{1}{2} \sum_\no \Bigl[ \gamma^2(1-c^2)\delta_\no + \gamma^2c^2 \Bigr] Y_\nu Y_\omega \nonum
  &\approx \exp\biggl\{ \frac{1}{2} \sum_\no \Bigl[ \gamma^2(1-c^2)\delta_\no + \gamma^2c^2 \Bigr] Y_\nu Y_\omega \biggr\}.
\end{align}
Averaging is performed first over $\nu$, then over $\mu$ [\cref{eq:overlaps}]. Combining the equations above, we obtain
\begin{align}
  \left\langle \prod_{\substack{\mu>1\\\nu\rho}} \exp\biggl[\beta m^\rho_\mn \sum_i (\eta^i_\mn + \zeta^i_\mn) S^\rho_i\biggr] \right\rangle 
  = \prod_{\mu>1} \exp\biggl\{ \frac{\beta N}{2} \beta\Gamma^2 \sum_{\no\rs} \bigl[(1-\kappa^2)\delta_\no + \kappa^2\bigr] q^\rs m^\rho_\mn m^\sigma_\mo \biggr\}
  \label{eq:uncondensed}
\end{align}
if we define
\begin{equation}
  \kappa^2 \equiv \frac{\gamma^2}{\Gamma^2}c^2
\end{equation}
and enforce 
\begin{equation}
  q^\rs = \frac{1}{N}\sum_i S^\rho_i S^\sigma_i.
\end{equation}
We will do so by introducing the following integrals over delta-function representations:
\begin{equation}
  \prod_\rs \int \dd q^\rs\,\delta\bigg(q^\rs - \frac{1}{N}\sum_i S^\rho_i S^\sigma_i\bigg) \propto \int \biggl[\prod_\rs \dd q^\rs\,\dd r^\rs\biggr] \exp\biggl[-\frac{\beta^2\alpha N}{2} \sum_\rs q^\rs r^\rs + \frac{\beta^2\alpha}{2} \sum_{i\rs} r^\rs S^\rho_i S^\sigma_i\biggr],
  \label{eq:delta}
\end{equation}
where $r^\rs$ are additional auxiliary variables whose integration limits extend from $-\ii\infty$ to $\ii\infty$, and the factor of $\beta^2\alpha N/2$ is introduced for later convenience.

We can now integrate over the uncondensed overlaps $m^\rho_\mn$:
\begin{alignat}{2}
  &\mathrlap{ \left\langle \prod_{\substack{\mu>1\\\nu\rho}} \int \dd m^\rho_\mn \biggl(\frac{\beta N}{2\pi}\biggr)^{\!\frac{1}{2}} \exp\biggl[-\frac{\beta N}{2} (m^\rho_\mn)^2 + \beta m^\rho_\mn \sum_i (\eta^i_\mn + \zeta^i_\mn) S^\rho_i\biggr] \right\rangle } \nonum
  &\qquad &&{}\propto \prod_{\mu>1} \int \biggl[\prod_{\nu\rho} \dd m^\rho_\mn \biggl(\frac{\beta N}{2\pi}\biggr)^{\!\frac{1}{2}}\biggr] \exp\biggl[-\frac{\beta N}{2} \sum_{\no\rs} \bigl[ \delta_\no\delta^\rs - \beta\Gamma^2 \bigl( (1-\kappa^2)\delta_\no + \kappa^2 \bigr) q^\rs \bigr] m^\rho_\mn m^\sigma_\mo\biggr] \nonum
  &&&{}= \biggl(\det\bigl[ \delta_\no\delta^\rs - \beta\Gamma^2 \bigl( (1-\kappa^2)\delta_\no + \kappa^2 \bigr) q^\rs \bigr]^{-\frac{1}{2}}\biggr)^{\!\alpha N-1} \nonum
  &&&{}\approx \exp \biggl\{-\frac{\alpha N}{2} \Tr\log\Bigl[ \delta_\no\delta^\rs - \beta\Gamma^2 \bigl((1-\kappa^2)\delta_\no + \kappa^2\bigr) q^\rs \Bigr] \biggr\},
\end{alignat}
where $\alpha N \gg 1$ is the total number of concepts.

Thus, so far, our partition function is
\begin{align}
  \langle Z^n \rangle \propto \int &\biggl[\prod_{\nu\rho} \dd m^\rho_{1\nu} \biggl(\frac{\beta N}{2\pi}\biggr)^{\frac{1}{2}}\biggr] \biggl[\prod_\rs \dd q^\rs\,\dd r^\rs\biggr] \nonum
  &{}\times \exp\Biggl\{ -\beta N \biggl[\frac{1}{2} \sum_{\nu\rho} (m^\rho_{1\nu})^2 + \frac{\alpha}{2\beta} \Tr\log\bigl[ \delta_\no\delta^\rs - \beta\Gamma^2 \bigl((1-\kappa^2)\delta_\no + \kappa^2\bigr) q^\rs \bigr] + \frac{\beta\alpha}{2} \sum_\rs q^\rs r^\rs \biggr]\Biggr\} \nonum
  &{}\times \biggl\langle \Tr_S \exp\biggl[\beta \sum_{i\nu\rho} m^\rho_{1\nu} (\eta^i_{1\nu} + \zeta^i_{1\nu}) S^\rho_i - \beta \biggl(\theta + \frac{\alpha s\Gamma^2}{2}\biggr) \sum_{i\rho}S^\rho_i + \frac{\beta^2\alpha}{2} \sum_{i\rs} r^\rs S^\rho_i S^\sigma_i\biggr] \biggr\rangle.
  \label{eq:Zn-uncondensed}
\end{align}

\subsection{Condensed patterns}

Now we consider the target patterns, whose large overlaps cannot be expanded into Gaussians and integrated away. When retrieving sparse examples, the network overlaps significantly with one stored pattern $\ve\eta_{11}$, but not $\ve\eta_{1\nu}$ for $\nu > 1$ and $\ve\zeta_{1\nu}$, which are nearly orthogonal to $\ve\eta_{11}$. When retrieving dense examples or concepts, the network overlaps significantly with all stored examples $\ve\zeta_{1\nu}$ within the target concept because they are correlated, but not $\ve\eta_{1\nu}$. Thus, either $\sum_i \eta^i_{11} S^\rho_i$ or $\sum_i \zeta^i_{1\nu} S^\rho_i$ is much larger than the other terms in $\sum_i (\eta^i_{1\nu} + \zeta^i_{1\nu}) S^\rho_i$, so we replace
\begin{equation}
  \sum_{i\nu\rho} m^\rho_{1\nu} (\eta^i_{1\nu} + \zeta^i_{1\nu}) S^\rho_i \approx \sum_{i\nu\rho} m^\rho_{1\nu} \chi^i_{1\nu} S^\rho_i,
\end{equation}
where $\chi^i_{1\nu} = \eta^i_{11}\delta_{1\nu}$ or $\zeta^i_{1\nu}$ depending on whether we are considering recovery of sparse or dense patterns. These patterns with significant overlaps are called \emph{condensed} patterns.

We now invoke self-averaging over the $i$ indices. For any function $G(\chi^i, S_i)$,
\begin{align}
  \Tr_S \exp\biggl[\sum_i G(\chi^i, S_i)\biggr]
  &= \prod_i \Tr_{S_i} \exp G(\chi^i, S_i) \nonum
  &= \exp\biggl[ \sum_i \log \Tr_{S_i} \exp G(\chi^i, S_i)\biggr] \nonum
  &= \exp\biggl[ N \bigl\langle \log \Tr_S \exp G(\chi, S) \bigr\rangle \biggr].
\end{align}
Now $\chi$ and $S$ represent the pattern entry and activity of a single neuron. This single neuron is representative of the entire network because pattern entries are generated independently for each neuron, so we can replace the average over neurons $i$ with an average over possible pattern entries $\chi$.  In doing so, we no longer need to pattern-average the trace of the exponential in \cref{eq:Zn-uncondensed}; critically, that average has been subsumed by a pattern average inside the exponential, which allows us to write
\begin{equation}
  \langle Z^n \rangle \propto \int \biggl[\prod_{\nu\rho} \dd m^\rho_{1\nu} \biggl(\frac{\beta N}{2\pi}\biggr)^{\!\frac{1}{2}}\biggr] \biggl[\prod_\rs \dd q^\rs\,\dd r^\rs\biggr] \exp[-\beta N f],
  \label{eq:Zn-condensed}
\end{equation}
where
\begin{align}
  f ={}& \frac{1}{2} \sum_{\nu\rho} (m^\rho_{1\nu})^2 + \frac{\alpha}{2\beta} \Tr\log\bigl[ \delta_\no\delta^\rs - \beta\Gamma^2 \bigl((1-\kappa^2)\delta_\no + \kappa^2\bigr) q^\rs \bigr] + \frac{\beta\alpha}{2} \sum_\rs q^\rs r^\rs \nonum
  &{}- \frac{1}{\beta} \left\langle \log\Tr_S \exp\Biggl\{ \beta \biggl[ \sum_{\nu\rho} m^\rho_{1\nu} \chi_{1\nu} S^\rho - \biggl(\theta + \frac{\alpha s \Gamma^2}{2}\biggr) \sum_\rho S^\rho + \frac{\beta\alpha}{2} \sum_\rs r^\rs S^\rho S^\sigma \biggr] \Biggr\} \right\rangle.
  \label{eq:f-condensed}
\end{align}
The replica partition function is now written in a form amenable to the saddle-point approximation. That is, in the $N \rightarrow \infty$ limit, we can replace integrals in \cref{eq:Zn-condensed} with the integrand evaluated where derivatives of $f$ with respect to the variables of integration equal 0.

\subsection{Saddle-point equations for interpretation}

Before proceeding with further simplifying $f$ by invoking replica symmetry, we seek to obtain physical interpretations for $m$, $q$, and $r$, which will serve as the order parameters of our system. To do so, we must recall several previously derived forms of the replica partition function and apply the saddle-point conditions to them.

Recall \cref{eq:uncondensed,eq:delta} obtained after introducing $q$ and $r$ but before integrating over the uncondensed patterns. Using those expressions in the partition function and performing self-averaging similarly to above, we can obtain
\begin{equation}
  \langle Z^n \rangle \propto \int \biggl[\prod_{\mn\rho} \dd m^\rho_\mn \biggl(\frac{\beta N}{2\pi}\biggr)^{\!\frac{1}{2}}\biggr] \biggl[\prod_\rs \dd q^\rs\,\dd r^\rs\biggr] \exp[-\beta N f],
  \label{eq:Zn-saddle1}
\end{equation}
where 
\begin{align}
  f ={}& \frac{1}{2} \sum_{\mn\rho} (m^\rho_\mn)^2 - \frac{\beta\Gamma^2 }{2} \sum_{\substack{\mu>1\\\no\rs}} \bigl[(1-\kappa^2)\delta_\no + \kappa^2\bigr] q^\rs m^\rho_\mn m^\sigma_\mo + \frac{\beta\alpha}{2} \sum_\rs q^\rs r^\rs \nonum
  &{}- \frac{1}{\beta} \left\langle \log\Tr_S \exp\Biggl\{ \beta \biggl[ \sum_{\nu\rho} m^\rho_{1\nu} \chi_{1\nu} S^\rho - \biggl(\theta + \frac{\alpha s \Gamma^2}{2}\biggr) \sum_\rho S^\rho + \frac{\beta\alpha}{2} \sum_\rs r^\rs S^\rho S^\sigma \biggr] \Biggr\} \right\rangle \nonum
  ={}& \frac{1}{2} \sum_{\mn\rho} (m^\rho_\mn)^2 - \frac{\beta\Gamma^2 }{2} \sum_{\substack{\mu>1\\\no\rs}} \bigl[(1-\kappa^2)\delta_\no + \kappa^2\bigr] q^\rs m^\rho_\mn m^\sigma_\mo + \frac{\beta\alpha}{2} \sum_\rs q^\rs r^\rs - \frac{1}{\beta} \bigl\langle \log\Tr_S \exp[ -\beta \mcH ] \bigr\rangle,
  \label{eq:f-saddle1}
\end{align}
and
\begin{equation}
  \mcH \equiv - \sum_{\nu\rho} m^\rho_{1\nu} \chi_{1\nu} S^\rho + \biggl(\theta + \frac{\alpha s \Gamma^2}{2}\biggr) \sum_\rho S^\rho - \frac{\beta\alpha}{2} \sum_\rs r^\rs S^\rho S^\sigma
  \label{eq:H-saddle1}
\end{equation}
is the effective single-neuron Hamiltonian across replicas.

At the saddle point, derivatives of $f$ with respect to variables of integration are 0, so
\begin{align}
  0 &= \parpar{f}{m^\rho_{1\nu}} = m^\rho_{1\nu} - \left\langle \frac{ \Tr_S\,\chi_{1\nu} S^\rho \exp[-\beta\mcH] }{ \Tr_S \exp[-\beta\mcH] } \right\rangle \nonum
  \Rightarrow m^\rho_{1\nu} &= \left\langle \chi_{1\nu} \overline{S^\rho} \right\rangle, 
  \label{eq:saddle-m} \\
  \nonum
  0 &= \parpar{f}{r^\rs} = \frac{\beta\alpha}{2} q^\rs - \frac{\beta\alpha}{2} \left\langle \frac{ \Tr_S\, S^\rho S^\sigma \exp[-\beta\mcH] }{ \Tr_S \exp[-\beta\mcH] } \right\rangle \nonum
  \Rightarrow q^\rs &= \left\langle \overline{S^\rho S^\sigma} \right\rangle = \begin{cases} \left\langle \overline{S^\rho} \cdot \overline{S^\sigma} \right\rangle & \rho \neq \sigma \\ \left\langle \overline{S^\rho} \right\rangle & \rho = \sigma, \end{cases}
  \label{eq:saddle-q} \\
  \nonum
  0 &= \parpar{f}{q^\rs} = \frac{\beta\alpha}{2} r^\rs - \frac{\beta}{2} \sum_{\substack{\mu>1\\\no}} \Gamma^2\bigl((1-\kappa^2)\delta_\no + \kappa^2\bigr) m^\rho_\mn m^\sigma_\mo \nonum
  \Rightarrow r^\rs &= \frac{1}{\alpha} \sum_{\substack{\mu>1\\\no}} \Gamma^2\bigl((1-\kappa^2)\delta_\no + \kappa^2\bigr) m^\rho_\mn m^\sigma_\mo.
  \label{eq:saddle-r}
\end{align}
Bars over variables represent the thermodynamic ensemble average. Thus, $m^\rho_{1\nu}$ is the overlap of the network with the condensed pattern to be recovered, $q^\rs$ is the Edwards-Anderson order parameter reflecting the overall neural activity, and $r^\rs$ represents interference from network overlap with uncondensed patterns $m^\rho_\mn$.

To explicitly see that $m^\rho_\mn$ describes the overlap of the network with uncondensed patterns for $\mu > 1$, recall \cref{eq:Zn-partition} obtained before introducing $q$ and $r$. By introducing $\chi$ and performing self-averaging similarly to above, we can obtain
\begin{equation}
  \langle Z^n \rangle = \int \biggl[\prod_{\mn\rho} \dd m^\rho_\mn \biggl(\frac{\beta N}{2\pi}\biggr)^{\!\frac{1}{2}}\biggr] \exp[-\beta N f],
  \label{eq:Zn-saddle2}
\end{equation}
where
\begin{align}
  f &= \frac{1}{2} \sum_{\mn\rho} (m^\rho_\mn)^2 - \frac{1}{\beta} \left\langle \log\Tr_S \exp\Biggl\{ \beta \biggl[ \sum_{\nu\rho} m^\rho_{1\nu} \chi_{1\nu} S^\rho - \biggl(\theta + \frac{\alpha s \Gamma^2}{2}\biggr) \sum_\rho S^\rho + \sum_{\substack{\mu>1\\\nu\rho}} m^\rho_\mn (\eta_\mn + \zeta_\mn) S^\rho \biggr] \Biggr\} \right\rangle \nonum
  &= \frac{1}{2} \sum_{\mn\rho} (m^\rho_\mn)^2 - \frac{1}{\beta} \bigl\langle \log\Tr_S \exp[ -\beta \mcH ] \bigr\rangle,
  \label{eq:f-saddle2}
\end{align}
and
\begin{equation}
  \mcH \equiv - \sum_{\nu\rho} m^\rho_{1\nu} \chi_{1\nu} S^\rho + \biggl(\theta + \frac{\alpha s \Gamma^2}{2}\biggr) \sum_\rho S^\rho - \sum_{\substack{\mu>1\\\nu\rho}} m^\rho_\mn (\eta_\mn + \zeta_\mn) S^\rho
  \label{eq:H-saddle2}
\end{equation}
is the effective single-neuron Hamiltonian. At the saddle point, this Hamiltonian is equivalent to the form in \cref{eq:H-saddle1} due to \cref{eq:uncondensed,eq:saddle-r}. The saddle-point condition applied to \cref{eq:Zn-saddle2,eq:f-saddle2} yields
\begin{align}
  0 &= \parpar{f}{m^\rho_\mn} = m^\rho_\mn - \left\langle \frac{ \Tr_S\,(\eta_\mn+\zeta_\mn) S^\rho \exp[-\beta\mcH] }{ \Tr_S \exp[-\beta\mcH] } \right\rangle \nonum
  \Rightarrow m^\rho_\mn &= \left\langle (\eta_\mn+\zeta_\mn) \overline{S^\rho} \right\rangle
  \cond{for} \mu > 1.
\end{align}
Thus $m^\rho_\mn$ is indeed the network overlap with $\eta_\mn+\zeta_\mn$ for $\mu > 1$. As asserted \emph{ex ante} to derive \cref{eq:Y-expand}, we expect it to be small.

\subsection{Replica-symmetry ansatz}

We are now finished with seeking physical interpretations for order parameters, and we return to the primary task of calculating the free energy \cref{eq:trick} using \cref{eq:Zn-condensed,eq:f-condensed}. To do so, we assume replica symmetry:
\begin{equation}
  m^\rho_\mn = m_\mn, \qquad q^\rs = q, \qquad q^{\rho\rho} = q_0, \qquad r^\rs = r, \qquad r^{\rho\rho} = r_0.
\end{equation}
Our expression for $f$ then becomes
\begin{align}
  f ={}& \frac{1}{2} n \sum_\nu (m_{1\nu})^2 + \frac{\alpha}{2\beta} \Tr\log\bigl[ \delta_\no\delta^\rs - \beta\Gamma^2 \bigl((1-\kappa^2)\delta_\no + \kappa^2\bigr) \bigl((q_0-q)\delta^\rs + q\bigr) \bigr] + \frac{\beta\alpha n}{2} q_0 r_0 + \frac{\beta\alpha n(n-1)}{2} qr \nonum
  &{}- \frac{1}{\beta} \left\langle \log\Tr_S \exp\Biggl\{ \beta\biggl[ \biggl(\sum_\nu m_{1\nu} \chi_{1\nu}  - \theta - \frac{\alpha s\Gamma^2}{2} + \frac{\beta\alpha}{2} (r_0 - r)\biggr) \sum_\rho S^\rho + \frac{\beta\alpha}{2} r \biggl(\sum_\rho S^\rho\biggr)^{\!\!2} \biggr]\Biggr\} \right\rangle.
  \label{eq:f-symmetric}
\end{align}

The eigenvalues of a constant $n \times n$ matrix with entries $A$ are $nA$ with multiplicity $1$ and $0$ with multiplicity $n-1$. Thus, the second term in \cref{eq:f-symmetric} under the limit in \cref{eq:trick} becomes
\begin{alignat}{3}
  &\mathrlap{ \lim_{n \rightarrow 0} \frac{1}{n} \Tr\log\bigl[ \delta_\no\delta^\rs - \beta\Gamma^2 \bigl((1-\kappa^2)\delta_\no + \kappa^2\bigr) \bigl((q_0-q)\delta^\rs + q\bigr) \bigr] } \nonum
  &\qquad{}={} && \lim_{n \rightarrow 0} \frac{1}{n} \biggl\{ &&\log\bigl[ 1 - \beta\Gamma^2 (1+s\kappa^2-\kappa^2) (q_0-q+nq) \bigr] + (n-1) \log\bigl[ 1 - \beta\Gamma^2 (1+s\kappa^2-\kappa^2) (q_0-q) \bigr]\nonum
  &&&&& + (s-1) \log\bigl[ 1 - \beta\Gamma^2 (1-\kappa^2) (q_0-q+nq) \bigr] + (s-1)(n-1) \log\bigl[ 1 - \beta\Gamma^2 (1-\kappa^2) (q_0-q) \bigr] \biggr\}\nonum
  &\qquad{}={} && \mathrlap{ (s-1) \biggl[\log\bigl[1 - Q (1-\kappa^2)\bigr] - \frac{\beta q \Gamma^2 (1-\kappa^2)}{1 - Q (1-\kappa^2)}\biggr] + \log\bigl[1 - Q (1+s\kappa^2-\kappa^2)\bigr] - \frac{\beta q \Gamma^2 (1+s\kappa^2-\kappa^2)}{1 - Q (1+s\kappa^2-\kappa^2)}, }\nonum
  &\qquad{}\equiv{} && \Lambda[q,q_0],
\end{alignat}
where
\begin{equation}
  Q \equiv \beta(q_0-q)\Gamma^2.
  \label{eq:Q}
\end{equation}

To evaluate the last term in \cref{eq:f-symmetric}, we can use another Gaussian integral [\cref{eq:gaussian}] to perform the trace over $S$ in the limit $n \rightarrow 0$:
\begin{align}
  &\left\langle \log\Tr_S \exp\Biggl\{ \beta \biggl[\biggl(\sum_\nu m_{1\nu} \chi_{1\nu} - \theta - \frac{\alpha s\Gamma^2}{2} + \frac{\beta\alpha}{2} (r_0-r)\biggr) \sum_\rho S^\rho + \frac{\beta\alpha}{2} r \biggl(\sum_\rho S^\rho\biggr)^{\!\!2} \biggr]\Biggr\} \right\rangle \nonum
  &\qquad {}= \left\langle \log\Tr_S \int\!\!\frac{\dd z}{\sqrt{2\pi}} \ee^{-z^2/2} \exp\Biggl\{ \beta \biggl[\biggl(\sum_\nu m_{1\nu} \chi_{1\nu} - \theta + \frac{\beta\alpha}{2} (r_0-r) - \frac{\alpha s\Gamma^2}{2} + \sqrt{\alpha r} z\biggr) \sum_\rho S^\rho \biggr]\Biggr\} \right\rangle \nonum
  &\qquad {}= \left\langle \log\int\!\!\frac{\dd z}{\sqrt{2\pi}} \ee^{-z^2/2} \Biggl\{1 + \exp\biggl[ \beta \biggl(\sum_\nu m_{1\nu} \chi_{1\nu} - \theta + \frac{\beta\alpha}{2} (r_0-r) - \frac{\alpha s\Gamma^2}{2} + \sqrt{\alpha r} z\biggr) \biggr] \Biggr\}^{\!n} \right\rangle \nonum
  &\qquad {}\approx \left\langle \log\int\!\!\frac{\dd z}{\sqrt{2\pi}} \ee^{-z^2/2} \Biggl\{1 + n \log \biggl\{1 + \exp\biggl[ \beta \biggl(\sum_\nu m_{1\nu} \chi_{1\nu} - \theta + \frac{\beta\alpha}{2} (r_0-r) - \frac{\alpha s\Gamma^2}{2} + \sqrt{\alpha r} z\biggr) \biggr] \biggl\} \Biggr\} \right\rangle \nonum
  &\qquad {}\approx n\,\left\langle \int\!\!\frac{\dd z}{\sqrt{2\pi}} \ee^{-z^2/2} \log \biggl\{1 + \exp\biggl[ \beta \biggl(\sum_\nu m_{1\nu} \chi_{1\nu} - \theta + \frac{\beta\alpha}{2} (r_0-r) - \frac{\alpha s\Gamma^2}{2} + \sqrt{\alpha r} z\biggr) \biggr] \biggr\} \right\rangle.
\end{align}

The free energy \cref{eq:trick} under replica symmetry becomes
\begin{align}
  \frac{F}{N} ={}& \frac{1}{2} \sum_\nu (m_{1\nu})^2 + \frac{\alpha}{2\beta} \Lambda[q,q_0] + \frac{\beta\alpha}{2} (q_0 r_0 - qr) \nonum
  &{}- \frac{1}{\beta} \left\llangle \log \biggl\{1 + \exp\biggl[ \beta \biggl(\sum_\nu m_{1\nu} \chi_{1\nu} - \theta + \frac{\beta\alpha}{2} (r_0-r) - \frac{\alpha s\Gamma^2}{2} + \sqrt{\alpha r} z\biggr) \biggr] \biggr\} \right\rrangle,
\end{align}
where now the double angle brackets indicate an average over $\chi_{1\nu}$ as well as the Gaussian variable $z$.

\subsection{Mean-field equations}

We can now minimize this free energy over the order parameters by setting derivatives of $F$ to zero, which yields the mean-field equations. This step is equivalent to applying the saddle-point approximation to replica-symmetric $f$ in the $n \rightarrow 0$ limit. We first note that
\begin{equation}
  \parpar{\Lambda}{q} = (s-1) \frac{\beta^2 q\Gamma^4 (1-\kappa^2)^2}{\bigl(1 - Q(1-\kappa^2)\bigr)^{\!2}} + \frac{\beta^2 q\Gamma^4 (1+s\kappa^2-\kappa^2)^2}{\bigl(1 - Q(1+s\kappa^2-\kappa^2)\bigr)^{\!2}}. 
\end{equation}
The combined fraction has numerator $\beta^2 q\Gamma^4$ multiplied by
\begin{align}
  &(s-1)\Bigl[\bigl(1-\kappa^2\bigr) \bigl(1 - Q(1+s\kappa^2-\kappa^2)\bigr)\Bigr]^2 + \Bigl[\bigl(1+s\kappa^2-\kappa^2\bigr) \bigl(1 - Q(1-\kappa^2)\bigr)\Bigr]^2 \nonum
  &\qquad{}= s \bigl(1 - Q(1-\kappa^2)(1+s\kappa^2-\kappa^2)\bigr)^{\!2} + s(s-1)\kappa^4.
\end{align}
Meanwhile,
\begin{align}
  \parpar{\Lambda}{q_0} &= (s-1)\biggl[ -\frac{\beta\Gamma^2 (1-\kappa^2)}{1 - Q(1-\kappa^2)} - \frac{\beta^2 q\Gamma^4 (1-\kappa^2)^2}{\bigl(1 - Q(1-\kappa^2)\bigr)^{\!2}} \biggr] - \frac{\beta (1+s\kappa^2-\kappa^2)}{1 - Q(1+s\kappa^2-\kappa^2)} - \frac{\beta^2 q\Gamma^4 (1+s\kappa^2-\kappa^2)^2}{\bigl(1 - Q(1+s\kappa^2-\kappa^2)\bigr)^{\!2}} \nonum
  &= -\parpar{\Lambda}{q} - \beta\Gamma^2\biggl[ \frac{(s-1)(1-\kappa^2)}{1 - Q(1-\kappa^2)} + \frac{1+s\kappa^2-\kappa^2}{1 - Q(1+s\kappa^2-\kappa^2)} \biggr].
\end{align}
The combined fraction inside the square brackets has numerator
\begin{align}
  (s-1)(1-\kappa^2)\bigl(1 - Q(1+s\kappa^2-\kappa^2)\bigr) + (1+s\kappa^2-\kappa^2)\bigl(1 - Q(1-\kappa^2)\bigr)= s\bigl(1 - Q(1-\kappa^2)(1+s\kappa^2-\kappa^2)\bigr).
\end{align}

Thus, derivatives of $F$ with respect to the order parameters are
\begin{align}
  0 &= \parpar{F}{q} = \frac{\alpha}{2 \beta} \biggl[\beta^2 qs\Gamma^4 \frac{\bigl(1 - Q(1-\kappa^2)(1+s\kappa^2-\kappa^2)\bigr)^{\!2} + (s-1)\kappa^4}{\bigl(1 - Q(1-\kappa^2)\bigr)^{\!2}\bigl(1 - Q(1+s\kappa^2-\kappa^2)\bigr)^{\!2}} \biggr] - \frac{\beta\alpha}{2} r \nonum
  \Rightarrow r &= qs\Gamma^4 \frac{\bigl(1 - Q(1-\kappa^2)(1+s_0\kappa^2)\bigr)^{\!2} + s_0\kappa^4}{\bigl(1 - Q(1-\kappa^2)\bigr)^{\!2}\bigl(1 - Q(1+s_0\kappa^2)\bigr)^{\!2}}
  \label{eq:r-symmetric} \\
  \nonum
  0 &= \parpar{F}{q_0} = \frac{\alpha}{2 \beta} \biggl[ -\parpar{\Lambda}{q} -\beta s\Gamma^2 \frac{1 - Q(1-\kappa^2)(1+s\kappa^2-\kappa^2)}{\bigl(1 - Q(1-\kappa^2)\bigr) \bigl(1 - Q(1+s\kappa^2-\kappa^2)\bigr)} \biggr] + \frac{\beta\alpha}{2} r_0 \nonum
  \Rightarrow r_0 &= r + \frac{s\Gamma^2}{\beta} \frac{1 - Q(1-\kappa^2)(1+s_0\kappa^2)}{\bigl(1 - Q(1-\kappa^2)\bigr) \bigl(1 - Q(1+s_0\kappa^2)\bigr)}
  \label{eq:r0-symmetric} \\
  \nonum
  0 &= \parpar{F}{m_{1\nu}} = m_{1\nu} - \left\llangle \chi_{1\nu} \sig[\beta h] \right\rrangle \nonum
  \Rightarrow m_{1\nu} &= \left\llangle \chi_{1\nu} \sig[\beta h] \right\rrangle,
  \label{eq:m-symmetric}
\end{align}
where $\sig(x) \equiv 1/(1+\ee^{-x})$ and
\begin{align}
  s_0 &\equiv s-1, \nonum
  h &\equiv \sum_\nu m_{1\nu} \chi_{1\nu} - \theta + \frac{\beta\alpha}{2} (r_0 - r) - \frac{\alpha s\Gamma^2}{2} + \sqrt\ar z.
\end{align}
$h$ is the local field under the mean-field approximation. We can simplify it via
\begin{align}
  \frac{\beta\alpha}{2} (r_0 - r) - \frac{\alpha s\Gamma^2}{2}
  &= \frac{\alpha s\Gamma^2}{2} \cdot \frac{1 - Q(1-\kappa^2)(1+s_0\kappa^2) - \bigl(1 - Q(1-\kappa^2)\bigr) \bigl(1 - Q(1+s_0\kappa^2)\bigr)}{\bigl(1 - Q(1-\kappa^2)\bigr) \bigl(1 - Q(1+s_0\kappa^2)\bigr)} \nonum
  &= \frac{Q\alpha s\Gamma^2}{2} \cdot \frac{1+s_0\kappa^4 - Q(1-\kappa^2)(1+s_0\kappa^2)}{\bigl(1 - Q(1-\kappa^2)\bigr) \bigl(1 - Q(1+s_0\kappa^2)\bigr)}.
\end{align}
Thus,
\begin{align}
  h &= \sum_\nu m_{1\nu} \chi_{1\nu} - \phi + \sqrt{\alpha r} z, \nonum
  \phi &\equiv \theta - \frac{Q\alpha s\Gamma^2}{2} \cdot \frac{1+s_0\kappa^4 - Q(1-\kappa^2)(1+s_0\kappa^2)}{\bigl(1 - Q(1-\kappa^2)\bigr) \bigl(1 - Q(1+s_0\kappa^2)\bigr)}
  \label{eq:hphi}
\end{align}
$\phi$ is the shifted threshold; we shall see that in retrieval regimes, it is almost identical to $\theta$.

Continuing, and using the identities $\int \dd z\,\ee^{-z^2/2} z f(z) = \int \dd z\,\ee^{-z^2/2}\ \dd f(z)/\dd z$ and $\dd\sig(x)/\dd x = \sig(x) - \sig(x)^2$,
\begin{align}
  0 &= \parpar{F}{r} = -\frac{\beta\alpha}{2} q - \frac{\sqrt{\alpha}}{2\sqrt{r}} \left\llangle z \sig[\beta h] \right\rrangle + \frac{\beta\alpha}{2} \left\llangle \sig[\beta h] \right\rrangle \nonum
  \Rightarrow q &= \mathrlap{ \left\llangle \sig[\beta h]^2 \right\rrangle }
  \label{eq:q-symmetric} \\
  \nonum
  0 &= \parpar{F}{r_0} = \frac{\beta\alpha}{2} q_0 - \frac{\beta\alpha}{2} \left\llangle \sig[\beta h] \right\rrangle \nonum
  \Rightarrow q_0 &= \mathrlap{ \left\llangle \sig[\beta h] \right\rrangle. }
  \label{eq:q0-symmetric}
  \end{align}
Thus, we recover the mean-field equations presented in \cref{eq:model-meanfield}.

\subsection{Zero-temperature limit}

From now on, we only consider the $T = 0$ limit with $\beta\rightarrow\infty$. In this limit,
\begin{equation}
  \int\!\!\frac{\dd z}{\sqrt{2\pi}} \ee^{-z^2/2}\, \sig[Az+B]
  \approx \int\!\!\frac{\dd z}{\sqrt{2\pi}} \ee^{-z^2/2}\, \Theta[Az+B]
  = \frac{1}{2} \biggl(1 + \erf\frac{B}{\sqrt{2}A}\biggr),
\end{equation}
where $\Theta$ is the Heaviside step function and $\erf$ is the error function. Thus, \cref{eq:m-symmetric,eq:q-symmetric,eq:r-symmetric} become
\begin{align}
  m_{1\nu} &\approx \biggl\llangle \chi_{1\nu} \,\Theta\biggl[\sum_\nu m_{1\nu}\chi_{1\nu}-\phi+\sqrt\ar z\biggr] \biggr\rrangle 
  = \frac{1}{2}\Biggl\langle \chi_{1\nu} \erf\frac{\sum_\nu m_{1\nu} \chi_{1\nu} - \phi}{\sqrt{2\ar}} \Biggr\rangle,
  \label{eq:m-zeroT} \\
  \nonum
  q &\approx \biggl\llangle \Theta\biggl[\sum_\nu m_{1\nu}\chi_{1\nu}-\phi+\sqrt\ar z\biggr]^2 \biggr\rrangle
  = \frac{1}{2} \Biggl[1 + \Biggl\langle \erf\frac{\sum_\nu m_{1\nu} \chi_{1\nu} - \phi}{\sqrt{2\ar}} \Biggr\rangle \Biggr],
  \label{eq:q-zeroT} \\
  \nonum
  r &\approx \frac{s\Gamma^4}{2} \cdot \frac{\bigl(1 - Q(1-\kappa^2)(1+s_0\kappa^2)\bigr)^{\!2} + s_0\kappa^4}{\bigl(1 - Q(1-\kappa^2)\bigr)^{\!2}\bigl(1 - Q(1+s_0\kappa^2)\bigr)^{\!2}} \Biggl[1 + \Biggl\langle \erf\frac{\sum_\nu m_{1\nu} \chi_{1\nu} - \phi}{\sqrt{2\ar}} \Biggr\rangle \Biggr].
  \label{eq:r-zeroT}
\end{align}
Here, single angle brackets again indicate an average over $\chi_{1\nu}$, with the average over $z$ performed. The formula for $m_{1\nu}$ was obtained using $\langle \chi_{1\nu} \rangle = 0$ for both sparse and dense patterns.

Also when $\beta \rightarrow \infty$, 
\begin{align}
  \int\!\!\frac{\dd z}{\sqrt{2\pi}} \ee^{-z^2/2} \Bigl\{ \sig[\beta(Az+B)] - \sig[\beta(Az+B)]^2 \Bigr\}
  &= \int\!\!\frac{\dd z}{\sqrt{2\pi}\beta A} \ee^{-z^2/2} \parpar{}{z}\sig[\beta(Az+B)] \nonum
  &\approx \int\!\!\frac{\dd z}{\sqrt{2\pi}\beta A} \ee^{-z^2/2} \parpar{}{z}\Theta[Az+B] \nonum
  &= \int\!\!\frac{\dd z}{\sqrt{2\pi}\beta} \ee^{-z^2/2} \delta[Az+B] \nonum
  &= \frac{1}{\sqrt{2\pi}\beta|A|} \ee^{-B^2/2A^2}.
\end{align}
We use this identity to simplify \cref{eq:Q} via \cref{eq:q-symmetric,eq:q0-symmetric}:
\begin{equation}
  Q \approx \Gamma^2 \biggl\llangle \delta\biggl[ \sum_\nu m_{1\nu}\chi_{1\nu}-\phi+\sqrt\ar z \biggr] \biggr\rrangle 
  = \frac{\Gamma^2}{\sqrt{2\pi\ar}} \Biggl\langle\exp\Biggl[ -\frac{\bigl(\sum_\nu m_{1\nu} \chi_{1\nu} - \phi\bigr)^{\!2}}{2\ar}\Biggr]\Biggr\rangle.
  \label{eq:Q-zeroT}
\end{equation}
\Cref{eq:m-zeroT,eq:r-zeroT,eq:Q-zeroT} are the zero-temperature mean-field equations connecting the order parameters $m_{1\nu}$, $r$, and $Q$ (we no longer need $q$, $q_0$, and $r_0$). All further derivations will start with these equations.

\section{\label{sec:se}Capacity for sparse examples}

\subsection{Sparse mean-field equations}

The mean-field equations \cref{eq:m-zeroT,eq:r-zeroT,eq:Q-zeroT} involve a generic target pattern $\chi_{1\nu}$. We now consider the case where the network recovers a sparse example, so $\chi_{1\nu} = \eta_{11} \delta_{1\nu}$. Using this expression, we can simplify the mean-field equations and find the critical example load $s_\rmc$ above which sparse examples can no longer be retrieved. In this section, we take the sparse limit with $a \ll 1$.

For convenience, we rename $m \equiv m_{11}$ and $\eta \equiv \eta_{11}$. From \cref{eq:patterns}, we have 
\begin{equation}
  \eta =
      \begin{cases} (1-2\gamma)(1-a) & \textrm{with probability } a   \\
                    -(1-2\gamma)a    & \textrm{with probability } 1-a \end{cases}
      \approx \begin{cases} 1-2\gamma & \textrm{with probability } a   \\
                            0         & \textrm{with probability } 1 \end{cases}
\end{equation}
Then, \cref{eq:m-zeroT,eq:r-zeroT,eq:Q-zeroT} become
\begin{align}
  m &= \frac{1}{2}\biggl\langle \eta \erf\frac{m\eta-\phi}{2\ar} \biggr\rangle = \frac{(1-2\gamma)a}{2} \Biggl\{\erf\frac{\phi}{\sqrt{2\ar}} + \erf\frac{(1-2\gamma)m-\phi}{\sqrt{2\ar}} \Biggr\}, \\
  \nonum
  r &= \frac{s\Gamma^4}{2} \cdot \frac{\bigl(1 - Q(1-\kappa^2)(1+s_0\kappa^2)\bigr)^{\!2} + s_0\kappa^4}{\bigl(1 - Q(1-\kappa^2)\bigr)^{\!2}\bigl(1 - Q(1+s_0\kappa^2)\bigr)^{\!2}} \biggl[1 + \biggl\langle \erf\frac{m\eta-\phi}{\sqrt{2\ar}} \biggr\rangle \biggr] \nonum
  &= \frac{s\Gamma^4}{2} \cdot \frac{\bigl(1 - Q(1-\kappa^2)(1+s_0\kappa^2)\bigr)^{\!2} + s_0\kappa^4}{\bigl(1 - Q(1-\kappa^2)\bigr)^{\!2}\bigl(1 - Q(1+s_0\kappa^2)\bigr)^{\!2}} \Biggl\{1 - \erf\frac{\phi}{\sqrt{2\ar}} + a \erf\frac{(1-2\gamma)m-\phi}{\sqrt{2\ar}} \Biggr\}, \\
  \nonum
  Q &= \frac{\Gamma^2}{\sqrt{2\pi\ar}} \biggl\langle\exp\biggl[ -\frac{(m\eta - \phi)^2}{2\ar}\biggr]\biggr\rangle = \frac{\Gamma^2}{\sqrt{2\pi\ar}} \Biggl\{\exp\biggl[-\frac{\phi^2}{2\ar}\biggr] + a\exp\biggl[-\frac{((1-2\gamma)m - \phi)^2}{2\ar}\biggr] \Biggr\}.
\end{align}

We will soon see that these equations yield $Q \ll 1$ in the retrieval regime. In that case,
\begin{align}
  m &= \frac{(1-2\gamma)a}{2} \Biggl\{\erf\frac{\phi}{\sqrt{2\ar}} + \erf\frac{(1-2\gamma)m-\phi}{\sqrt{2\ar}} \Biggr\},
  \label{eq:A-m} \\
  r &= \frac{s (1+s_0\kappa^4) \Gamma^4}{2} \Biggl\{1 - \erf\frac{\phi}{\sqrt{2\ar}} + a \erf\frac{(1-2\gamma)m-\phi}{\sqrt{2\ar}} \Biggr\}.
  \label{eq:A-r}
\end{align}
These mean-field equations for sparse examples are presented in \cref{eq:model-A} with $m$ replaced by its original name $m_{11}$. They can be numerically solved to find regimes of successful retrieval, but we will analyze them further in search of formulas for the capacity $s_\rmc$. 

In the limit that the network only stores sparse patterns with $2\gamma = 0$, these mean-field equations simplify to those of the sparse Hopfield network \cite{Tsodyks.1988}. Note that their error function obeys $\erf x \rightarrow -1$ as $x \rightarrow \infty$, which is commonly called the complementary error function. To match our equations to theirs, make the replacements $\gamma \rightarrow 0$, $\Gamma \rightarrow a(1-a)$, $s \rightarrow 1$, and $\erf x \rightarrow 1 - \erf x$ in our equations, and eliminate higher orders of $a \ll 1$.

Instead of invoking this limit to match the equations exactly, we can rewrite the mean-field equations \cref{eq:A-m,eq:A-r} in the form of Ref.~\onlinecite{Tsodyks.1988} with the rescalings
\begin{align}
  m &= (1-2\gamma)a \cdot m', \nonum
  \phi &= (1-2\gamma)^2a \cdot \theta', \nonum
  r &= s \bigl(1+s_0\kappa^4\bigr) \Gamma^4 \cdot r', \nonum
  \alpha &= \frac{{(1-2\gamma)^4}a^2}{s \bigl(1+s_0\kappa^4\bigr)\Gamma^4} \cdot \alpha'.
  \label{eq:A-rescaled}
\end{align}
Then,
\begin{align}
  m' &= \frac{1}{2}\biggl[\erf\frac{\theta'}{\sqrt{2\arp}} + \erf\frac{m' - \theta'}{\sqrt{2\arp}}\biggr] \\
  r' &= \frac{1}{2}\biggl[1 - \erf\frac{\theta'}{\sqrt{2\arp}} + a \erf\frac{m' - \theta'}{\sqrt{2\arp}}\biggr].
\end{align}

Successful retrieval means that $m' \approx 1$, which requires $\theta'/\sqrt{2\arp} \gg 1$ and $(m'-\theta')/\sqrt{2\arp} \gg 1$. Under these limits, $0 < \theta' < 1$ and $Q \ll 1$, which validates our previous assumption. We can use asymptotic forms of the error function to obtain
\begin{align}
  m' &= 1 - \frac{1}{\sqrt{2\pi}} \frac{\sqrt\arp}{\theta'}\ee^{-\theta'^2/2\arp} - \frac{1}{\sqrt{2\pi}} \frac{\sqrt\arp}{m'-\theta'}\ee^{-(m'-\theta')^2/2\arp}
  \label{eq:m-A} \\
  r' &= \frac{1}{\sqrt{2\pi}} \frac{\sqrt\arp}{\theta'}\ee^{-\theta'^2/2\arp} + \frac{a}{2} - \frac{a}{\sqrt{2\pi}} \frac{\sqrt\arp}{m'-\theta'}\ee^{-(m'-\theta')^2/2\arp}.
  \label{eq:r-A}
\end{align}

\subsection{Capacity formula for \texorpdfstring{$\theta' \rightarrow 0$}{θ' → 0}}

To derive capacity formulas, we need to make further assumptions about $\theta'$. First, we consider small $\theta'$. Because we still require $m' \approx 1$, the third term in \cref{eq:r-A} becomes much smaller than the second, so
\begin{equation}
  r' \approx \frac{1}{\sqrt{2\pi}} \frac{\sqrt\arp}{\theta'}\ee^{-\theta'^2/2\arp} + \frac{a}{2}.
\end{equation}
This equation no longer depends on $m'$. If we take $y \equiv \theta'/\sqrt\arp \gg 1$, it becomes
\begin{align}
  \frac{\theta'^2}{\alpha'} &= \frac{1}{\sqrt{2\pi}}y\ee^{-y^2/2} + \frac{a}{2}y^2.
  \label{eq:A-thetaalpha}
\end{align}
The capacity is the maximum example load $s$ for which this equation still admits a solution. Note that $s$ is proportional to $\alpha'$ according to \cref{eq:A-rescaled}. Thus, we maximize $\alpha'$ by minimizing the right-hand side of \cref{eq:A-thetaalpha} over $y$:
\begin{align}
  0 &= \frac{1}{\sqrt{2\pi}}(1-y^2)\ee^{-y^2/2} + ay \nonum
  y\ee^{-y^2/2} &\approx \sqrt{2\pi}a \nonum
  y &= \sqrt{-W_{-1}(-2\pi a^2)} \approx \sqrt{2|\!\log a|},
  \label{eq:A-y}
\end{align}
where $W_{-1}$ is the negative branch of the Lambert $W$ function, which is also known as the product logarithm. Substituting \cref{eq:A-y} back into \cref{eq:A-thetaalpha}, we obtain the maximal value
\begin{equation}
  \alpha'_\rmc \sim \frac{\theta'^2}{a |\!\log a|}.
\end{equation}
This expression implicitly defines the capacity $s_\rmc$ for $\theta' \rightarrow 0$.

We can use this expression to obtain critical values for $m'_\rmc$ and $r'_\rmc$:
\begin{align}
  m'_\rmc &\approx 1 - \frac{1}{2\sqrt{\pi|\!\log a|}} \frac{\theta'}{1-\theta'} a^{(1-\theta')^2/\theta'^2}, \\
  r'_\rmc &\approx \frac{a}{2}.
\end{align}
Note that $m'_\rmc \approx 1$, which confirms that our solution is self-consistent.

\subsection{Capacity formula for \texorpdfstring{$\theta' \rightarrow 1$}{θ' → 1}}

Next, we derive a capacity formula for large $\theta'$. In this case, \cref{eq:m-A,eq:r-A} become
\begin{align}
  m' &= 1 - \frac{1}{\sqrt{2\pi}} \frac{\sqrt\arp}{m'-\theta'}\ee^{-(m'-\theta')^2/2\arp}
  \label{eq:m-A2} \\
  r' &= \frac{a}{2} - \frac{a}{\sqrt{2\pi}} \frac{\sqrt\arp}{m'-\theta'}\ee^{-(m'-\theta')^2/2\arp},
\end{align}
which yields
\begin{equation}
  r' = \frac{a}{2} - a(1-m') \approx \frac{a}{2}.
  \label{eq:r-A3}
\end{equation}
If we define $y \equiv (m'-\theta')/\sqrt{\alpha' r'} \gg 1$ and use \cref{eq:r-A3}, we can write \cref{eq:m-A2} as
\begin{equation}
  \sqrt{\frac{a\alpha'}{2}} = \frac{1 - \theta'}{y} - \frac{1}{\sqrt{2\pi}y^2} \ee^{-y^2/2}.
  \label{eq:A-aalpha}
\end{equation}
Again, the example load $s$ is proportional to $\alpha'$ [\cref{eq:A-rescaled}], so we maximize $\alpha'$ by maximizing the right-hand size of \cref{eq:A-aalpha} with respect to $y$:
\begin{align}
  0 &= -\frac{1-\theta'}{y^2} + \frac{1}{\sqrt{2\pi}} \biggl(\frac{2}{y^3} + \frac{1}{y}\biggr) \ee^{-y^2/2} \nonum
  y\ee^{-y^2/2} &\approx \sqrt{2\pi}(1-\theta') \nonum
  y &\approx \sqrt{-W_{-1}\bigl[-2\pi (1-\theta')^2\bigr]} \approx \sqrt{2|\!\log (1-\theta')|}.
  \label{eq:A-y2}
\end{align}
Substituting \cref{eq:A-y2} into \cref{eq:A-aalpha}, we obtain
\begin{equation}
  \alpha'_\rmc \sim \frac{(1-\theta')^2}{a |\!\log (1-\theta')|}.
\end{equation}
This expression implicitly defines the capacity $s_\rmc$ for $\theta' \rightarrow 1$.

Similarly to before, we use this expression to obtain the critical value for $m'_\rmc$:
\begin{equation}
  m'_\rmc \approx 1 - \frac{1-\theta'}{y^2} 
      = 1 - \frac{1-\theta'}{2|\!\log (1-\theta')|}.
\end{equation}
$m'_\rmc \approx 1$, which confirms that our solution was obtained self-consistently.

\subsection{Maximizing capacity over \texorpdfstring{$\theta'$}{θ'}}

We have derived two expressions for $\alpha'_\rmc$, which is proportional to the capacity $s_\rmc$, in different regimes of the rescaled threshold $\theta'$:
\begin{equation}
  \alpha'_\rmc \sim
    \begin{dcases}
      \frac{\theta'^2}{a |\!\log a|}               & \theta' \rightarrow 0, \\
      \frac{(1-\theta')^2}{a |\!\log (1-\theta')|} & \theta' \rightarrow 1.
    \end{dcases}
  \label{eq:alpha2}
\end{equation}
We now take $\theta'$ to be a free parameter and maximize the capacity over it. The first expression for $\alpha'_\rmc$ grows from 0 as $\theta'$ increases from 0, and the second one grows from 0 as $\theta'$ decreases from 1. Thus, the optimum value should lie somewhere in between, and we estimate its location by finding the crossover point where the two expressions meet.

We assume that $\theta'$ is sufficiently far from 1 such that $|\!\log (1-\theta')| \sim 1$. Then, the crossover point is given by
\begin{align}
  \frac{\theta'^2}{a |\!\log a|} &\approx \frac{(1-\theta')^2}{a} \nonum
  \theta' &= \frac{\sqrt{|\!\log a|}}{1 + \sqrt{|\!\log a|}}.
\end{align}
Substituting this optimal threshold back into \cref{eq:alpha2}, we obtain
\begin{equation}
  \alpha'_\rmc \sim \frac{1}{a |\!\log a|},
\end{equation}
By converting $\alpha'$ back to $\alpha$ with \cref{eq:A-rescaled}, we recover the capacity formula \cref{eq:results-A1} at optimal threshold.

\section{\label{sec:de}Capacity for dense examples}

\subsection{Dense asymmetric mean-field equations}

We return to the generic mean-field equations \cref{eq:m-zeroT,eq:r-zeroT,eq:Q-zeroT} and consider the case where the network recovers a dense example $\zeta_{11}$. Due to correlations, the network will overlap with all dense patterns $\zeta_{1\nu}$, so $\chi_{1\nu} = \zeta_{1\nu}$. Using this expression, we can simplify the mean-field equations and find the critical example load $s_\rmc$ above which dense examples can no longer be retrieved. 

Recall from \cref{eq:patterns} that
\begin{align}
  \zeta_{1\nu} &=
      \begin{cases}  \zeta_1 & \textrm{with probability } \frac{1+c}{2} \\
                    -\zeta_1 & \textrm{with probability } \frac{1-c}{2} \end{cases} \nonum
  \zeta_1 &=
      \begin{cases}  \gamma & \textrm{with probability } \frac{1}{2}  \\
                    -\gamma & \textrm{with probability } \frac{1}{2}. \end{cases}
  \label{eq:0-patterns}
\end{align}
To help us in our calculations, we note the integrals
\begin{align}
  \int_{-\infty}^\infty \dd x\,\ee^{-(x-A)^2/\rho^2} \ee^{-(x-B)^2/\sigma^2} &= \sqrt\frac{\pi}{\rho^{-2}+\sigma^{-2}} \exp\biggl[ -\frac{(a-B)^2}{\rho^2+\sigma^2} \biggr], \nonum
  \int_{-\infty}^\infty \dd x\,\ee^{-(x-A)^2/\rho^2} \erf\biggl[ \frac{x-B}{\sigma} \biggr] &= \sqrt\pi \rho \erf\biggl[ \frac{A-B}{\sqrt{\rho^2+\sigma^2}} \biggr], \nonum
  \int_{-\infty}^\infty \dd x\,\ee^{-(x-A)^2/\rho^2} x \erf\biggl[ \frac{x-B}{\sigma} \biggr] &= \rho\,\Biggl\{ \frac{\rho^2}{\sqrt{\rho^2+\sigma^2}} \exp\biggl[ -\frac{(A-B)^2}{\rho^2+\sigma^2} \biggr] + \sqrt\pi A \erf\biggl[ \frac{A-B}{\sqrt{\rho^2+\sigma^2}} \biggr] \Biggr\}.
  \label{eq:0-integrals}
\end{align}

During successful retrieval, the network overlaps strongly with the target pattern $\zeta_{11}$. It will also overlap with other examples $\zeta_{1\nu}$ for $\nu > 1$ to a degree governed by the correlation parameter $c$ [\cref{eq:overlaps}]. As $N \rightarrow \infty$, these other overlaps converge towards one another due to the law of large numbers; we call this asymptotic value $m_0 \equiv m_{1\nu}$ for $\nu > 1$. Thus, we can write
\begin{equation}
  \sum_\nu m_{1\nu}\zeta_{1\nu} = m_{11}\zeta_{11} + m_0 \sum_{\nu>1} \zeta_{1\nu} = m\zeta + s_0m_0x_0.
\end{equation}
We rename $m \equiv m_{11}$ and $\zeta \equiv \zeta_{11}$ for convenience. $x_0$ is the average over the $s_0 = s-1$ other examples in concept 1, and it follows a binomial distribution with mean $c\zeta_1$ variance $\gamma^2(1-c^2)/s_0$ according to \cref{eq:0-patterns}. In the large $s$ limit, it can be approximated by a Gaussian random variable with the same moments. We also introduce $m_1$, which is the network overlap with the concept pattern $\zeta_1$.

With these considerations, \cref{eq:m-zeroT,eq:r-zeroT,eq:Q-zeroT} yield
\begin{align}
  m &= \frac{1}{2}\biggl\llangle \zeta \erf\frac{m\zeta+s_0m_0x_0-\phi}{\sqrt{2\ar}} \biggr\rrangle,
  \label{eq:m-0x} \\
  m_0 &= \frac{1}{2}\biggl\llangle x_0 \erf\frac{m\zeta+s_0m_0x_0-\phi}{\sqrt{2\ar}} \biggr\rrangle, \\
  m_1 &= \frac{1}{2}\biggl\llangle \zeta_1 \erf\frac{m\zeta+s_0m_0x_0-\phi}{\sqrt{2\ar}} \biggr\rrangle, \\
  r &= \frac{s\Gamma^4}{2} \cdot \frac{\bigl(1 - Q(1-\kappa^2)(1+s_0\kappa^2)\bigr)^{\!2} + s_0\kappa^4}{\bigl(1 - Q(1-\kappa^2)\bigr)^{\!2}\bigl(1 - Q(1+s_0\kappa^2)\bigr)^{\!2}} \Biggl[1 + \biggl\llangle \erf\frac{m\zeta+s_0m_0x_0-\phi}{\sqrt{2\ar}} \biggr\rrangle \Biggr], \\
  Q &= \frac{\Gamma^2}{\sqrt{2\pi\ar}} \biggl\llangle\exp\biggl[ -\frac{(m\zeta+s_0m_0x_0-\phi)^2}{2\ar}\biggr]\biggr\rrangle.
  \label{eq:Q-0x} 
\end{align}
The double angle brackets indicate averages over $\zeta$ and $x_0$, which is a Gaussian random variable with mean and variance listed above. We define the following variables
\begin{align}
  \sigma_0^2 &\equiv s_0\gamma^2(1-c^2)m_0^2 + \ar \nonum
  Y_{\pm\pm} &\equiv \frac{\gamma m \pm s_0\gamma cm_0 \pm \phi}{\sqrt{2}\sigma_0},
\end{align}
with choices for $+$ and $-$ in $Y_{\pm\pm}$ corresponding to signs in the right-hand side. Now we come to the task of performing the averages in \crefrange{eq:m-0x}{eq:Q-0x}. For each variable, we average successively over $\zeta$, $x_0$, and $\zeta_1$.

First,
\begin{equation}
  Q = \frac{\Gamma^2}{\sqrt{2\pi\ar}} \Biggl\{ \frac{1+c}{2} \biggl\llangle\exp\biggl[ -\frac{(m\zeta_1+s_0m_0x_0-\phi)^2}{2\ar}\biggr]\biggr\rrangle + \frac{1-c}{2} \biggl\llangle\exp\biggl[ -\frac{(m\zeta_1-s_0m_0x_0+\phi)^2}{2\ar}\biggr]\biggr\rrangle \Biggr\}.
\end{equation}
Then,
\begin{align}
  &\frac{1}{\sqrt{2\pi\ar}} \biggl\llangle\exp\biggl[ -\frac{(m\zeta_1+s_0m_0 x_0-\phi)^2}{2\ar}\biggr]\biggr\rrangle \nonum
  &\qquad= \frac{1}{\sqrt{2\pi\ar}} \sqrt\frac{s_0}{2\pi\gamma^2(1-c^2)} \biggl\langle \int\dd x_0\,\ee^{-s_0(x_0-c\zeta_1)^2/2\gamma^2(1-c^2)} \ee^{-s_0^2m_0^2(x_0+(m\zeta_1-\phi)/s_0m_0)^2/2\ar} \biggr\rangle \nonum
  &\qquad= \frac{1}{\sqrt{2\pi\bigl(s_0\gamma^2(1-c^2)m_0^2 + \ar\bigr)}} \biggl\langle \exp\biggl[-\frac{(m\zeta_1+s_0cm_0\zeta_1-\phi)^2}{2\bigl(s_0\gamma^2(1-c^2)m_0^2 + \ar\bigr)}\biggr] \biggr\rangle \nonum
  &\qquad= \frac{1}{\sqrt{2\pi}\sigma_0}\cdot\frac{1}{2} \biggl\{ \ee^{-Y_{++}^2}+\ee^{-Y_{+-}^2} \biggr\}.
\end{align}
Thus,
\begin{equation}
  Q = \frac{\Gamma^2}{\sqrt{2\pi}\sigma_0} \biggl\{ \frac{1+c}{4} \Bigl[\ee^{-Y_{++}^2}+\ee^{-Y_{+-}^2}\Bigr] + \frac{1-c}{4} \Bigl[\ee^{-Y_{-+}^2}+\ee^{-Y_{--}^2}\Bigr] \biggr\}.
\end{equation}

Next,
\begin{equation}
  \biggl\llangle \erf\frac{m\zeta+s_0m_0x_0-\phi}{\sqrt{2\ar}} \biggr\rrangle = \Biggl\{ \frac{1+c}{2} \biggl\llangle\erf\frac{m\zeta_1+s_0m_0x_0-\phi}{2\ar}\biggr\rrangle - \frac{1-c}{2} \biggl\llangle\erf\frac{m\zeta_1-s_0m_0x_0+\phi}{2\ar}\biggr\rrangle \Biggr\}.
\end{equation}
Then,
\begin{align}
  \biggl\llangle\erf\frac{m\zeta_1+s_0m_0x_0-\phi}{2\ar}\biggr\rrangle 
  &= \sqrt\frac{s_0}{2\pi\gamma^2(1-c^2)} \Biggl\langle \int\dd x_0\,\ee^{-s_0(x_0 - c\zeta_1)^2/2\gamma^2(1-c^2)} \erf\biggl[\frac{s_0m_0}{\sqrt{2\ar}} \biggl(x_0+\frac{m\zeta_1-\phi}{s_0m_0}\biggr)\biggr] \Biggr\rangle \nonum
  &= \Biggl\langle \erf\frac{m\zeta_1+s_0cm_0\zeta_1-\phi}{\sqrt{2\bigl(s_0\gamma^2(1-c^2)m_0^2 + \ar\bigr)}} \Biggr\rangle \nonum
  &= -\frac{1}{2} \biggl\{ \erf Y_{++}-\erf Y_{+-} \biggr\}.
\end{align}
Thus,
\begin{equation}
  r = \frac{s\Gamma^4}{2} \cdot \frac{\bigl(1 - Q(1-\kappa^2)(1+s_0\kappa^2)\bigr)^{\!2} + s_0\kappa^4}{\bigl(1 - Q(1-\kappa^2)\bigr)^{\!2}\bigl(1 - Q(1+s_0\kappa^2)\bigr)^{\!2}} \biggl\{1 - \frac{1+c}{4} \Bigl[\erf Y_{++}-\erf Y_{+-}\Bigr] - \frac{1-c}{4} \Bigl[\erf Y_{-+}-\erf Y_{--}\Bigr] \biggr\}.
\end{equation}

Next,
\begin{equation}
  m = \frac{1}{2} \Biggl\{ \frac{1+c}{2} \biggl\llangle \zeta_1 \erf\frac{m\zeta_1+s_0m_0x_0-\phi}{2\ar}\biggr\rrangle + \frac{1-c}{2} \biggl\llangle \zeta_1 \erf\frac{m\zeta_1-s_0m_0x_0+\phi}{2\ar}\biggr\rrangle \Biggr\}.
\end{equation}
Then,
\begin{align}
  \biggl\llangle \zeta_1 \erf\frac{m\zeta_1+s_0m_0x_0-\phi}{2\ar}\biggr\rrangle
  &= \sqrt\frac{s_0}{2\pi\gamma^2(1-c^2)} \Biggl\langle \zeta_1 \int\dd x_0\,\ee^{-s_0(x_0 - c\zeta_1)^2/2\gamma^2(1-c^2)} \erf\biggl[\frac{s_0m_0}{\sqrt{2\ar}} \biggl(x_0+\frac{m\zeta_1-\phi}{s_0m_0}\biggr)\biggr] \Biggr\rangle \nonum
  &= \Biggl\langle \zeta_1 \erf\frac{m\zeta_1+s_0cm_0\zeta_1-\phi}{\sqrt{2\bigl(s_0\gamma^2(1-c^2)m_0^2 + \ar\bigr)}} \Biggr\rangle \nonum
  &= \frac{\gamma}{2} \biggl\{ \erf Y_{++}+\erf Y_{+-} \biggr\}.
\end{align}
Thus,
\begin{equation}
  m = \frac{\gamma}{2} \biggl\{\frac{1+c}{4} \Bigl[\erf Y_{++}+\erf Y_{+-}\Bigr] + \frac{1-c}{4} \Bigl[\erf Y_{-+}+\erf Y_{--}\Bigr] \biggr\}.
\end{equation}
Similarly,
\begin{align}
  m_1 &= \frac{1}{2} \Biggl\{ \frac{1+c}{2} \biggl\llangle \zeta_1 \erf\frac{m\zeta_1+s_0m_0x_0-\phi}{2\ar}\biggr\rrangle - \frac{1-c}{2} \biggl\llangle \zeta_1 \erf\frac{m\zeta_1-s_0m_0x_0+\phi}{2\ar}\biggr\rrangle \Biggr\} \nonum
  &= \frac{\gamma}{2} \biggl\{\frac{1+c}{4} \Bigl[\erf Y_{++}+\erf Y_{+-}\Bigr] - \frac{1-c}{4} \Bigl[\erf Y_{-+}+\erf Y_{--}\Bigr] \biggr\}.
\end{align}

Finally,
\begin{equation}
  m_0 = \frac{1}{2} \Biggl\{ \frac{1+c}{2} \biggl\llangle x_0 \erf\frac{m\zeta_1+s_0m_0x_0-\phi}{2\ar}\biggr\rrangle - \frac{1-c}{2} \biggl\llangle x_0 \erf\frac{m\zeta_1-s_0m_0x_0+\phi}{2\ar}\biggr\rrangle \Biggr\}.
\end{equation}
Then,
\begin{align}
  \biggl\llangle x_0 \erf\frac{m\zeta_1+s_0m_0x_0-\phi}{\sqrt{2\ar}}\biggr\rrangle
  &= \sqrt\frac{s_0}{2\pi\gamma^2(1-c^2)} \Biggl\langle \int\dd x_0\,\ee^{-s_0(x_0-c\zeta_1)^2/2\gamma^2(1-c^2)} x_0 \erf\biggl[\frac{s_0m_0}{\sqrt{2\ar}} \biggl(x_0+\frac{m\zeta_1-\phi}{s_0m_0}\biggr)\biggr] \Biggr\rangle \nonum
  &= \gamma^2(1-c^2)m_0 \sqrt\frac{2}{\pi\bigl(s_0\gamma^2(1-c^2)m_0^2 + \ar\bigr)} \biggl\langle \exp\biggl[-\frac{(m\zeta_1+s_0cm_0\zeta_1-\phi)^2}{2\bigl(s_0\gamma^2(1-c^2)m_0^2 + \ar\bigr)}\biggr] \biggr\rangle \nonum
  &\phantom{{}={}}{}+ c \Biggl\langle \zeta_1 \erf\frac{m\zeta_1+s_0cm_0\zeta_1-\phi}{\sqrt{2\bigl(s_0\gamma^2(1-c^2)m_0^2 + \ar\bigr)}} \Biggr\rangle \nonum
  &= \frac{\gamma^2(1-c^2)m_0}{\sqrt{2\pi}\sigma_0} \biggl\{ \ee^{-Y_{++}^2}+\ee^{-Y_{+-}^2} \biggr\} + \frac{\gamma c}{2} \biggl\{ \erf Y_{++}+\erf Y_{+-} \biggr\}.
\end{align}
Thus,
\begin{align}
  m_0 &= \frac{\gamma c}{2} \biggl\{ \frac{1+c}{4} \Bigl[\erf Y_{++}+\erf Y_{+-}\Bigr] - \frac{1-c}{4} \Bigl[\erf Y_{-+}+\erf Y_{--}\Bigr] \biggr\} + Q\tfrac{\gamma^2}{\Gamma^2}(1-c^2)m_0 \nonum
  &= \frac{\gamma c}{2\Bigl(1-Q\frac{\gamma^2}{\Gamma^2}(1-c^2)\Bigr)} \biggl\{ \frac{1+c}{4} \Bigl[\erf Y_{++}+\erf Y_{+-}\Bigr] - \frac{1-c}{4} \Bigl[\erf Y_{-+}+\erf Y_{--}\Bigr] \biggr\}.
\end{align}

These mean-field equations are presented in \cref{eq:model-0} with $m$ replaced by its original name $m_{11}$. They can be numerically solved to find regimes of successful retrieval, but we will analyze them further in search of a formula for the capacity $s_\rmc$.

In the limit that the network only stores dense, uncorrelated patterns with $2\gamma = 1$ and $c = 0$, these mean-field equations simplify to those of the Hopfield network with 0/1 neurons \cite{Bruce.1987}, which has half the capacity of the original Hopfield network with ${+}1$/${-}1$ neurons. Note that the pattern storage strength in Ref.~\onlinecite{Bruce.1987} is twice that of ours; in other words, their connectivity weights are scaled by a factor of 4 in comparison to ours.  To match our equations to theirs, make the replacements $\gamma \rightarrow 1$, $\Gamma \rightarrow 1$, $s \rightarrow 1$, $c \rightarrow 0$, and $\kappa \rightarrow 0$ in our equations, and recall \crefrange{eq:r-symmetric}{eq:hphi}.

\subsection{Simplified mean-field equations}

To derive a capacity formula, we make three further assumptions. First, we assume $c^2 \ll 1$, which implies $\kappa^2 \ll 1$ as well. Second, we assume that the rescaled threshold $\phi = 0$. This assumption is justified empirically, for we find that the capacity is maximized at $|\phi| < 10^{-6}$ over all parameter ranges in \cref{fig:0}. It is also justified theoretically, since we will derive that $Q \ll 1$, which means $\phi \approx \theta$. For dense patterns in classic Hopfield network, retrieval is maximized at threshold $\theta = 0$ \cite{Weisbuch.1985}. Finally, we assume $s \gg 1$, so $s_0 = s$; this is not necessary, but it makes the expressions simpler.

We rescale the order parameters with
\begin{align}
  m &= \frac{\gamma}{2} \cdot m', \nonum
  m_0 &= \frac{\gamma}{2} \cdot m'_0, \nonum
  r &= \frac{s\Gamma^4}{2} \cdot r', \nonum
  \alpha &= \frac{\gamma^4}{2\Gamma^4} \cdot \alpha'.
  \label{eq:0-rescaled}
\end{align}
The mean-field equations then become
\begin{align}
  m' &= \frac{1+c}{2} \erf Y'_+ + \frac{1-c}{2} \erf Y'_-,
  \label{eq:m-0rescaled} \\
  m'_0 &= \frac{c}{1-Q\frac{\gamma^2}{\Gamma^2}} \biggl\{ \frac{1+c}{2} \erf Y'_+ - \frac{1-c}{2} \erf Y'_- \biggr\},
  \label{eq:m0-0rescaled} \\
  r' &= \frac{1}{(1 - Q)^2}, \\
  Q &= \sqrt\frac{2}{\pi} \frac{\Gamma^2}{\sigma'_0\gamma^2} \biggl\{ \frac{1+c}{2} \ee^{-(Y'_+)^2} + \frac{1-c}{2} \ee^{-(Y'_-)^2} \biggr\},
  \label{eq:Q-0rescaled}
\end{align}
where
\begin{align}
  {\sigma_0'}^{\!2} &\equiv s\bigl({m'_0}^{\!2} + \arp\bigr) \nonum
  Y'_\pm &\equiv \frac{m'}{\sqrt{2}\sigma'_0} \biggl(1 \pm \frac{scm'_0}{m'}\biggr).
\end{align}

Successful retrieval means that $m' \approx 1$, which requires $Y'_\pm \gg 1$. This condition in turn yields $m'_0 \approx c^2$ and $Q \ll 1$ through \cref{eq:m0-0rescaled,eq:Q-0rescaled}, which confirms our previous assumption. Thus, $Y'_\pm \approx (y/\sqrt{2})(1 \pm sc^3)$, where
\begin{equation}
  y \equiv \frac{m'}{\sigma'_0}.
\end{equation}
For $Y'_\pm \gg 1$, we need $sc^3 \lesssim 1$ and $y \gg 1$, which we use to boldly simplify \crefrange{eq:m-0rescaled}{eq:Q-0rescaled}:
\begin{align}
  m' &= 1 - \frac{1}{y^2}\expyy,
  \label{eq:m-0simple} \\
  m'_0 &= \frac{c^2}{1-Q\frac{\gamma^2}{\Gamma^2}} \Biggl\{ 1 - \biggl(\frac{1}{y^2}-sc^2\biggr) \expyy \Biggr\}, \\
  \alpha' &= \biggl(\frac{{m'}^2}{sy^2} - {m'_0}^{\!2}\biggr) (1 - Q)^2, \\
  Q &= \frac{\Gamma^2}{\gamma^2m'} \biggl(1 - sc^4y^2 + \frac{1}{2}s^2c^6y^4\biggr) \expyy.
  \label{eq:Q-0simple}
\end{align}
For mathematical tractability, we have expanded in $sc^3$ and $1/y$, even though the former is not strictly small and the latter can be empirically close to 1.

\subsection{Capacity formula}

In \crefrange{eq:m-0simple}{eq:Q-0simple}, we substitute formulas for $m'$, $m'_0$, and $Q$ into the equation for $\alpha'$ and keep only leading terms in $1/y$ and $c$. After much simplification, we obtain
\begin{equation}
  s(\alpha' + c^4) \approx \frac{1}{y^2} - \frac{\Gamma^2}{\gamma^2} \biggl(\frac{1}{y} + \frac{1}{2}s^2c^6y^3\biggr) \sqrt\frac{8}{\pi} \expy.
  \label{eq:0-y}
\end{equation}

At the critical value of $s$ above which \cref{eq:0-y} equation cannot be satisfied by any $y$, derivatives with respect to $y$ on both sides of the equation must be equal. In other words, we expect the critical $s_\rmc$ to be a saddle-node bifurcation point. For mathematical tractability, we ignore the term proportional to $s^2$. This simplification is rather arbitrary, but it can be empirically justified by comparing the resulting formula with numerical analysis of the full mean-field equations [\cref{fig:0}]. We also eliminate higher orders in $1/y$ to obtain
\begin{equation}
  0 \approx -\frac{2}{y^3} + \frac{\Gamma^2}{\gamma^2} \sqrt\frac{8}{\pi} \expy.
  \label{eq:0-dy}
\end{equation}
Solving for $y$, we obtain
\begin{equation}
  y = \sqrt{-3 W_{-1}\Bigl[-\tfrac{1}{3} \bigl(\tfrac{\pi}{2}\bigr)^{\!1/3} \bigl(\tfrac{\gamma}{\Gamma}\bigr)^{\!4/3} \Bigr]} \approx \sqrt{3 \log \Bigl[3\bigl(\tfrac{2}{\pi}\bigr)^{\!1/3} \bigl(\tfrac{\Gamma}{\gamma}\bigr)^{\!4/3} \Bigr]},
\end{equation}
where $W_{-1}$ is the negative branch of the Lambert $W$ function. Since this function involves a logarithm, it varies very slowly as a function of $\gamma/\Gamma$. For $\gamma = 0.1$ and $a$ between $0.001$ and $0.1$, this expression for $y$ ranges from $1.7$ and $3.3$. Within this range, $m' > 0.88$ according to \cref{eq:m-0simple}, which confirms that our earlier simplifications using $m' \approx 1$ yield self-consistent results.

We can use \cref{eq:0-dy} to simplify \cref{eq:0-y} to leading order in $1/y$:
\begin{equation}
  s_\rmc(\alpha' + c^4) \approx \frac{1}{y^2} - s_\rmc^2c^6.
\end{equation}
Solving for $s_\rmc$,
\begin{equation}
  s_\rmc = \frac{\sqrt{(\alpha'+c^4)^2 + \frac{4}{y^2} c^6} - (\alpha'+c^4)}{2c^6}.
\end{equation}
To heuristically obtain a simpler equation, we note that $s_\rmc \rightarrow 1/yc^3$ when $\alpha' \rightarrow 0$ and $s_\rmc \rightarrow 1/y^2\alpha'$ when $\alpha' \rightarrow \infty$. We simply capture both these behaviors with 
\begin{equation}
  s_\rmc \sim \frac{1}{yc^3 + y^2 \alpha'}.
\end{equation}
Again, $y$ varies slowly within its range, so we simplify this equation by simply setting $y \sim 3$. After converting $\alpha'$ back to $\alpha$ with \cref{eq:0-rescaled}, we obtain \cref{eq:results-0}.

\section{\label{sec:dc}Critical load for dense concepts}

\subsection{Dense symmetric mean-field equations}

We return to the generic mean-field equations \cref{eq:m-zeroT,eq:r-zeroT,eq:Q-zeroT} and consider the case where the network recovers a dense concept $\zeta_1$. Due to correlations, the network will overlap with all dense patterns $\zeta_{1\nu}$, so $\chi_{1\nu} = \zeta_{1\nu}$. Using this expression, we can simplify the mean-field equations and find the critical example load $s_\rmc$ below which dense concepts cannot be retrieved. Recall the dense pattern statistics \cref{eq:0-patterns} and Gaussian integrals \cref{eq:0-integrals}, which will aid us in our derivations.

Successful retrieval means that the network overlaps strongly with the target concept $\zeta_1$. The correlation parameter $c$ produces overlaps with all example patterns $\zeta_{1\nu}$ [\cref{eq:overlaps}], which converge to an asymptotic value $m_\rms \equiv m_{1\nu}$ as $N \rightarrow \infty$. The ``s'' signifies ``symmetric'', i.e.\@ equal overlap with all examples in concept 1. Thus, we can write
\begin{equation}
  \sum_\nu m_{1\nu}\zeta_{1\nu} = m_\rms \sum_\nu\zeta_{1\nu} = sm_\rms x_\rms.
\end{equation}
$x_\rms$ is the average over the $s$ examples in concept 1, and it follows a binomial distribution with mean $c\zeta_1$ and variance $\gamma^2(1-c^2)/s$ according to \cref{eq:0-patterns}. In the large $s$ limit, it can be approximated by a Gaussian random variable with the same moments. We explicitly introduce $m_1$, which is the network overlap with the target concept $\zeta_1$.

With these considerations, \cref{eq:m-zeroT,eq:r-zeroT,eq:Q-zeroT} yield
\begin{align}
  m_\rms &= \frac{1}{2}\biggl\llangle x_\rms \erf\frac{sm_\rms x_\rms-\phi}{\sqrt{2\ar}}\biggr\rrangle,
  \label{eq:ms-Sx} \\
  m_1 &= \frac{1}{2}\biggl\llangle \zeta_1 \erf\frac{sm_\rms x_\rms-\phi}{\sqrt{2\ar}}\biggr\rrangle, \\
  r &= \frac{s\Gamma^4}{2} \cdot \frac{\bigl(1 - Q(1-\kappa^2)(1+s_0\kappa^2)\bigr)^{\!2} + s_0\kappa^4}{\bigl(1 - Q(1-\kappa^2)\bigr)^{\!2}\bigl(1 - Q(1+s_0\kappa^2)\bigr)^{\!2}} \Biggl[1 + \biggl\llangle \erf\frac{sm_\rms x_\rms-\phi}{\sqrt{2\ar}} \biggr\rrangle \Biggr], \\
  Q &= \frac{\Gamma^2}{\sqrt{2\pi\ar}} \biggl\llangle\exp\biggl[ -\frac{(sm_\rms x_\rms-\phi)^2}{2\ar}\biggr]\biggr\rrangle.
  \label{eq:Q-Sx}
\end{align}
The double angle brackets indicate averages over $\zeta$ and $x_\rms$, which is a Gaussian random variable with mean and variance listed above. We define the following variables
\begin{align}
  \sigma_\rms^2 &\equiv s\gamma^2(1-c^2)m_\rms^2 + \ar \nonum
  Y_\pm &\equiv \frac{s\gamma cm_\rms \pm \phi}{\sqrt{2}\sigma_\rms},
\end{align}
with choices for $+$ and $-$ in $Y_\pm$ corresponding the sign in the right-hand side. Now we come to the task of performing the averages in \crefrange{eq:ms-Sx}{eq:Q-Sx}. For each variable, we average successively over $\zeta_1$ and $x_\rms$.

First,
\begin{align}
  Q &= \frac{\Gamma^2}{\sqrt{2\pi\ar}} \biggl\llangle\exp\biggl[ -\frac{(sm_\rms x_\rms-\phi)^2}{2\ar}\biggr]\biggr\rrangle \nonum
  &= \frac{\Gamma^2}{\sqrt{2\pi\ar}} \sqrt\frac{s}{2\pi\gamma^2(1-c^2)} \biggl\langle \int\dd x_\rms\,\ee^{-s( x_\rms-c\zeta_1)^2/2\gamma^2(1-c^2)} \ee^{-s^2m_\rms^2( x_\rms-\phi/sm_\rms)^2/2\ar} \biggr\rangle \nonum
  &= \frac{\Gamma^2}{\sqrt{2\pi\bigl(s\gamma^2(1-c^2)m_\rms^2 + \ar\bigr)}} \biggl\langle \exp\biggl[-\frac{(scm_\rms\zeta_1-\phi)^2}{2\bigl(s\gamma^2(1-c^2)m_\rms^2 + \ar\bigr)}\biggr] \biggr\rangle \nonum
  &= \frac{\Gamma^2}{\sqrt{8\pi}\sigma_\rms} \biggl\{ \ee^{-Y_+^2}+\ee^{-Y_-^2} \biggr\}.
\end{align}

Next,
\begin{align}
  \biggl\llangle \erf\frac{sm_\rms x_\rms-\phi}{\sqrt{2\ar}} \biggr\rrangle
  &= \sqrt\frac{s}{2\pi\gamma^2(1-c^2)} \Biggl\langle \int\dd x_\rms\,\ee^{-s( x_\rms - c\zeta_1)^2/2\gamma^2(1-c^2)} \erf\biggl[\frac{sm_\rms}{\sqrt{2\ar}} \biggl( x_\rms-\frac{\phi}{sm_\rms}\biggr)\biggr] \Biggr\rangle \nonum
  &= \Biggl\langle \erf\frac{scm_\rms\zeta_1-\phi}{\sqrt{2\bigl(s\gamma^2(1-c^2)m_\rms^2 + \ar\bigr)}} \Biggr\rangle \nonum
  &= -\frac{1}{2} \biggl\{ \erf Y_+-\erf Y_- \biggr\}.
\end{align}
Thus,
\begin{equation}
  r = \frac{s\Gamma^4}{2} \cdot \frac{\bigl(1 - Q(1-\kappa^2)(1+s_0\kappa^2)\bigr)^{\!2} + s_0\kappa^4}{\bigl(1 - Q(1-\kappa^2)\bigr)^{\!2}\bigl(1 - Q(1+s_0\kappa^2)\bigr)^{\!2}} \biggl\{ 1 - \frac{1}{2} \Bigl[\erf Y_+-\erf Y_-\Bigr] \biggr\}.
\end{equation}

Next,
\begin{align}
  m_1 &= \frac{1}{2} \biggl\llangle \zeta_1 \erf\frac{sm_\rms x_\rms-\phi}{2\ar}\biggr\rrangle \nonum
  &= \frac{1}{2}\sqrt\frac{s}{2\pi\gamma^2(1-c^2)} \Biggl\langle \zeta_1 \int\dd x_\rms\,\ee^{-s( x_\rms - c\zeta_1)^2/2\gamma^2(1-c^2)} \erf\biggl[\frac{sm_\rms}{\sqrt{2\ar}} \biggl( x_\rms-\frac{\phi}{sm_\rms}\biggr)\biggr] \Biggr\rangle \nonum
  &= \frac{1}{2}\Biggl\langle \zeta_1 \erf\frac{scm_\rms\zeta_1-\phi}{\sqrt{2\bigl(s\gamma^2(1-c^2)m_\rms^2 + \ar\bigr)}} \Biggr\rangle \nonum
  &= \frac{\gamma}{4} \biggl\{ \erf Y_++\erf Y_- \biggr\}.
\end{align}

Finally,
\begin{align}
  m_\rms &= \frac{1}{2} \biggl\llangle  x_\rms \erf\frac{sm_\rms x_\rms-\phi}{2\ar}\biggr\rrangle \nonum
  &= \frac{1}{2} \sqrt\frac{s}{2\pi\gamma^2(1-c^2)} \Biggl\langle \int\dd  x_\rms\,\ee^{-s( x_\rms-c\zeta_1)^2/2\gamma^2(1-c^2)}  x_\rms \erf\biggl[\frac{sm_\rms}{\sqrt{2\ar}} \biggl( x_\rms-\frac{\phi}{sm_\rms}\biggr)\biggr] \Biggr\rangle \nonum
  &= \frac{\gamma^2(1-c^2)m_\rms}{2} \sqrt\frac{2}{\pi\bigl(s\gamma^2(1-c^2)m_\rms^2 + \ar\bigr)} \biggl\langle \exp\biggl[-\frac{(scm_\rms\zeta_1-\phi)^2}{2\bigl(s\gamma^2(1-c^2)m_\rms^2 + \ar\bigr)}\biggr] \biggr\rangle \nonum
  &\phantom{{}={}}{}+ \frac{c}{2} \Biggl\langle \zeta_1 \erf\frac{scm_\rms\zeta_1-\phi}{\sqrt{2\bigl(s\gamma^2(1-c^2)m_\rms^2 + \ar\bigr)}} \Biggr\rangle \nonum
  &= Q\tfrac{\gamma^2}{\Gamma^2}(1-c^2)m_\rms + \frac{\gamma c}{4} \biggl\{ \erf Y_++\erf Y_- \biggr\} \nonum
  &= \frac{\gamma c}{4\Bigl(1-Q\frac{\gamma^2}{\Gamma^2}(1-c^2)\Bigr)} \biggl\{ \erf Y_++\erf Y_- \biggr\}.
\end{align}

These mean-field equations are presented in \cref{eq:model-S}.

\subsection{Simplified mean-field equations}

To derive a formula for the critical example load $s_\rmc$, we make three further assumptions. First, we assume $c^2 \ll 1$, which implies $\kappa^2 \ll 1$ as well. Second, we assume that rescaled threshold $\phi = 0$. This assumption is justified empirically. We find that $s_\rmc$ is minimized at $|\phi| < 0.5$ over all parameter ranges in \cref{fig:S}; moreover, these values are very close to that obtained by enforcing $\phi = 0$ [\cref{fig:S-extra}(a)]. Finally, we assume $s \gg 1$, so $s_0 = s$; this is not necessary, but it makes the expressions simpler.

\begin{figure*}[h!]
  \centering
  \includegraphics{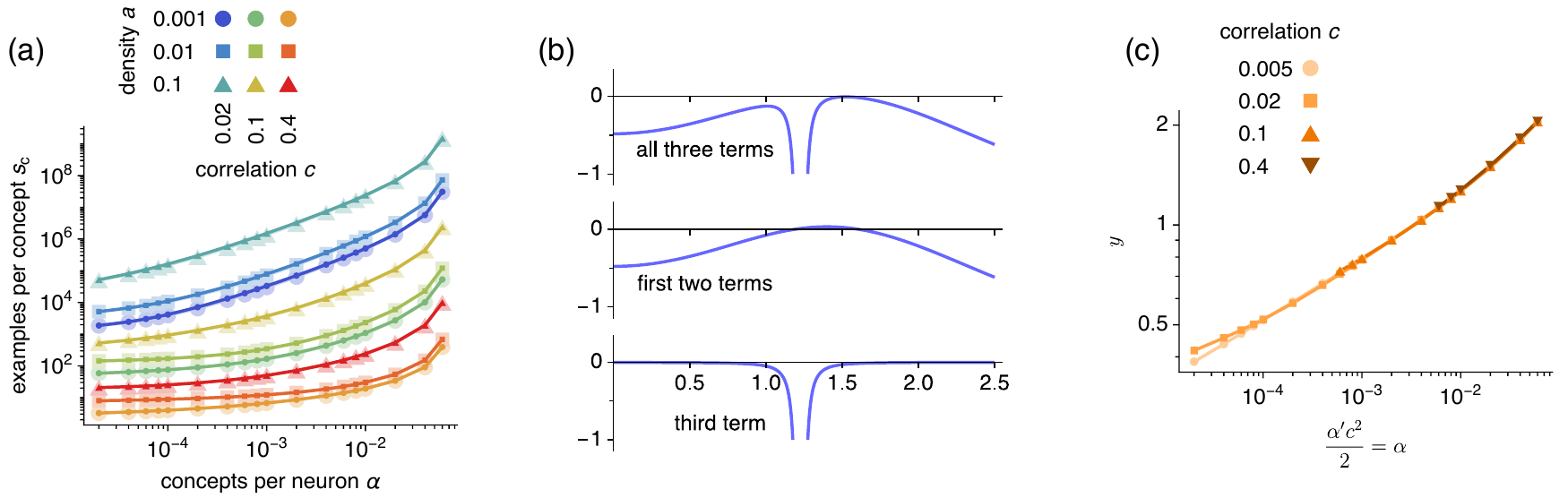}
  \caption{
    \label{fig:S-extra}
    (a) Critical example load $s_\rmc$ for dense concepts obtained through numerical analysis of \cref{eq:model-S}. We either set $\phi = 0$ (dark, thin lines) or maximize over $\phi$ (light, thick lines).
    (b) Right-hand side of \cref{eq:S-y} and its terms plotted separately.
    (c) $y$ as a function of $\alpha$ for density $a = 0$ obtained by numerically solving \cref{eq:Sa-alpha}.
  }
\end{figure*}

We rescale the order parameters with
\begin{align}
  m_\rms &= \frac{\gamma c}{2} \cdot m'_\rms, \nonum
  r &= \frac{s\Gamma^4}{2} \cdot m'_\rms r', \nonum
  \alpha &= \frac{\gamma^4c^2}{2\Gamma^4} \cdot \alpha'.
  \label{eq:S-rescaled}
\end{align}
We also define
\begin{equation}
  y \equiv \sqrt\frac{sc^2}{1+\arp},
\end{equation}
so $Y_\pm \approx y/\sqrt{2}$. The mean-field equations then become
\begin{align}
  m'_\rms - Qm'_\rms \frac{\gamma^2}{\Gamma^2} &= \erfy,
  \label{eq:m-Ssimple} \\
  \frac{1}{\alpha'}\biggl(\frac{sc^2}{y^2} - 1\biggr) &= \frac{\bigl(m'_\rms - Qm'_\rms(1+s\kappa^2)\bigr)^{\!2} + (m'_\rms)^2s\kappa^4}{\bigl(m'_\rms - Qm'_\rms\bigr)^{\!2}\bigl(m'_\rms - Qm'_\rms(1+s\kappa^2)\bigr)^{\!2}},
  \label{eq:r-Ssimple} \\
  Qm'_\rms &= \frac{\Gamma^2}{\gamma^2sc^2}\expyy.
  \label{eq:Q-Ssimple}
\end{align}

We can now substitute expressions for $m'_\rms$ and $Qm'_\rms$ into \cref{eq:r-Ssimple} to obtain
\begin{equation}
  0 = \frac{1}{\alpha'}\biggl(\frac{sc^2}{y^2}-1\biggr) \biggl[{\textstyle\erfy}-\frac{(1-2\gamma)^2a(1-a)}{\gamma^2sc^2}{\textstyle\expyy}\biggr]^2 - 1 - s\kappa^4 \frac{\biggl[\erfy+\frac{1}{sc^2}\expyy\biggr]^2}{\biggl[\erfy-\Bigl(\frac{(1-2\gamma)^2a(1-a)}{\gamma^2sc^2} + 1\Bigr)\expyy\biggr]^2}.
  \label{eq:S-y}
\end{equation}
At the critical value of $s$ above which \cref{eq:S-y} equation cannot be satisfied by any $y$, its derivative with respect to $y$ must be 0. In other words, we expect the critical $s_\rmc$ to be a saddle-node bifurcation point.

\subsection{Critical load relations for \texorpdfstring{$a \gtrsim \gamma^2$}{a ≳ γ²}}

To derive formulas for $s_\rmc$, we need to make further assumptions about $a$. First, we consider the case where $a$ is not too small. In \cref{fig:S-extra}(b), we plot the right-hand side (RHS) of \cref{eq:S-y}, along with its first two terms and third term separately. The first two terms generally capture the behavior of the RHS. The third term contributes a pole, whose location approximately sets the position of the local maximum of the RHS where its derivative equals 0. Thus, we use the first two terms to satisfy \cref{eq:S-y} and the denominator of the third term to satisfy its derivative:
\begin{align}
  \alpha' &\approx \biggl(\frac{s_\rmc c^2}{y^2}-1\biggr) \biggl[\erfy-\frac{(1-2\gamma)^2a(1-a)}{\gamma^2s_\rmc c^2}\expyy\biggr]^2 \nonum
  0 &\approx \erfy - \biggl(\frac{(1-2\gamma)^2a(1-a)}{\gamma^2s_\rmc c^2} + 1\biggr)\expyy.
\end{align}
We can manipulate these equations to obtain \cref{eq:results-S1} if we convert $\alpha'$ back to $\alpha$ with \cref{eq:S-rescaled}.

\subsection{Critical load formula for \texorpdfstring{$a \ll \gamma^2$}{a ≪ γ}}

Next we consider $a \rightarrow 0$. In this case, the pole location $y \rightarrow 0$ in \cref{eq:S-y}, which does not correspond to a retrieval solution $m'_\rms \approx 1$ according to \cref{eq:m-Ssimple}. Thus the pole location cannot be used to satisfy the derivative of \cref{eq:S-y}. To proceed, we instead set $a = 0$ in \cref{eq:S-y} and obtain
\begin{equation}
  0 = \frac{1}{\alpha'}\biggl(\frac{sc^2}{y^2}-1\biggr)\biggl[\erfy\biggr]^2 \biggl[\erfy-\expyy\biggr]^2 - \biggl[\erfy-\expyy\biggr]^2 - sc^4 \biggl[\erfy + \frac{1}{sc^2}\expyy\biggr]^2.
  \label{eq:S-a0}
\end{equation}
We then directly calculate its derivative with respect to $y$. Along with the original \cref{eq:S-a0}, this gives
\begin{align}
  \alpha' &= \frac{\Bigl(\frac{s_\rmc c^2}{y^2}-1\Bigr) \bigl[\erfy\bigr]^2 \Bigr[\erfy-\expyy\Bigr]^2}{\Bigr[\erfy-\expyy\Bigr]^2 + s_\rmc c^4 \Bigl[\erfy + \frac{1}{s_\rmc c^2}\expyy\Bigr]^2} \nonum
  \alpha'c^2 &= \frac{\erfy \biggl\{\frac{s_\rmc c^2}{y^2}\Bigr[\erfy-\expyy\Bigr] + \expyy\biggr\} \Bigr[\erfy-\expyy\Bigr]^3}{(s_\rmc c^2+1)\expyy \Bigl[\erfy + \frac{1}{s_\rmc c^2}\expyy\Bigr] \Bigl[(y^2-1)\erfy + \expyy\Bigr]}.
  \label{eq:Sa-alpha}
\end{align}

To find a formula for $s_\rmc$, we boldly expand these equations in leading powers of $y$ while preserving extra powers of $c^4$. By solving \cref{eq:Sa-alpha} numerically, we see that $y \sim 1$, so this simplification is not strictly valid [\cref{fig:S-extra}(b)]; nevertheless, our ultimately derived formula matches reasonably well with numerical results [\cref{fig:S}(e)]. The equations become
\begin{align}
  \alpha' &\approx \frac{2s_\rmc c^2y^4 \bigl(3s_\rmc c^2 - (3+s_\rmc c^2)y^2\bigr)}{3\pi\bigl(s_\rmc c^2y^4 + 9c^2(1+s_\rmc c^2)^2\bigr)} \nonum
  \alpha'c^2 &\approx \frac{s_\rmc c^2 (3+s_\rmc c^2) y^6}{27\pi(1+s_\rmc c^2)^2}.
  \label{eq:Sa-alpha2}
\end{align}
Equating these two expressions for $\alpha'$, we get
\begin{equation}
  s_\rmc c^2(3+s_\rmc c^2)y^6 + 27(1+s_\rmc c^2)^2(3+s_\rmc c^2)c^2y^2 = 54s_\rmc c^2(1+s_\rmc c^2)^2c^2.
\end{equation}
We can solve this equation for $y$ using the cubic formula to obtain
\begin{equation}
  y^2 = \bigl(\sqrt{A^3+B^2}+B\bigr)^{1/3} - \bigl(\sqrt{A^3+B^2}-B\bigr)^{1/3}, \quad\textrm{ where }\quad A = \frac{9(1+s_\rmc c^2)^2c^2}{s_\rmc c^2}, \quad B = \frac{27(1+s_\rmc c^2)^2c^2}{3+s_\rmc c^2}.
\end{equation}
Substituting this expression into \cref{eq:Sa-alpha2}, we find an equation for $\alpha'$ in terms of $s_\rmc$:
\begin{equation}
  \alpha' = \frac{s_\rmc c^2}{\pi B} \Bigl\{ 2B - 3A\Bigl[ \bigl(\sqrt{A^3+B^2}+B\bigr)^{1/3} - \bigl(\sqrt{A^3+B^2}-B\bigr)^{1/3} \Bigr] \Bigr\}
\end{equation}
Finally, we can solve for $s_\rmc$ as a series in $\alpha'$. We keep only the leading term in $\alpha'$ and the leading term in $c$ to obtain
\begin{equation}
  s_\rmc \approx 3 \biggl(\frac{3\pi}{8}\biggr)^{\!\!1/4} \frac{(\alpha')^{1/4}}{c^{3/2}} + \frac{3\pi}{8} \frac{\alpha'}{c^2}.
\end{equation}
This yields \cref{eq:results-S2} if we convert $\alpha'$ back to $\alpha$ with \cref{eq:S-rescaled}.

\clearpage
\end{widetext}


\begin{thebibliography}{53}%
\makeatletter
\providecommand \@ifxundefined [1]{%
 \@ifx{#1\undefined}
}%
\providecommand \@ifnum [1]{%
 \ifnum #1\expandafter \@firstoftwo
 \else \expandafter \@secondoftwo
 \fi
}%
\providecommand \@ifx [1]{%
 \ifx #1\expandafter \@firstoftwo
 \else \expandafter \@secondoftwo
 \fi
}%
\providecommand \natexlab [1]{#1}%
\providecommand \enquote  [1]{``#1''}%
\providecommand \bibnamefont  [1]{#1}%
\providecommand \bibfnamefont [1]{#1}%
\providecommand \citenamefont [1]{#1}%
\providecommand \href@noop [0]{\@secondoftwo}%
\providecommand \href [0]{\begingroup \@sanitize@url \@href}%
\providecommand \@href[1]{\@@startlink{#1}\@@href}%
\providecommand \@@href[1]{\endgroup#1\@@endlink}%
\providecommand \@sanitize@url [0]{\catcode `\\12\catcode `\$12\catcode
  `\&12\catcode `\#12\catcode `\^12\catcode `\_12\catcode `\%12\relax}%
\providecommand \@@startlink[1]{}%
\providecommand \@@endlink[0]{}%
\providecommand \url  [0]{\begingroup\@sanitize@url \@url }%
\providecommand \@url [1]{\endgroup\@href {#1}{\urlprefix }}%
\providecommand \urlprefix  [0]{URL }%
\providecommand \Eprint [0]{\href }%
\providecommand \doibase [0]{https://doi.org/}%
\providecommand \selectlanguage [0]{\@gobble}%
\providecommand \bibinfo  [0]{\@secondoftwo}%
\providecommand \bibfield  [0]{\@secondoftwo}%
\providecommand \translation [1]{[#1]}%
\providecommand \BibitemOpen [0]{}%
\providecommand \bibitemStop [0]{}%
\providecommand \bibitemNoStop [0]{.\EOS\space}%
\providecommand \EOS [0]{\spacefactor3000\relax}%
\providecommand \BibitemShut  [1]{\csname bibitem#1\endcsname}%
\let\auto@bib@innerbib\@empty
\bibitem [{\citenamefont {McNaughton}\ and\ \citenamefont
  {Morris}(1987)}]{McNaughton.1987}%
  \BibitemOpen
  \bibfield  {author} {\bibinfo {author} {\bibfnamefont {B.}~\bibnamefont
  {McNaughton}}\ and\ \bibinfo {author} {\bibfnamefont {R.}~\bibnamefont
  {Morris}},\ }\bibfield  {title} {\bibinfo {title} {{Hippocampal synaptic
  enhancement and information storage within a distributed memory system}},\
  }\href {https://doi.org/10.1016/0166-2236(87)90011-7} {\bibfield  {journal}
  {\bibinfo  {journal} {Trends Neurosci.}\ }\textbf {\bibinfo {volume} {10}},\
  \bibinfo {pages} {408} (\bibinfo {year} {1987})}\BibitemShut {NoStop}%
\bibitem [{\citenamefont {O'Reilly}\ and\ \citenamefont
  {Rudy}(2001)}]{O'Reilly.2001}%
  \BibitemOpen
  \bibfield  {author} {\bibinfo {author} {\bibfnamefont {R.~C.}\ \bibnamefont
  {O'Reilly}}\ and\ \bibinfo {author} {\bibfnamefont {J.~W.}\ \bibnamefont
  {Rudy}},\ }\bibfield  {title} {\bibinfo {title} {{Conjunctive representations
  in learning and memory: Principles of cortical and hippocampal function}},\
  }\href {https://doi.org/10.1037/0033-295x.108.2.311} {\bibfield  {journal}
  {\bibinfo  {journal} {Psychol. Rev.}\ }\textbf {\bibinfo {volume} {108}},\
  \bibinfo {pages} {311} (\bibinfo {year} {2001})}\BibitemShut {NoStop}%
\bibitem [{\citenamefont {Rolls}\ and\ \citenamefont
  {Kesner}(2006)}]{Rolls.20069yq}%
  \BibitemOpen
  \bibfield  {author} {\bibinfo {author} {\bibfnamefont {E.~T.}\ \bibnamefont
  {Rolls}}\ and\ \bibinfo {author} {\bibfnamefont {R.~P.}\ \bibnamefont
  {Kesner}},\ }\bibfield  {title} {{\selectlanguage {English}\bibinfo {title}
  {{A computational theory of hippocampal function, and empirical tests of the
  theory}}},\ }\href {https://doi.org/10.1016/j.pneurobio.2006.04.005}
  {\bibfield  {journal} {\bibinfo  {journal} {Prog. Neurobiol.}\ }\textbf
  {\bibinfo {volume} {79}},\ \bibinfo {pages} {1} (\bibinfo {year}
  {2006})}\BibitemShut {NoStop}%
\bibitem [{\citenamefont {Hopfield}\ and\ \citenamefont
  {Tank}(1985)}]{Hopfield.1985}%
  \BibitemOpen
  \bibfield  {author} {\bibinfo {author} {\bibfnamefont {J.~J.}\ \bibnamefont
  {Hopfield}}\ and\ \bibinfo {author} {\bibfnamefont {D.~W.}\ \bibnamefont
  {Tank}},\ }\bibfield  {title} {\bibinfo {title} {{“Neural” computation of
  decisions in optimization problems}},\ }\href
  {https://doi.org/10.1007/bf00339943} {\bibfield  {journal} {\bibinfo
  {journal} {Biol. Cybern.}\ }\textbf {\bibinfo {volume} {52}},\ \bibinfo
  {pages} {141} (\bibinfo {year} {1985})}\BibitemShut {NoStop}%
\bibitem [{\citenamefont {Barra}\ \emph {et~al.}(2012)\citenamefont {Barra},
  \citenamefont {Bernacchia}, \citenamefont {Santucci},\ and\ \citenamefont
  {Contucci}}]{Barra.2012}%
  \BibitemOpen
  \bibfield  {author} {\bibinfo {author} {\bibfnamefont {A.}~\bibnamefont
  {Barra}}, \bibinfo {author} {\bibfnamefont {A.}~\bibnamefont {Bernacchia}},
  \bibinfo {author} {\bibfnamefont {E.}~\bibnamefont {Santucci}},\ and\
  \bibinfo {author} {\bibfnamefont {P.}~\bibnamefont {Contucci}},\ }\bibfield
  {title} {\bibinfo {title} {{On the equivalence of Hopfield networks and
  Boltzmann Machines}},\ }\href {https://doi.org/10.1016/j.neunet.2012.06.003}
  {\bibfield  {journal} {\bibinfo  {journal} {Neural Networks}\ }\textbf
  {\bibinfo {volume} {34}},\ \bibinfo {pages} {1} (\bibinfo {year}
  {2012})}\BibitemShut {NoStop}%
\bibitem [{\citenamefont {Marr}(1971)}]{Marr.1971}%
  \BibitemOpen
  \bibfield  {author} {\bibinfo {author} {\bibfnamefont {D.}~\bibnamefont
  {Marr}},\ }\bibfield  {title} {{\selectlanguage {English}\bibinfo {title}
  {{Simple memory: a theory for archicortex}}},\ }\href
  {https://doi.org/10.1098/rstb.1971.0078} {\bibfield  {journal} {\bibinfo
  {journal} {Philos. Trans. R. Soc. B}\ }\textbf {\bibinfo {volume} {262}},\
  \bibinfo {pages} {23} (\bibinfo {year} {1971})}\BibitemShut {NoStop}%
\bibitem [{\citenamefont {Tsodyks}\ and\ \citenamefont
  {Feigel'man}(1988)}]{Tsodyks.1988}%
  \BibitemOpen
  \bibfield  {author} {\bibinfo {author} {\bibfnamefont {M.~V.}\ \bibnamefont
  {Tsodyks}}\ and\ \bibinfo {author} {\bibfnamefont {M.~V.}\ \bibnamefont
  {Feigel'man}},\ }\bibfield  {title} {{\selectlanguage {English}\bibinfo
  {title} {{The enhanced storage capacity in neural networks with low activity
  level}}},\ }\href {https://doi.org/10.1209/0295-5075/6/2/002} {\bibfield
  {journal} {\bibinfo  {journal} {Europhys. Lett.}\ }\textbf {\bibinfo {volume}
  {6}},\ \bibinfo {pages} {101} (\bibinfo {year} {1988})}\BibitemShut {NoStop}%
\bibitem [{\citenamefont {Kanerva}(1988)}]{Kanerva.1988}%
  \BibitemOpen
  \bibfield  {author} {\bibinfo {author} {\bibfnamefont {P.}~\bibnamefont
  {Kanerva}},\ }\href@noop {} {\emph {\bibinfo {title} {{Sparse distributed
  memory}}}}\ (\bibinfo  {publisher} {MIT press},\ \bibinfo {address}
  {Cambridge, Massachusetts},\ \bibinfo {year} {1988})\BibitemShut {NoStop}%
\bibitem [{\citenamefont {Nadal}\ and\ \citenamefont
  {Toulouse}(1990)}]{Nadal.1990}%
  \BibitemOpen
  \bibfield  {author} {\bibinfo {author} {\bibfnamefont {J.-P.}\ \bibnamefont
  {Nadal}}\ and\ \bibinfo {author} {\bibfnamefont {G.}~\bibnamefont
  {Toulouse}},\ }\bibfield  {title} {\bibinfo {title} {{Information storage in
  sparsely coded memory nets}},\ }\href
  {https://doi.org/10.1088/0954-898x\_1\_1\_005} {\bibfield  {journal}
  {\bibinfo  {journal} {Netw. Comput. Neural Syst.}\ }\textbf {\bibinfo
  {volume} {1}},\ \bibinfo {pages} {61} (\bibinfo {year} {1990})}\BibitemShut
  {NoStop}%
\bibitem [{\citenamefont {Rolls}\ and\ \citenamefont
  {Treves}(1990)}]{Rolls.1990}%
  \BibitemOpen
  \bibfield  {author} {\bibinfo {author} {\bibfnamefont {E.~T.}\ \bibnamefont
  {Rolls}}\ and\ \bibinfo {author} {\bibfnamefont {A.}~\bibnamefont {Treves}},\
  }\bibfield  {title} {\bibinfo {title} {{The relative advantages of sparse
  versus distributed encoding for associative neuronal networks in the
  brain}},\ }\href {https://doi.org/10.1088/0954-898x\_1\_4\_002} {\bibfield
  {journal} {\bibinfo  {journal} {Netw. Comput. Neural Syst.}\ }\textbf
  {\bibinfo {volume} {1}},\ \bibinfo {pages} {407} (\bibinfo {year}
  {1990})}\BibitemShut {NoStop}%
\bibitem [{\citenamefont {Treves}\ and\ \citenamefont
  {Rolls}(1991)}]{Treves.1991}%
  \BibitemOpen
  \bibfield  {author} {\bibinfo {author} {\bibfnamefont {A.}~\bibnamefont
  {Treves}}\ and\ \bibinfo {author} {\bibfnamefont {E.~T.}\ \bibnamefont
  {Rolls}},\ }\bibfield  {title} {{\selectlanguage {English}\bibinfo {title}
  {{What determines the capacity of autoassociative memories in the brain?}}},\
  }\href {https://doi.org/10.1088/0954-898x\_2\_4\_004} {\bibfield  {journal}
  {\bibinfo  {journal} {Netw. Comput. Neural Syst.}\ }\textbf {\bibinfo
  {volume} {2}},\ \bibinfo {pages} {371} (\bibinfo {year} {1991})}\BibitemShut
  {NoStop}%
\bibitem [{\citenamefont {Palm}(2013)}]{Palm.2013}%
  \BibitemOpen
  \bibfield  {author} {\bibinfo {author} {\bibfnamefont {G.}~\bibnamefont
  {Palm}},\ }\bibfield  {title} {\bibinfo {title} {{Neural associative memories
  and sparse coding}},\ }\href {https://doi.org/10.1016/j.neunet.2012.08.013}
  {\bibfield  {journal} {\bibinfo  {journal} {Neural Networks}\ }\textbf
  {\bibinfo {volume} {37}},\ \bibinfo {pages} {165} (\bibinfo {year}
  {2013})}\BibitemShut {NoStop}%
\bibitem [{\citenamefont {Fontanari}(1990)}]{Fontanari.1990}%
  \BibitemOpen
  \bibfield  {author} {\bibinfo {author} {\bibfnamefont {J.~F.}\ \bibnamefont
  {Fontanari}},\ }\bibfield  {title} {{\selectlanguage {English}\bibinfo
  {title} {{Generalization in a Hopfield network}}},\ }\href
  {https://doi.org/10.1051/jphys:0199000510210242100} {\bibfield  {journal}
  {\bibinfo  {journal} {J. Phys.}\ }\textbf {\bibinfo {volume} {51}},\ \bibinfo
  {pages} {2421} (\bibinfo {year} {1990})}\BibitemShut {NoStop}%
\bibitem [{\citenamefont {Stariolo}\ and\ \citenamefont
  {Tamarit}(1992)}]{Stariolo.1992}%
  \BibitemOpen
  \bibfield  {author} {\bibinfo {author} {\bibfnamefont {D.~A.}\ \bibnamefont
  {Stariolo}}\ and\ \bibinfo {author} {\bibfnamefont {F.~A.}\ \bibnamefont
  {Tamarit}},\ }\bibfield  {title} {{\selectlanguage {English}\bibinfo {title}
  {{Generalization in an analog neural network}}},\ }\href
  {https://doi.org/10.1103/physreva.46.5249} {\bibfield  {journal} {\bibinfo
  {journal} {Phys. Rev. A}\ }\textbf {\bibinfo {volume} {46}},\ \bibinfo
  {pages} {5249} (\bibinfo {year} {1992})}\BibitemShut {NoStop}%
\bibitem [{\citenamefont {Dominguez}(1998)}]{Dominguez.1998}%
  \BibitemOpen
  \bibfield  {author} {\bibinfo {author} {\bibfnamefont {D.~R.~C.}\
  \bibnamefont {Dominguez}},\ }\bibfield  {title} {{\selectlanguage
  {English}\bibinfo {title} {{Information capacity of a hierarchical neural
  network}}},\ }\href {https://doi.org/10.1103/physreve.58.4811} {\bibfield
  {journal} {\bibinfo  {journal} {Phys. Rev. E}\ }\textbf {\bibinfo {volume}
  {58}},\ \bibinfo {pages} {4811} (\bibinfo {year} {1998})}\BibitemShut
  {NoStop}%
\bibitem [{\citenamefont {Kang}\ and\ \citenamefont
  {Toyoizumi}(2023)}]{Kang.2023b}%
  \BibitemOpen
  \bibfield  {author} {\bibinfo {author} {\bibfnamefont {L.}~\bibnamefont
  {Kang}}\ and\ \bibinfo {author} {\bibfnamefont {T.}~\bibnamefont
  {Toyoizumi}},\ }\bibfield  {title} {\bibinfo {title} {{Distinguishing
  examples while building concepts in hippocampal and artificial networks}},\
  }\href {https://doi.org/10.1101/2023.02.21.529365} {\bibfield  {journal}
  {\bibinfo  {journal} {bioRxiv}\ ,\ \bibinfo {pages} {2023.02.21.529365}}
  (\bibinfo {year} {2023})}\BibitemShut {NoStop}%
\bibitem [{\citenamefont {Amaral}\ and\ \citenamefont
  {Pierre}(2006)}]{Amaral.2006}%
  \BibitemOpen
  \bibfield  {author} {\bibinfo {author} {\bibfnamefont {D.}~\bibnamefont
  {Amaral}}\ and\ \bibinfo {author} {\bibfnamefont {L.}~\bibnamefont
  {Pierre}},\ }\bibfield  {title} {\bibinfo {title} {{Hippocampal
  neuroanatomy}},\ }in\ \href
  {https://doi.org/10.1093/acprof:oso/9780195100273.003.0003} {\emph {\bibinfo
  {booktitle} {The Hippocampus Book}}},\ \bibinfo {series and number} {The
  Hippocampus Book},\ \bibinfo {editor} {edited by\ \bibinfo {editor}
  {\bibfnamefont {P.}~\bibnamefont {Andersen}}, \bibinfo {editor}
  {\bibfnamefont {R.}~\bibnamefont {Morris}}, \bibinfo {editor} {\bibfnamefont
  {D.}~\bibnamefont {Amaral}}, \bibinfo {editor} {\bibfnamefont
  {T.}~\bibnamefont {Bliss}},\ and\ \bibinfo {editor} {\bibfnamefont
  {J.}~\bibnamefont {O'Keefe}}}\ (\bibinfo  {publisher} {Oxford University
  Press},\ \bibinfo {year} {2006})\ pp.\ \bibinfo {pages} {37--114}\BibitemShut
  {NoStop}%
\bibitem [{\citenamefont {Hopfield}(1982)}]{Hopfield.1982}%
  \BibitemOpen
  \bibfield  {author} {\bibinfo {author} {\bibfnamefont {J.~J.}\ \bibnamefont
  {Hopfield}},\ }\bibfield  {title} {{\selectlanguage {English}\bibinfo {title}
  {{Neural networks and physical systems with emergent collective computational
  abilities}}},\ }\href@noop {} {\bibfield  {journal} {\bibinfo  {journal}
  {Proc. Natl. Acad. Sci. U.S.A.}\ }\textbf {\bibinfo {volume} {79}},\ \bibinfo
  {pages} {2554} (\bibinfo {year} {1982})}\BibitemShut {NoStop}%
\bibitem [{\citenamefont {Amit}\ \emph {et~al.}(1985)\citenamefont {Amit},
  \citenamefont {Gutfreund},\ and\ \citenamefont {Sompolinsky}}]{Amit.1985mb}%
  \BibitemOpen
  \bibfield  {author} {\bibinfo {author} {\bibfnamefont {D.~J.}\ \bibnamefont
  {Amit}}, \bibinfo {author} {\bibfnamefont {H.}~\bibnamefont {Gutfreund}},\
  and\ \bibinfo {author} {\bibfnamefont {H.}~\bibnamefont {Sompolinsky}},\
  }\bibfield  {title} {\bibinfo {title} {{Spin-glass models of neural
  networks}},\ }\href {https://doi.org/10.1103/physreva.32.1007} {\bibfield
  {journal} {\bibinfo  {journal} {Phys. Rev. A}\ }\textbf {\bibinfo {volume}
  {32}},\ \bibinfo {pages} {1007} (\bibinfo {year} {1985})}\BibitemShut
  {NoStop}%
\bibitem [{\citenamefont {Hertz}\ \emph {et~al.}(2018)\citenamefont {Hertz},
  \citenamefont {Krogh},\ and\ \citenamefont {Palmer}}]{Hertz.2018}%
  \BibitemOpen
  \bibfield  {author} {\bibinfo {author} {\bibfnamefont {J.}~\bibnamefont
  {Hertz}}, \bibinfo {author} {\bibfnamefont {A.}~\bibnamefont {Krogh}},\ and\
  \bibinfo {author} {\bibfnamefont {R.}~\bibnamefont {Palmer}},\ }\href
  {https://books.google.co.jp/books?id=dI2rDnN\_eZEC} {\emph {\bibinfo {title}
  {{Introduction To The Theory Of Neural Computation}}}},\ \bibinfo {series}
  {Santa Fe Institute studies in the sciences of complexity: Lecture notes}\
  No.~\bibinfo {number} {1}\ (\bibinfo  {publisher} {CRC Press},\ \bibinfo
  {address} {Boca Raton},\ \bibinfo {year} {2018})\BibitemShut {NoStop}%
\bibitem [{\citenamefont {Weisbuch}\ and\ \citenamefont
  {Fogelman-Soulie}(1985)}]{Weisbuch.1985}%
  \BibitemOpen
  \bibfield  {author} {\bibinfo {author} {\bibfnamefont {G.}~\bibnamefont
  {Weisbuch}}\ and\ \bibinfo {author} {\bibfnamefont {F.}~\bibnamefont
  {Fogelman-Soulie}},\ }\bibfield  {title} {\bibinfo {title} {{Scaling laws for
  the attractors of Hopfield networks}},\ }\href
  {https://doi.org/10.1051/jphyslet:019850046014062300} {\bibfield  {journal}
  {\bibinfo  {journal} {Journal de Physique Lettres}\ }\textbf {\bibinfo
  {volume} {46}},\ \bibinfo {pages} {623} (\bibinfo {year} {1985})}\BibitemShut
  {NoStop}%
\bibitem [{\citenamefont {Mézard}\ and\ \citenamefont
  {Virasoro}(1985)}]{Mezard.1985}%
  \BibitemOpen
  \bibfield  {author} {\bibinfo {author} {\bibfnamefont {M.}~\bibnamefont
  {Mézard}}\ and\ \bibinfo {author} {\bibfnamefont {M.~A.}\ \bibnamefont
  {Virasoro}},\ }\bibfield  {title} {\bibinfo {title} {{The microstructure of
  ultrametricity}},\ }\href {https://doi.org/10.1051/jphys:019850046080129300}
  {\bibfield  {journal} {\bibinfo  {journal} {J. Phys.}\ }\textbf {\bibinfo
  {volume} {46}},\ \bibinfo {pages} {1293} (\bibinfo {year}
  {1985})}\BibitemShut {NoStop}%
\bibitem [{\citenamefont {Dotsenko}(1985)}]{Dotsenko.1985}%
  \BibitemOpen
  \bibfield  {author} {\bibinfo {author} {\bibfnamefont {V.~S.}\ \bibnamefont
  {Dotsenko}},\ }\bibfield  {title} {\bibinfo {title} {{‘Ordered’ spin
  glass: a hierarchical memory machine}},\ }\href
  {https://doi.org/10.1088/0022-3719/18/31/008} {\bibfield  {journal} {\bibinfo
   {journal} {J. Phys. C: Solid State Phys.}\ }\textbf {\bibinfo {volume}
  {18}},\ \bibinfo {pages} {L1017} (\bibinfo {year} {1985})}\BibitemShut
  {NoStop}%
\bibitem [{\citenamefont {Cortes}\ \emph {et~al.}(1987)\citenamefont {Cortes},
  \citenamefont {Krogh},\ and\ \citenamefont {Hertz}}]{Cortes.1987}%
  \BibitemOpen
  \bibfield  {author} {\bibinfo {author} {\bibfnamefont {C.}~\bibnamefont
  {Cortes}}, \bibinfo {author} {\bibfnamefont {A.}~\bibnamefont {Krogh}},\ and\
  \bibinfo {author} {\bibfnamefont {J.~A.}\ \bibnamefont {Hertz}},\ }\bibfield
  {title} {\bibinfo {title} {{Hierarchical associative networks}},\ }\href
  {https://doi.org/10.1088/0305-4470/20/13/044} {\bibfield  {journal} {\bibinfo
   {journal} {J. Phys. A: Math. Gen.}\ }\textbf {\bibinfo {volume} {20}},\
  \bibinfo {pages} {4449} (\bibinfo {year} {1987})}\BibitemShut {NoStop}%
\bibitem [{\citenamefont {Virasoro}(1988)}]{Virasoro.1988}%
  \BibitemOpen
  \bibfield  {author} {\bibinfo {author} {\bibfnamefont {M.~A.}\ \bibnamefont
  {Virasoro}},\ }\bibfield  {title} {\bibinfo {title} {{The effect of synapses
  destruction on categorization by neural networks}},\ }\href
  {https://doi.org/10.1209/0295-5075/7/4/002} {\bibfield  {journal} {\bibinfo
  {journal} {EPL}\ }\textbf {\bibinfo {volume} {7}},\ \bibinfo {pages} {293}
  (\bibinfo {year} {1988})}\BibitemShut {NoStop}%
\bibitem [{\citenamefont {Gutfreund}(1988)}]{Gutfreund.1988}%
  \BibitemOpen
  \bibfield  {author} {\bibinfo {author} {\bibfnamefont {H.}~\bibnamefont
  {Gutfreund}},\ }\bibfield  {title} {{\selectlanguage {English}\bibinfo
  {title} {{Neural networks with hierarchically correlated patterns}}},\ }\href
  {https://doi.org/10.1103/physreva.37.570} {\bibfield  {journal} {\bibinfo
  {journal} {Phys. Rev. A}\ }\textbf {\bibinfo {volume} {37}},\ \bibinfo
  {pages} {570} (\bibinfo {year} {1988})}\BibitemShut {NoStop}%
\bibitem [{\citenamefont {Krogh}\ and\ \citenamefont
  {Hertz}(1988)}]{Krogh.1988}%
  \BibitemOpen
  \bibfield  {author} {\bibinfo {author} {\bibfnamefont {A.}~\bibnamefont
  {Krogh}}\ and\ \bibinfo {author} {\bibfnamefont {J.~A.}\ \bibnamefont
  {Hertz}},\ }\bibfield  {title} {{\selectlanguage {English}\bibinfo {title}
  {{Mean-field analysis of hierarchical associative networks with
  ‘magnetisation’}}},\ }\href {https://doi.org/10.1088/0305-4470/21/9/033}
  {\bibfield  {journal} {\bibinfo  {journal} {J. Phys. A: Math. Gen.}\ }\textbf
  {\bibinfo {volume} {21}},\ \bibinfo {pages} {2211} (\bibinfo {year}
  {1988})}\BibitemShut {NoStop}%
\bibitem [{\citenamefont {Treves}\ and\ \citenamefont
  {Rolls}(1992)}]{Treves.1992}%
  \BibitemOpen
  \bibfield  {author} {\bibinfo {author} {\bibfnamefont {A.}~\bibnamefont
  {Treves}}\ and\ \bibinfo {author} {\bibfnamefont {E.~T.}\ \bibnamefont
  {Rolls}},\ }\bibfield  {title} {{\selectlanguage {English}\bibinfo {title}
  {{Computational constraints suggest the need for two distinct input systems
  to the hippocampal CA3 network}}},\ }\href
  {https://doi.org/10.1088/0305-4470/24/11/029} {\bibfield  {journal} {\bibinfo
   {journal} {Hippocampus}\ }\textbf {\bibinfo {volume} {2}},\ \bibinfo {pages}
  {189} (\bibinfo {year} {1992})}\BibitemShut {NoStop}%
\bibitem [{\citenamefont {O'Reilly}\ and\ \citenamefont
  {McClelland}(1994)}]{O'Reilly.1994}%
  \BibitemOpen
  \bibfield  {author} {\bibinfo {author} {\bibfnamefont {R.~C.}\ \bibnamefont
  {O'Reilly}}\ and\ \bibinfo {author} {\bibfnamefont {J.~L.}\ \bibnamefont
  {McClelland}},\ }\bibfield  {title} {{\selectlanguage {English}\bibinfo
  {title} {{Hippocampal conjunctive encoding, storage, and recall: Avoiding a
  trade-off}}},\ }\href {https://doi.org/10.1002/hipo.450040605} {\bibfield
  {journal} {\bibinfo  {journal} {Hippocampus}\ }\textbf {\bibinfo {volume}
  {4}},\ \bibinfo {pages} {661} (\bibinfo {year} {1994})}\BibitemShut {NoStop}%
\bibitem [{\citenamefont {Vinje}\ and\ \citenamefont
  {Gallant}(2000)}]{Vinje.2000}%
  \BibitemOpen
  \bibfield  {author} {\bibinfo {author} {\bibfnamefont {W.~E.}\ \bibnamefont
  {Vinje}}\ and\ \bibinfo {author} {\bibfnamefont {J.~L.}\ \bibnamefont
  {Gallant}},\ }\bibfield  {title} {\bibinfo {title} {{Sparse coding and
  decorrelation in primary visual cortex during natural vision}},\ }\href
  {https://doi.org/10.1126/science.287.5456.1273} {\bibfield  {journal}
  {\bibinfo  {journal} {Science}\ }\textbf {\bibinfo {volume} {287}},\ \bibinfo
  {pages} {1273} (\bibinfo {year} {2000})}\BibitemShut {NoStop}%
\bibitem [{\citenamefont {Wiechert}\ \emph {et~al.}(2010)\citenamefont
  {Wiechert}, \citenamefont {Judkewitz}, \citenamefont {Riecke},\ and\
  \citenamefont {Friedrich}}]{Wiechert.2010}%
  \BibitemOpen
  \bibfield  {author} {\bibinfo {author} {\bibfnamefont {M.~T.}\ \bibnamefont
  {Wiechert}}, \bibinfo {author} {\bibfnamefont {B.}~\bibnamefont {Judkewitz}},
  \bibinfo {author} {\bibfnamefont {H.}~\bibnamefont {Riecke}},\ and\ \bibinfo
  {author} {\bibfnamefont {R.~W.}\ \bibnamefont {Friedrich}},\ }\bibfield
  {title} {\bibinfo {title} {{Mechanisms of pattern decorrelation by recurrent
  neuronal circuits}},\ }\href {https://doi.org/10.1038/nn.2591} {\bibfield
  {journal} {\bibinfo  {journal} {Nat. Neurosci.}\ }\textbf {\bibinfo {volume}
  {13}},\ \bibinfo {pages} {1003} (\bibinfo {year} {2010})}\BibitemShut
  {NoStop}%
\bibitem [{\citenamefont {Pitkow}\ and\ \citenamefont
  {Meister}(2012)}]{Pitkow.2012}%
  \BibitemOpen
  \bibfield  {author} {\bibinfo {author} {\bibfnamefont {X.}~\bibnamefont
  {Pitkow}}\ and\ \bibinfo {author} {\bibfnamefont {M.}~\bibnamefont
  {Meister}},\ }\bibfield  {title} {\bibinfo {title} {{Decorrelation and
  efficient coding by retinal ganglion cells}},\ }\href
  {https://doi.org/10.1038/nn.3064} {\bibfield  {journal} {\bibinfo  {journal}
  {Nat. Neurosci.}\ }\textbf {\bibinfo {volume} {15}},\ \bibinfo {pages} {628}
  (\bibinfo {year} {2012})}\BibitemShut {NoStop}%
\bibitem [{\citenamefont {Cayco-Gajic}\ \emph {et~al.}(2017)\citenamefont
  {Cayco-Gajic}, \citenamefont {Clopath},\ and\ \citenamefont
  {Silver}}]{Cayco-Gajic.2017}%
  \BibitemOpen
  \bibfield  {author} {\bibinfo {author} {\bibfnamefont {N.~A.}\ \bibnamefont
  {Cayco-Gajic}}, \bibinfo {author} {\bibfnamefont {C.}~\bibnamefont
  {Clopath}},\ and\ \bibinfo {author} {\bibfnamefont {R.~A.}\ \bibnamefont
  {Silver}},\ }\bibfield  {title} {\bibinfo {title} {{Sparse synaptic
  connectivity is required for decorrelation and pattern separation in
  feedforward networks}},\ }\href {https://doi.org/10.1038/s41467-017-01109-y}
  {\bibfield  {journal} {\bibinfo  {journal} {Nat. Commun.}\ }\textbf {\bibinfo
  {volume} {8}},\ \bibinfo {pages} {1116} (\bibinfo {year} {2017})}\BibitemShut
  {NoStop}%
\bibitem [{\citenamefont {Kim}\ \emph {et~al.}(2012)\citenamefont {Kim},
  \citenamefont {Guzman}, \citenamefont {Hu},\ and\ \citenamefont
  {Jonas}}]{Kim.2012}%
  \BibitemOpen
  \bibfield  {author} {\bibinfo {author} {\bibfnamefont {S.}~\bibnamefont
  {Kim}}, \bibinfo {author} {\bibfnamefont {S.~J.}\ \bibnamefont {Guzman}},
  \bibinfo {author} {\bibfnamefont {H.}~\bibnamefont {Hu}},\ and\ \bibinfo
  {author} {\bibfnamefont {P.}~\bibnamefont {Jonas}},\ }\bibfield  {title}
  {\bibinfo {title} {{Active dendrites support efficient initiation of
  dendritic spikes in hippocampal CA3 pyramidal neurons}},\ }\href
  {https://doi.org/10.1038/nn.3060} {\bibfield  {journal} {\bibinfo  {journal}
  {Nat. Neurosci.}\ }\textbf {\bibinfo {volume} {15}},\ \bibinfo {pages} {600}
  (\bibinfo {year} {2012})}\BibitemShut {NoStop}%
\bibitem [{\citenamefont {Makara}\ and\ \citenamefont
  {Magee}(2013)}]{Makara.2013}%
  \BibitemOpen
  \bibfield  {author} {\bibinfo {author} {\bibfnamefont {J.}~\bibnamefont
  {Makara}}\ and\ \bibinfo {author} {\bibfnamefont {J.}~\bibnamefont {Magee}},\
  }\bibfield  {title} {\bibinfo {title} {{Variable dendritic integration in
  hippocampal CA3 pyramidal neurons}},\ }\href
  {https://doi.org/10.1016/j.neuron.2013.10.033} {\bibfield  {journal}
  {\bibinfo  {journal} {Neuron}\ }\textbf {\bibinfo {volume} {80}},\ \bibinfo
  {pages} {1438} (\bibinfo {year} {2013})}\BibitemShut {NoStop}%
\bibitem [{\citenamefont {Kaifosh}\ and\ \citenamefont
  {Losonczy}(2016)}]{Kaifosh.2016}%
  \BibitemOpen
  \bibfield  {author} {\bibinfo {author} {\bibfnamefont {P.}~\bibnamefont
  {Kaifosh}}\ and\ \bibinfo {author} {\bibfnamefont {A.}~\bibnamefont
  {Losonczy}},\ }\bibfield  {title} {{\selectlanguage {English}\bibinfo {title}
  {{Mnemonic functions for nonlinear dendritic integration in hippocampal
  pyramidal circuits}}},\ }\href {https://doi.org/10.1016/j.neuron.2016.03.019}
  {\bibfield  {journal} {\bibinfo  {journal} {Neuron}\ }\textbf {\bibinfo
  {volume} {90}},\ \bibinfo {pages} {622} (\bibinfo {year} {2016})}\BibitemShut
  {NoStop}%
\bibitem [{\citenamefont {Scoville}\ and\ \citenamefont
  {Milner}(1957)}]{Scoville.1957}%
  \BibitemOpen
  \bibfield  {author} {\bibinfo {author} {\bibfnamefont {W.~B.}\ \bibnamefont
  {Scoville}}\ and\ \bibinfo {author} {\bibfnamefont {B.}~\bibnamefont
  {Milner}},\ }\bibfield  {title} {\bibinfo {title} {{Loss of recent memory
  after bilateral hippocampal lesions}},\ }\href
  {https://doi.org/10.1136/jnnp.20.1.11} {\bibfield  {journal} {\bibinfo
  {journal} {J. Neurol. Neurosurg. Psychiatry}\ }\textbf {\bibinfo {volume}
  {20}},\ \bibinfo {pages} {11} (\bibinfo {year} {1957})}\BibitemShut {NoStop}%
\bibitem [{\citenamefont {Squire}(1992)}]{Squire.1992}%
  \BibitemOpen
  \bibfield  {author} {\bibinfo {author} {\bibfnamefont {L.~R.}\ \bibnamefont
  {Squire}},\ }\bibfield  {title} {\bibinfo {title} {{Memory and the
  hippocampus: A synthesis from findings with rats, monkeys, and humans}},\
  }\href@noop {} {\bibfield  {journal} {\bibinfo  {journal} {Psychol. Rev.}\
  }\textbf {\bibinfo {volume} {99}},\ \bibinfo {pages} {195} (\bibinfo {year}
  {1992})}\BibitemShut {NoStop}%
\bibitem [{\citenamefont {Leutgeb}\ \emph {et~al.}(2007)\citenamefont
  {Leutgeb}, \citenamefont {Leutgeb}, \citenamefont {Moser},\ and\
  \citenamefont {Moser}}]{Leutgeb.2007}%
  \BibitemOpen
  \bibfield  {author} {\bibinfo {author} {\bibfnamefont {J.~K.}\ \bibnamefont
  {Leutgeb}}, \bibinfo {author} {\bibfnamefont {S.}~\bibnamefont {Leutgeb}},
  \bibinfo {author} {\bibfnamefont {M.-B.}\ \bibnamefont {Moser}},\ and\
  \bibinfo {author} {\bibfnamefont {E.~I.}\ \bibnamefont {Moser}},\ }\bibfield
  {title} {{\selectlanguage {English}\bibinfo {title} {{Pattern separation in
  the dentate gyrus and CA3 of the hippocampus}}},\ }\href
  {https://doi.org/10.1126/science.1135801} {\bibfield  {journal} {\bibinfo
  {journal} {Science}\ }\textbf {\bibinfo {volume} {315}},\ \bibinfo {pages}
  {961} (\bibinfo {year} {2007})}\BibitemShut {NoStop}%
\bibitem [{\citenamefont {Aimone}\ \emph {et~al.}(2011)\citenamefont {Aimone},
  \citenamefont {Deng},\ and\ \citenamefont {Gage}}]{Aimone.2011}%
  \BibitemOpen
  \bibfield  {author} {\bibinfo {author} {\bibfnamefont {J.~B.}\ \bibnamefont
  {Aimone}}, \bibinfo {author} {\bibfnamefont {W.}~\bibnamefont {Deng}},\ and\
  \bibinfo {author} {\bibfnamefont {F.~H.}\ \bibnamefont {Gage}},\ }\bibfield
  {title} {{\selectlanguage {English}\bibinfo {title} {{Resolving new memories:
  A critical look at the dentate gyrus, adult neurogenesis, and pattern
  separation}}},\ }\href {https://doi.org/10.1016/j.neuron.2011.05.010}
  {\bibfield  {journal} {\bibinfo  {journal} {Neuron}\ }\textbf {\bibinfo
  {volume} {70}},\ \bibinfo {pages} {589} (\bibinfo {year} {2011})}\BibitemShut
  {NoStop}%
\bibitem [{\citenamefont {Knowlton}\ and\ \citenamefont
  {Squire}(1993)}]{Knowlton.1993}%
  \BibitemOpen
  \bibfield  {author} {\bibinfo {author} {\bibfnamefont {B.~J.}\ \bibnamefont
  {Knowlton}}\ and\ \bibinfo {author} {\bibfnamefont {L.~R.}\ \bibnamefont
  {Squire}},\ }\bibfield  {title} {\bibinfo {title} {{The learning of
  categories: Parallel brain systems for item memory and category knowledge}},\
  }\href {https://doi.org/10.1126/science.8259522} {\bibfield  {journal}
  {\bibinfo  {journal} {Science}\ }\textbf {\bibinfo {volume} {262}},\ \bibinfo
  {pages} {1747} (\bibinfo {year} {1993})}\BibitemShut {NoStop}%
\bibitem [{\citenamefont {Zeithamova}\ \emph {et~al.}(2008)\citenamefont
  {Zeithamova}, \citenamefont {Maddox},\ and\ \citenamefont
  {Schnyer}}]{Zeithamova.2008}%
  \BibitemOpen
  \bibfield  {author} {\bibinfo {author} {\bibfnamefont {D.}~\bibnamefont
  {Zeithamova}}, \bibinfo {author} {\bibfnamefont {W.~T.}\ \bibnamefont
  {Maddox}},\ and\ \bibinfo {author} {\bibfnamefont {D.~M.}\ \bibnamefont
  {Schnyer}},\ }\bibfield  {title} {\bibinfo {title} {{Dissociable prototype
  learning systems: Evidence from brain imaging and behavior}},\ }\href
  {https://doi.org/10.1523/jneurosci.2915-08.2008} {\bibfield  {journal}
  {\bibinfo  {journal} {J. Neurosci.}\ }\textbf {\bibinfo {volume} {28}},\
  \bibinfo {pages} {13194} (\bibinfo {year} {2008})}\BibitemShut {NoStop}%
\bibitem [{\citenamefont {Schapiro}\ \emph {et~al.}(2014)\citenamefont
  {Schapiro}, \citenamefont {Gregory}, \citenamefont {Landau}, \citenamefont
  {McCloskey},\ and\ \citenamefont {Turk-Browne}}]{Schapiro.2014}%
  \BibitemOpen
  \bibfield  {author} {\bibinfo {author} {\bibfnamefont {A.~C.}\ \bibnamefont
  {Schapiro}}, \bibinfo {author} {\bibfnamefont {E.}~\bibnamefont {Gregory}},
  \bibinfo {author} {\bibfnamefont {B.}~\bibnamefont {Landau}}, \bibinfo
  {author} {\bibfnamefont {M.}~\bibnamefont {McCloskey}},\ and\ \bibinfo
  {author} {\bibfnamefont {N.~B.}\ \bibnamefont {Turk-Browne}},\ }\bibfield
  {title} {\bibinfo {title} {{The necessity of the medial temporal lobe for
  statistical learning}},\ }\href {https://doi.org/10.1162/jocn\_a\_00578}
  {\bibfield  {journal} {\bibinfo  {journal} {J. Cognit. Neurosci.}\ }\textbf
  {\bibinfo {volume} {26}},\ \bibinfo {pages} {1736} (\bibinfo {year}
  {2014})}\BibitemShut {NoStop}%
\bibitem [{\citenamefont {Mack}\ \emph {et~al.}(2016)\citenamefont {Mack},
  \citenamefont {Love},\ and\ \citenamefont {Preston}}]{Mack.2016}%
  \BibitemOpen
  \bibfield  {author} {\bibinfo {author} {\bibfnamefont {M.~L.}\ \bibnamefont
  {Mack}}, \bibinfo {author} {\bibfnamefont {B.~C.}\ \bibnamefont {Love}},\
  and\ \bibinfo {author} {\bibfnamefont {A.~R.}\ \bibnamefont {Preston}},\
  }\bibfield  {title} {\bibinfo {title} {{Dynamic updating of hippocampal
  object representations reflects new conceptual knowledge}},\ }\href
  {https://doi.org/10.1073/pnas.1614048113} {\bibfield  {journal} {\bibinfo
  {journal} {Proc. Natl. Acad. Sci. U.S.A.}\ }\textbf {\bibinfo {volume}
  {113}},\ \bibinfo {pages} {13203} (\bibinfo {year} {2016})}\BibitemShut
  {NoStop}%
\bibitem [{\citenamefont {Covington}\ \emph {et~al.}(2018)\citenamefont
  {Covington}, \citenamefont {Brown-Schmidt},\ and\ \citenamefont
  {Duff}}]{Covington.2018}%
  \BibitemOpen
  \bibfield  {author} {\bibinfo {author} {\bibfnamefont {N.~V.}\ \bibnamefont
  {Covington}}, \bibinfo {author} {\bibfnamefont {S.}~\bibnamefont
  {Brown-Schmidt}},\ and\ \bibinfo {author} {\bibfnamefont {M.~C.}\
  \bibnamefont {Duff}},\ }\bibfield  {title} {\bibinfo {title} {{The necessity
  of the hippocampus for statistical learning}},\ }\href
  {https://doi.org/10.1162/jocn\_a\_01228} {\bibfield  {journal} {\bibinfo
  {journal} {J. Cognit. Neurosci.}\ }\textbf {\bibinfo {volume} {30}},\
  \bibinfo {pages} {680} (\bibinfo {year} {2018})}\BibitemShut {NoStop}%
\bibitem [{\citenamefont {Quian Quiroga}\ \emph {et~al.}(2005)\citenamefont
  {Quian Quiroga}, \citenamefont {Reddy}, \citenamefont {Kreiman},
  \citenamefont {Koch},\ and\ \citenamefont {Fried}}]{Quiroga.2005}%
  \BibitemOpen
  \bibfield  {author} {\bibinfo {author} {\bibfnamefont {R.}~\bibnamefont
  {Quian Quiroga}}, \bibinfo {author} {\bibfnamefont {L.}~\bibnamefont
  {Reddy}}, \bibinfo {author} {\bibfnamefont {G.}~\bibnamefont {Kreiman}},
  \bibinfo {author} {\bibfnamefont {C.}~\bibnamefont {Koch}},\ and\ \bibinfo
  {author} {\bibfnamefont {I.}~\bibnamefont {Fried}},\ }\bibfield  {title}
  {{\selectlanguage {English}\bibinfo {title} {{Invariant visual representation
  by single neurons in the human brain.}}},\ }\href
  {https://doi.org/10.1038/nature03687} {\bibfield  {journal} {\bibinfo
  {journal} {Nature}\ }\textbf {\bibinfo {volume} {435}},\ \bibinfo {pages}
  {1102} (\bibinfo {year} {2005})}\BibitemShut {NoStop}%
\bibitem [{\citenamefont {Quian Quiroga}\ \emph {et~al.}(2009)\citenamefont
  {Quian Quiroga}, \citenamefont {Kraskov}, \citenamefont {Koch},\ and\
  \citenamefont {Fried}}]{Quiroga.2009}%
  \BibitemOpen
  \bibfield  {author} {\bibinfo {author} {\bibfnamefont {R.}~\bibnamefont
  {Quian Quiroga}}, \bibinfo {author} {\bibfnamefont {A.}~\bibnamefont
  {Kraskov}}, \bibinfo {author} {\bibfnamefont {C.}~\bibnamefont {Koch}},\ and\
  \bibinfo {author} {\bibfnamefont {I.}~\bibnamefont {Fried}},\ }\bibfield
  {title} {{\selectlanguage {English}\bibinfo {title} {{Explicit encoding of
  multimodal percepts by single neurons in the human brain}}},\ }\href
  {https://doi.org/10.1016/j.cub.2009.06.060} {\bibfield  {journal} {\bibinfo
  {journal} {Curr. Biol.}\ }\textbf {\bibinfo {volume} {19}},\ \bibinfo {pages}
  {1308} (\bibinfo {year} {2009})}\BibitemShut {NoStop}%
\bibitem [{\citenamefont {Buzsáki}(2002)}]{Buzsaki.2002}%
  \BibitemOpen
  \bibfield  {author} {\bibinfo {author} {\bibfnamefont {G.}~\bibnamefont
  {Buzsáki}},\ }\bibfield  {title} {{\selectlanguage {English}\bibinfo {title}
  {{Theta oscillations in the hippocampus}}},\ }\href
  {https://doi.org/10.1016/s0896-6273(02)00586-x} {\bibfield  {journal}
  {\bibinfo  {journal} {Neuron}\ }\textbf {\bibinfo {volume} {33}},\ \bibinfo
  {pages} {325} (\bibinfo {year} {2002})}\BibitemShut {NoStop}%
\bibitem [{\citenamefont {Litwin-Kumar}\ \emph {et~al.}(2017)\citenamefont
  {Litwin-Kumar}, \citenamefont {Harris}, \citenamefont {Axel}, \citenamefont
  {Sompolinsky},\ and\ \citenamefont {Abbott}}]{Litwin-Kumar.2017}%
  \BibitemOpen
  \bibfield  {author} {\bibinfo {author} {\bibfnamefont {A.}~\bibnamefont
  {Litwin-Kumar}}, \bibinfo {author} {\bibfnamefont {K.~D.}\ \bibnamefont
  {Harris}}, \bibinfo {author} {\bibfnamefont {R.}~\bibnamefont {Axel}},
  \bibinfo {author} {\bibfnamefont {H.}~\bibnamefont {Sompolinsky}},\ and\
  \bibinfo {author} {\bibfnamefont {L.~F.}\ \bibnamefont {Abbott}},\ }\bibfield
   {title} {{\selectlanguage {English}\bibinfo {title} {{Optimal degrees of
  synaptic connectivity}}},\ }\href
  {https://doi.org/10.1016/j.neuron.2017.01.030} {\bibfield  {journal}
  {\bibinfo  {journal} {Neuron}\ }\textbf {\bibinfo {volume} {93}},\ \bibinfo
  {pages} {1153} (\bibinfo {year} {2017})}\BibitemShut {NoStop}%
\bibitem [{\citenamefont {Jeanne}\ \emph {et~al.}(2018)\citenamefont {Jeanne},
  \citenamefont {Fişek},\ and\ \citenamefont {Wilson}}]{Jeanne.2018}%
  \BibitemOpen
  \bibfield  {author} {\bibinfo {author} {\bibfnamefont {J.~M.}\ \bibnamefont
  {Jeanne}}, \bibinfo {author} {\bibfnamefont {M.}~\bibnamefont {Fişek}},\
  and\ \bibinfo {author} {\bibfnamefont {R.~I.}\ \bibnamefont {Wilson}},\
  }\bibfield  {title} {\bibinfo {title} {{The organization of projections from
  olfactory glomeruli onto higher-order neurons}},\ }\href
  {https://doi.org/10.1016/j.neuron.2018.05.011} {\bibfield  {journal}
  {\bibinfo  {journal} {Neuron}\ }\textbf {\bibinfo {volume} {98}},\ \bibinfo
  {pages} {1198} (\bibinfo {year} {2018})}\BibitemShut {NoStop}%
\bibitem [{\citenamefont {Medina}\ \emph {et~al.}(2002)\citenamefont {Medina},
  \citenamefont {Repa}, \citenamefont {Mauk},\ and\ \citenamefont
  {LeDoux}}]{Medina.2002}%
  \BibitemOpen
  \bibfield  {author} {\bibinfo {author} {\bibfnamefont {J.~F.}\ \bibnamefont
  {Medina}}, \bibinfo {author} {\bibfnamefont {J.~C.}\ \bibnamefont {Repa}},
  \bibinfo {author} {\bibfnamefont {M.~D.}\ \bibnamefont {Mauk}},\ and\
  \bibinfo {author} {\bibfnamefont {J.~E.}\ \bibnamefont {LeDoux}},\ }\bibfield
   {title} {\bibinfo {title} {{Parallels between cerebellum- and
  amygdala-dependent conditioning}},\ }\href {https://doi.org/10.1038/nrn728}
  {\bibfield  {journal} {\bibinfo  {journal} {Nat. Rev. Neurosci.}\ }\textbf
  {\bibinfo {volume} {3}},\ \bibinfo {pages} {122} (\bibinfo {year}
  {2002})}\BibitemShut {NoStop}%
\bibitem [{\citenamefont {Hige}\ \emph {et~al.}(2015)\citenamefont {Hige},
  \citenamefont {Aso}, \citenamefont {Rubin},\ and\ \citenamefont
  {Turner}}]{Hige.2015}%
  \BibitemOpen
  \bibfield  {author} {\bibinfo {author} {\bibfnamefont {T.}~\bibnamefont
  {Hige}}, \bibinfo {author} {\bibfnamefont {Y.}~\bibnamefont {Aso}}, \bibinfo
  {author} {\bibfnamefont {G.~M.}\ \bibnamefont {Rubin}},\ and\ \bibinfo
  {author} {\bibfnamefont {G.~C.}\ \bibnamefont {Turner}},\ }\bibfield  {title}
  {\bibinfo {title} {{Plasticity-driven individualization of olfactory coding
  in mushroom body output neurons}},\ }\href
  {https://doi.org/10.1038/nature15396} {\bibfield  {journal} {\bibinfo
  {journal} {Nature}\ }\textbf {\bibinfo {volume} {526}},\ \bibinfo {pages}
  {258} (\bibinfo {year} {2015})}\BibitemShut {NoStop}%
\bibitem [{\citenamefont {Bruce}\ \emph {et~al.}(1987)\citenamefont {Bruce},
  \citenamefont {Gardner},\ and\ \citenamefont {Wallace}}]{Bruce.1987}%
  \BibitemOpen
  \bibfield  {author} {\bibinfo {author} {\bibfnamefont {A.~D.}\ \bibnamefont
  {Bruce}}, \bibinfo {author} {\bibfnamefont {E.~J.}\ \bibnamefont {Gardner}},\
  and\ \bibinfo {author} {\bibfnamefont {D.~J.}\ \bibnamefont {Wallace}},\
  }\bibfield  {title} {\bibinfo {title} {{Dynamics and statistical mechanics of
  the Hopfield model}},\ }\href {https://doi.org/10.1088/0305-4470/20/10/035}
  {\bibfield  {journal} {\bibinfo  {journal} {J. Phys. A: Math. Gen.}\ }\textbf
  {\bibinfo {volume} {20}},\ \bibinfo {pages} {2909} (\bibinfo {year}
  {1987})}\BibitemShut {NoStop}%
\end{thebibliography}
\end{document}